\input harvmac
\overfullrule=0pt

\def\frac#1#2{{#1\over #2}}
\def\coeff#1#2{{\textstyle{#1\over #2}}}
\def\half{\frac12}
\def\sst{\scriptscriptstyle}

\def\L{Liouville}
\def\M{M}

\def\ap{\alpha'}
\def\One{{1\hskip -3pt {\rm l}}}
\def\ch{{\rm ch}}
\def\sh{{\rm sh}}
\def\RR{{R\!R}}
\def\NS{{N\!S}}
\def\chat{\hat c}
\def\IZ{Z}
\def\IR{{\bf R}}
\def\FL{{{\bf F}_{\!L}}}

\def\mubare{\mu_0}
\def\mub{\mu^{ }_{\!B}}
\def\mubsq{\mu^{2}_{\!B}}
\def\mukpz{\mu}

\lref\fzz{V.~Fateev, A.~B.~Zamolodchikov and A.~B.~Zamolodchikov,
``Boundary Liouville field theory. I: Boundary state and boundary
two-point function,'' arXiv:hep-th/0001012.
}

\lref\ponsot{B.~Ponsot and J.~Teschner, ``Boundary Liouville field
theory: Boundary three point function,'' Nucl.\ Phys.\ B {\bf
622}, 309 (2002) [arXiv:hep-th/0110244].
}

\lref\KostovXW{ I.~K.~Kostov, ``Solvable statistical models on a
random lattice,'' Nucl.\ Phys.\ Proc.\ Suppl.\  {\bf 45A}, 13
(1996) [arXiv:hep-th/9509124].
}

\lref\ArvisKD{ J.~F.~Arvis, ``Spectrum Of The Supersymmetric
Liouville Theory,'' Nucl.\ Phys.\ B {\bf 218}, 309 (1983).
}

\lref\DHokerZY{ E.~D'Hoker, ``Classical And Quantal Supersymmetric
Liouville Theory,'' Phys.\ Rev.\ D {\bf 28}, 1346 (1983).
}

\lref\DijkgraafPP{ R.~Dijkgraaf, S.~Gukov, V.~A.~Kazakov and
C.~Vafa, ``Perturbative analysis of gauged matrix models,''
arXiv:hep-th/0210238.
}

\lref\AharonyUB{ O.~Aharony, M.~Berkooz, D.~Kutasov and
N.~Seiberg, ``Linear dilatons, NS5-branes and holography,'' JHEP
{\bf 9810}, 004 (1998) [arXiv:hep-th/9808149].
}

\lref\Pol{ A.~M.~Polyakov, ``Gauge fields and space-time,'' Int.\
J.\ Mod.\ Phys.\ A {\bf 17S1}, 119 (2002) [arXiv:hep-th/0110196].
}

\lref\KutasovUA{ D.~Kutasov and N.~Seiberg, ``Noncritical
Superstrings,'' Phys.\ Lett.\ B {\bf 251}, 67 (1990).
}

\lref\ZamolodchikovVX{ A.~B.~Zamolodchikov,
``On The Entropy Of Random Surfaces,''
Phys.\ Lett.\ B {\bf 117}, 87 (1982).
}

\lref\zz{
A.~B.~Zamolodchikov and A.~B.~Zamolodchikov,
``Liouville field theory on a pseudosphere,''
arXiv:hep-th/0101152.
}

\lref\gk{ D.~J.~Gross and I.~R.~Klebanov, ``Fermionic String Field
Theory Of $c = 1$ Two-Dimensional Quantum Gravity,'' Nucl.\ Phys.\ B
{\bf 352}, 671 (1991).
}

\lref\igor{ I.~R.~Klebanov, ``String theory in two-dimensions,''
arXiv:hep-th/9108019.
}

\lref\dj{ S.~R.~Das and A.~Jevicki, ``String Field Theory And
Physical Interpretation of $d = 1$ Strings,'' Mod.\ Phys.\ Lett.\ A
{\bf 5}, 1639 (1990).
}

\lref\jp{ J.~Polchinski, ``Classical Limit Of (1+1)-Dimensional
String Theory,'' Nucl.\ Phys.\ B {\bf 362}, 125 (1991).
}

\lref\mgv{ J.~McGreevy and H.~Verlinde, ``Strings from tachyons:
The $c = 1$ matrix reloaded,'' arXiv:hep-th/0304224.
}

\lref\llm{ N.~Lambert, H.~Liu and J.~Maldacena, ``Closed strings
from decaying D-branes,'' arXiv:hep-th/0303139.
}

\lref\BIPZ{
E.~Brezin, C.~Itzykson, G.~Parisi and J.~B.~Zuber,
``Planar Diagrams,''
Commun.\ Math.\ Phys.\  {\bf 59}, 35 (1978).
}

\lref\joe{J.~Polchinski, ``What is String Theory?''
arXiv:hep-th/9411028
}

\lref\ginsparg{ P.~Ginsparg and G.~W.~Moore, ``Lectures On 2-D
Gravity And 2-D String Theory,'' arXiv:hep-th/9304011.
}

 \lref\Pol{ A.~M.~Polyakov, ``Gauge fields and space-time,''
Int.\ J.\ Mod.\ Phys.\ A {\bf 17S1}, 119 (2002)
[arXiv:hep-th/0110196].
}

\lref\ortin{ P.~Meessen and T.~Ortin, ``Type 0 T-duality and the
tachyon coupling,'' Phys.\ Rev.\ D {\bf 64}, 126005 (2001)
[arXiv:hep-th/0103244].
}
\lref\shenker{ S.~H.~Shenker, ``The Strength Of Nonperturbative
Effects In String Theory,'' RU-90-47
{\it Presented at the Cargese Workshop on Random Surfaces,
Quantum Gravity and Strings, Cargese, France, May 28 - Jun 1, 1990}
}

\lref\gaiotto{ D.~Gaiotto, N.~Itzhaki and L.~Rastelli, ``Closed
strings as imaginary D-branes,'' arXiv:hep-th/0304192.
}

\lref\Polch{ J.~Polchinski, ``Combinatorics Of Boundaries In
String Theory,'' Phys.\ Rev.\ D {\bf 50}, 6041 (1994)
[arXiv:hep-th/9407031].
}

 \lref\Green{ M.~B.~Green, ``A Gas of D instantons,'' Phys.\
Lett.\ B {\bf 354}, 271 (1995) [arXiv:hep-th/9504108].
}

\lref\SenNX{ A.~Sen and B.~Zwiebach, ``Tachyon condensation in
string field theory,'' JHEP {\bf 0003}, 002 (2000)
[arXiv:hep-th/9912249].
}

 \lref\gregk{ D.~Kutasov, M.~Marino and G.~W.~Moore, ``Some exact
results on tachyon condensation in string field theory,'' JHEP
{\bf 0010}, 045 (2000) [arXiv:hep-th/0009148].
}

 \lref\GerasimovZP{ A.~A.~Gerasimov and S.~L.~Shatashvili, ``On
exact tachyon potential in open string field theory,'' JHEP {\bf
0010}, 034 (2000) [arXiv:hep-th/0009103].
}

\lref\GutperleAI{
M.~Gutperle and A.~Strominger,
``Spacelike branes,''
JHEP {\bf 0204}, 018 (2002)
[arXiv:hep-th/0202210].
}

\lref\PolchinskiJP{ J.~Polchinski, ``On the nonperturbative
consistency of $d = 2$ string theory,'' Phys.\ Rev.\ Lett.\  {\bf
74}, 638 (1995) [arXiv:hep-th/9409168].
}

\lref\ZwiebachIE{
B.~Zwiebach,
``Closed string field theory: Quantum action and the B-V master equation,''
Nucl.\ Phys.\ B {\bf 390}, 33 (1993)
[arXiv:hep-th/9206084].
}

\lref\DiVecchiaPR{
P.~Di Vecchia, M.~Frau, I.~Pesando, S.~Sciuto, A.~Lerda and R.~Russo,
``Classical p-branes from boundary state,''
Nucl.\ Phys.\ B {\bf 507}, 259 (1997)
[arXiv:hep-th/9707068].
}

\lref\zuber{
C.~Itzykson and J.~B.~Zuber,
``Quantum Field Theory,''
}

\lref\GarousiTR{
M.~R.~Garousi,
``Tachyon couplings on non-BPS D-branes and Dirac-Born-Infeld action,''
Nucl.\ Phys.\ B {\bf 584}, 284 (2000)
[arXiv:hep-th/0003122].
}

\lref\SenMD{
A.~Sen,
``Supersymmetric world-volume action for non-BPS D-branes,''
JHEP {\bf 9910}, 008 (1999)
[arXiv:hep-th/9909062].
}

\lref\MoellerVX{
N.~Moeller and B.~Zwiebach,
``Dynamics with infinitely many time derivatives and rolling tachyons,''
JHEP {\bf 0210}, 034 (2002)
[arXiv:hep-th/0207107].
}

\lref\korean{
C.~Ahn, C.~Rim and M.~Stanishkov,
``Exact one-point function of $N = 1$ super-Liouville theory with boundary,''
Nucl.\ Phys.\ B {\bf 636}, 497 (2002)
[arXiv:hep-th/0202043].
}

\lref\gsw{
M.~B.~Green, J.~H.~Schwarz and E.~Witten,
``Superstring Theory. Vol. 1: Introduction,''}

\lref\SenNU{ A.~Sen, ``Rolling tachyon,'' JHEP {\bf 0204}, 048
(2002) [arXiv:hep-th/0203211].
}

\lref\SenIN{ A.~Sen, ``Tachyon matter,'' JHEP {\bf 0207}, 065
(2002) [arXiv:hep-th/0203265].
}

\lref\SenTM{ A.~Sen, ``Dirac-Born-Infeld Action on the Tachyon
Kink and Vortex,'' [arXiv:hep-th/0303057].
}

\lref\SenQA{ A.~Sen, ``Time and tachyon,'' [arXiv:hep-th/0209122].
}

\lref\SenAN{ A.~Sen, ``Field theory of tachyon matter,'' Mod.\
Phys.\ Lett.\ A {\bf 17}, 1797 (2002) [arXiv:hep-th/0204143].
}

\lref\PolyakovJU{ A.~M.~Polyakov, ``The wall of the cave,'' Int.\
J.\ Mod.\ Phys.\ A {\bf 14}, 645 (1999) [arXiv:hep-th/9809057].
}

\lref\SenVV{ A.~Sen, ``Time evolution in open string theory,''
JHEP {\bf 0210}, 003 (2002) [arXiv:hep-th/0207105].
}

\lref\ChenFP{ B.~Chen, M.~Li and F.~L.~Lin, ``Gravitational
radiation of rolling tachyon,'' JHEP {\bf 0211}, 050 (2002)
[arXiv:hep-th/0209222].
}

\lref\SenMS{
P.~Mukhopadhyay and A.~Sen,
``Decay of unstable D-branes with electric field,''
JHEP {\bf 0211}, 047 (2002)
[arXiv:hep-th/0208142].
}

\lref\Rey{S.~J.~Rey and S.~Sugimoto,
``Rolling Tachyon with Electric and Magnetic Fields -- T-duality approach,''
[arXiv:hep-th/0301049].
}

\lref\GutperleAI{ M.~Gutperle and A.~Strominger, ``Spacelike
branes,'' JHEP {\bf 0204}, 018 (2002) [arXiv:hep-th/0202210].
}

\lref\StromingerPC{ A.~Strominger, ``Open string creation by
S-branes,'' [arXiv:hep-th/0209090].
}

\lref\GutperleBL{M.~Gutperle and A.~Strominger,
``Timelike Boundary Liouville Theory,''
[arXiv:hep-th/0301038].
}

\lref\WaldWT{
R.~M.~Wald,
``Existence Of The S Matrix In Quantum Field Theory In Curved Space-Time,''
Annals Phys.\  {\bf 118}, 490 (1979).
}

\lref\Larsen{F.~Larsen, A.~Naqvi and S.~Terashima,
``Rolling tachyons and decaying branes,''
[arXiv:hep-th/0212248].
}
\lref\krauslarsen{
B.~Craps, P.~Kraus and F.~Larsen,
``Loop corrected tachyon condensation,''
JHEP {\bf 0106}, 062 (2001)
[arXiv:hep-th/0105227].
}

\lref\weinberg{
S. Weinberg,  ``Gravitation and Cosmology''.
}

\lref\Callan{
C.~G.~Callan, I.~R.~Klebanov, A.~W.~Ludwig and J.~M.~Maldacena,
``Exact solution of a boundary conformal field theory,''
Nucl.\ Phys.\ B {\bf 422}, 417 (1994)
[arXiv:hep-th/9402113].
}

 \lref\larus{ J.~Polchinski and L.~Thorlacius, ``Free Fermion
Representation Of A Boundary Conformal Field Theory,'' Phys.\
Rev.\ D {\bf 50}, 622 (1994) [arXiv:hep-th/9404008].
}

 \lref\teschner{ J.~Teschner, ``Remarks on Liouville theory with
boundary,'' arXiv:hep-th/0009138.
}

\lref\RS{
A.~Recknagel and V.~Schomerus,
``Boundary deformation theory and moduli spaces of D-branes,''
Nucl.\ Phys.\ B {\bf 545}, 233 (1999)
[arXiv:hep-th/9811237].
}

\lref\SenMG{
A.~Sen,
``Non-BPS states and branes in string theory,''
[arXiv:hep-th/9904207].
}

\lref\juan{J.~Maldacena, Unpublished}

\lref\shiraz{S.~Minwalla and K.~Pappododimas, Unpublished}

\lref\hwang{
S.~Hwang,
``Cosets as gauge slices in SU(1,1) strings,''
Phys.\ Lett.\ B {\bf 276}, 451 (1992)
[arXiv:hep-th/9110039];
J.~M.~Evans, M.~R.~Gaberdiel and M.~J.~Perry,
``The no-ghost theorem for AdS(3) and the stringy exclusion principle,''
Nucl.\ Phys.\ B {\bf 535}, 152 (1998)
[arXiv:hep-th/9806024].
}

\lref\PolchinskiRQ{
J.~Polchinski,
``String Theory. Vol. 1: An Introduction To The Bosonic String,''
}

\lref\polchinskitorus{
J.~Polchinski,
``Evaluation Of The One Loop String Path Integral,''
Commun.\ Math.\ Phys.\  {\bf 104}, 37 (1986).
}

\lref\PolchinskiRR{
J.~Polchinski,
``String Theory. Vol. 2: Superstring Theory And Beyond,''
Cambridge University Press (1998).
}

\lref\GurarieXQ{
V.~Gurarie,
``Logarithmic operators in conformal field theory,''
Nucl.\ Phys.\ B {\bf 410}, 535 (1993)
[arXiv:hep-th/9303160].
}

\lref\fs{
W.~Fischler and L.~Susskind,
``Dilaton Tadpoles, String Condensates And Scale Invariance,''
Phys.\ Lett.\ B {\bf 171}, 383 (1986);
``Dilaton Tadpoles, String Condensates And Scale Invariance. 2,''
Phys.\ Lett.\ B {\bf 173}, 262 (1986).
}

\lref\MaloneyCK{
A.~Maloney, A.~Strominger and X.~Yin,
``S-brane thermodynamics,''
[arXiv:hep-th/0302146].
}

\lref\BuchelTJ{
A.~Buchel, P.~Langfelder and J.~Walcher,
``Does the tachyon matter?,''
Annals Phys.\  {\bf 302}, 78 (2002)
[arXiv:hep-th/0207235];
A.~Buchel and J.~Walcher,
``The tachyon does matter,''
arXiv:hep-th/0212150.
}

\lref\LeblondDB{ F.~Leblond and A.~W.~Peet,
``SD-brane gravity fields and rolling tachyons,''
[arXiv:hep-th/0303035].
}

\lref\FischlerJA{
W.~Fischler, S.~Paban and M.~Rozali,
``Collective Coordinates for D-branes,''
Phys.\ Lett.\ B {\bf 381}, 62 (1996)
[arXiv:hep-th/9604014];
``Collective coordinates in string theory,''
Phys.\ Lett.\ B {\bf 352}, 298 (1995)
[arXiv:hep-th/9503072].
}

\lref\PO{
V.~Periwal and O.~Tafjord,
``D-brane recoil,''
Phys.\ Rev.\ D {\bf 54}, 3690 (1996), arXiv:hep-th/9603156.
}

\lref\KutasovER{ D.~Kutasov and V.~Niarchos, ``Tachyon effective
actions in open string theory,'' arXiv:hep-th/0304045.
}

\lref\OkuyamaWM{ K.~Okuyama, ``Wess-Zumino term in tachyon
effective action,'' arXiv:hep-th/0304108.
}

\lref\KMW{
J.~S.~Caux, I.~I.~Kogan and A.~M.~Tsvelik,
``Logarithmic Operators and Hidden Continuous Symmetry
in Critical Disordered Models,''
Nucl.\ Phys.\ B {\bf 466}, 444 (1996)
[arXiv:hep-th/9511134];
I.~I.~Kogan and N.~E.~Mavromatos,
``World-Sheet Logarithmic Operators
and Target Space Symmetries in String Theory,''
Phys.\ Lett.\ B {\bf 375}, 111 (1996)
[arXiv:hep-th/9512210];
I.~I.~Kogan, N.~E.~Mavromatos and J.~F.~Wheater,
``D-brane recoil and logarithmic operators,''
Phys.\ Lett.\ B {\bf 387}, 483 (1996),
arXiv:hep-th/9606102.
}

\lref\FlohrZS{
M.~Flohr,
``Bits and pieces in logarithmic conformal field theory,''
arXiv:hep-th/0111228.
}

\lref\GaberdielTR{
M.~R.~Gaberdiel,
``An algebraic approach to logarithmic conformal field theory,''
arXiv:hep-th/0111260.
}

\lref\GaiottoRM{ D.~Gaiotto, N.~Itzhaki and L.~Rastelli, ``Closed
strings as imaginary D-branes,'' arXiv:hep-th/0304192.
}

\lref\OkudaYD{ T.~Okuda and S.~Sugimoto, ``Coupling of rolling
tachyon to closed strings,'' Nucl.\ Phys.\ B {\bf 647}, 101 (2002)
[arXiv:hep-th/0208196].
}

\lref\Aref{
I.~Y.~Aref'eva, L.~V.~Joukovskaya and A.~S.~Koshelev,
``Time evolution in superstring field theory on non-BPS brane.
I: Rolling  tachyon and energy-momentum
conservation,''
arXiv:hep-th/0301137.
}

\lref\IshidaCJ{
A.~Ishida and S.~Uehara,
``Rolling down to D-brane and tachyon matter,''
JHEP {\bf 0302}, 050 (2003)
[arXiv:hep-th/0301179].
}

\lref\KlusonAV{
J.~Kluson,
``Time dependent solution in open Bosonic string field theory,''
arXiv:hep-th/0208028;
``Exact solutions in open Bosonic string field theory and
marginal  deformation in CFT,''
[arXiv:hep-th/0209255].
}

\lref\MinahanIF{
J.~A.~Minahan,
``Rolling the tachyon in super BSFT,''
JHEP {\bf 0207}, 030 (2002)
[arXiv:hep-th/0205098].
}

\lref\SugimotoFP{
S.~Sugimoto and S.~Terashima,
``Tachyon matter in boundary string field theory,''
JHEP {\bf 0207}, 025 (2002)
[arXiv:hep-th/0205085].
}

\lref\WittenYJ{ E.~Witten and B.~Zwiebach, ``Algebraic structures
and differential geometry in $2-D$ string theory,'' Nucl.\ Phys.\
B {\bf 377}, 55 (1992) [arXiv:hep-th/9201056].
}

\lref\KlebanovUI{ I.~R.~Klebanov and A.~Pasquinucci, ``Correlation
functions from two-dimensional string ward identities,'' Nucl.\
Phys.\ B {\bf 393}, 261 (1993) [arXiv:hep-th/9204052].
I.~R.~Klebanov and A.~M.~Polyakov, ``Interaction of discrete
states in two-dimensional string theory,'' Mod.\ Phys.\ Lett.\ A
{\bf 6}, 3273 (1991) [arXiv:hep-th/9109032].
}

\lref\ReyXS{
S.~J.~Rey and S.~Sugimoto,
``Rolling tachyon with electric and magnetic fields: T-duality approach,''
arXiv:hep-th/0301049.
}

\lref\rastelli{  D.~Gaiotto, N.~Itzhaki and L.~Rastelli, ``Closed
strings as imaginary D-branes,'' arXiv:hep-th/0304192.
}

\lref\LambertHK{
N.~D.~Lambert and I.~Sachs,
``Tachyon dynamics and the effective action approximation,''
Phys.\ Rev.\ D {\bf 67}, 026005 (2003)
[arXiv:hep-th/0208217].
}

\lref\HarveyQU{ J.~A.~Harvey, P.~Horava and P.~Kraus,
``D-sphalerons and the topology of string configuration space,''
JHEP {\bf 0003}, 021 (2000) [arXiv:hep-th/0001143].
}

\lref\DrukkerWX{ N.~Drukker, D.~J.~Gross and N.~Itzhaki,
``Sphalerons, merons and unstable branes in AdS,'' Phys.\ Rev.\ D
{\bf 62}, 086007 (2000) [arXiv:hep-th/0004131].
}

\lref\yang{ Z.~Yang, ``Dynamical Loops In $D = 1$ Random Matrix
Models,'' Phys.\ Lett.\ B {\bf 257}, 40 (1991).
}

\lref\kk{ V.~A.~Kazakov and I.~K.~Kostov, ``Loop gas model for
open strings,'' Nucl.\ Phys.\ B {\bf 386}, 520 (1992)
[arXiv:hep-th/9205059].
}

\lref\dgh{
L.~J.~Dixon, P.~Ginsparg and J.~A.~Harvey,
``$\hat c = 1$ Superconformal Field Theory,''
Nucl.\ Phys.\ B {\bf 306}, 470 (1988).
}

\lref\BerKleb{
M.~Bershadsky and I.~R.~Klebanov,
``Genus One Path Integral In Two-Dimensional Quantum Gravity,''
Phys.\ Rev.\ Lett.\  {\bf 65}, 3088 (1990);
N.~Sakai and Y.~Tanii,
``Compact Boson Coupled To Two-Dimensional Gravity,''
Int.\ J.\ Mod.\ Phys.\ A {\bf 6}, 2743 (1991).
}

\lref\GrossKleb{
D.~J.~Gross and I.~R.~Klebanov,
``One-Dimensional String Theory On A Circle,''
Nucl.\ Phys.\ B {\bf 344}, 475 (1990).
}

\lref\KlebL{
I.~R.~Klebanov and D.~A.~Lowe,
``Correlation functions in two-dimensional quantum gravity
coupled to a compact scalar field,''
Nucl.\ Phys.\ B {\bf 363}, 543 (1991).
}

\lref\fran{
P.~Di Francesco, H.~Saleur and J.~B.~Zuber,
``Generalized Coulomb Gas Formalism
For Two-Dimensional Critical Models Based On SU(2) Coset Construction,''
Nucl.\ Phys.\ B {\bf 300}, 393 (1988).
}

\lref\PeriwalGF{
V.~Periwal and D.~Shevitz,
``Unitary Matrix Models As Exactly Solvable String Theories,''
Phys.\ Rev.\ Lett.\  {\bf 64}, 1326 (1990).
}

\lref\GrossHE{
D.~J.~Gross and E.~Witten,
``Possible Third Order Phase Transition In The Large N Lattice Gauge Theory,''
Phys.\ Rev.\ D {\bf 21}, 446 (1980).
}

\lref\BerKlebnew{
M.~Bershadsky and I.~R.~Klebanov,
``Partition functions and physical states
in two-dimensional quantum gravity and supergravity,''
Nucl.\ Phys.\ B {\bf 360}, 559 (1991).
}

\lref\AlvarezGaumeJD{
L.~Alvarez-Gaume, H.~Itoyama, J.~L.~Manes and A.~Zadra,
``Superloop equations and two-dimensional supergravity,''
Int.\ J.\ Mod.\ Phys.\ A {\bf 7}, 5337 (1992)
[arXiv:hep-th/9112018].
}

\lref\KT{
I.~R.~Klebanov and A.~A.~Tseytlin,
``D-branes and dual gauge theories in type 0 strings,''
Nucl.\ Phys.\ B {\bf 546}, 155 (1999)
[arXiv:hep-th/9811035].
}

\lref\Dudas{
E.~Dudas, J.~Mourad and A.~Sagnotti,
``Charged and uncharged D-branes in various string theories,''
Nucl.\ Phys.\ B {\bf 620}, 109 (2002)
[arXiv:hep-th/0107081].
}

\lref\Morris{
T.~R.~Morris,
``Checkered Surfaces And Complex Matrices,''
Nucl.\ Phys.\ B {\bf 356}, 703 (1991).
}

\lref\Dalley{
S.~Dalley,
``The Weingarten model a la Polyakov,''
Mod.\ Phys.\ Lett.\ A {\bf 7}, 1651 (1992)
[arXiv:hep-th/9203019].
}

\lref\KlebanovKM{
I.~R.~Klebanov, J.~Maldacena and N.~Seiberg,
``D-brane decay in two-dimensional string theory,''
arXiv:hep-th/0305159.
}

\lref\McGreevyEP{
J.~McGreevy, J.~Teschner and H.~Verlinde,
``Classical and quantum D-branes in 2D string theory,''
arXiv:hep-th/0305194.
}

\lref\DiFrancescoUD{
P.~Di Francesco and D.~Kutasov,
``World sheet and space-time physics in two-dimensional (Super)string theory,''
Nucl.\ Phys.\ B {\bf 375}, 119 (1992)
[arXiv:hep-th/9109005].
}

\lref\unpub{ M. Douglas, D. Kutasov, E. Martinec, N. Seiberg,
``A Noncritical Fermionic String Ansatz,'' unpublished (1992).}

\lref\MartinecKA{
E.~J.~Martinec,
``The annular report on non-critical string theory,''
arXiv:hep-th/0305148.
}

\lref\AldazabalAE{ G.~Aldazabal, M.~Bonini and J.~M.~Maldacena,
``Factorization And Discrete States In C = 1 Superliouville
Theory,'' Int.\ J.\ Mod.\ Phys.\ A {\bf 9}, 3969 (1994)
[arXiv:hep-th/9209010].
}

\lref\AlexandrovNN{
S.~Y.~Alexandrov, V.~A.~Kazakov and D.~Kutasov,
``Non-Perturbative Effects in Matrix Models and D-branes,''
arXiv:hep-th/0306177.
}

\lref\PeriwalQB{ V.~Periwal and D.~Shevitz, ``Exactly Solvable
Unitary Matrix Models: Multicritical Potentials And
Correlations,'' Nucl.\ Phys.\ B {\bf 344}, 731 (1990).
}
\lref\CrnkovicMS{
C.~Crnkovic, M.~R.~Douglas and G.~W.~Moore,
``Physical Solutions For Unitary Matrix Models,''
Nucl.\ Phys.\ B {\bf 360}, 507 (1991).
}
\lref\Crn{ C.~Crnkovic, M.~R.~Douglas and G.~W.~Moore,
``Loop equations and the topological phase of multi-cut matrix
models,'' Int.\ J.\ Mod.\ Phys.\ A {\bf 7}, 7693 (1992)
[arXiv:hep-th/9108014].
}
\lref\teschner{ J.~Teschner, ``Remarks on Liouville theory with
boundary,'' arXiv:hep-th/0009138.
}
\lref\GubKleb{
F.~Sugino and O.~Tsuchiya,
``Critical behavior in $c = 1$ matrix model with branching interactions,''
Mod.\ Phys.\ Lett.\ A {\bf 9}, 3149 (1994)
[arXiv:hep-th/9403089];
S.~S.~Gubser and I.~R.~Klebanov,
``A Modified $c = 1$ matrix model with new critical behavior,''
Phys.\ Lett.\ B {\bf 340}, 35 (1994)
[arXiv:hep-th/9407014];
I.~R.~Klebanov and A.~Hashimoto,
``Nonperturbative solution of matrix models modified by trace squared terms,''
Nucl.\ Phys.\ B {\bf 434}, 264 (1995)
[arXiv:hep-th/9409064].
}

\lref\DHokerZY{
E.~D'Hoker,
``Classical And Quantal Supersymmetric Liouville Theory,''
Phys.\ Rev.\ D {\bf 28}, 1346 (1983).
}

\lref\RashkovJX{
R.~C.~Rashkov and M.~Stanishkov,
``Three-point correlation functions in $N=1$ Super Liouville Theory,''
Phys.\ Lett.\ B {\bf 380}, 49 (1996)
[arXiv:hep-th/9602148].
}

\lref\PoghosianDW{
R.~H.~Poghosian,
``Structure constants in the $N=1$ super-Liouville field theory,''
Nucl.\ Phys.\ B {\bf 496}, 451 (1997)
[arXiv:hep-th/9607120].
}

\lref\MartinecHT{
E.~J.~Martinec, G.~W.~Moore and N.~Seiberg,
``Boundary operators in 2-D gravity,''
Phys.\ Lett.\ B {\bf 263}, 190 (1991).
}

\lref\DornXN{
H.~Dorn and H.~J.~Otto,
``Two and three point functions in Liouville theory,''
Nucl.\ Phys.\ B {\bf 429}, 375 (1994)
[arXiv:hep-th/9403141].
}

\lref\ZamolodchikovAA{
A.~B.~Zamolodchikov and A.~B.~Zamolodchikov,
``Structure constants and conformal bootstrap in Liouville field theory,''
Nucl.\ Phys.\ B {\bf 477}, 577 (1996)
[arXiv:hep-th/9506136].
}

\lref\WittenZD{
E.~Witten,
``Ground ring of two-dimensional string theory,''
Nucl.\ Phys.\ B {\bf 373}, 187 (1992)
[arXiv:hep-th/9108004].
}

\lref\BouwknegtYG{
P.~Bouwknegt, J.~G.~McCarthy and K.~Pilch,
``BRST analysis of physical states for 2-D gravity coupled to
$c \le 1$ matter,''
Commun.\ Math.\ Phys.\  {\bf 145}, 541 (1992);
C.~Imbimbo, S.~Mahapatra and S.~Mukhi,
``Construction of physical states of nontrivial ghost number
in $c < 1$ string theory,''
Nucl.\ Phys.\ B {\bf 375}, 399 (1992).
}

\lref\BouwknegtAM{
P.~Bouwknegt, J.~G.~McCarthy and K.~Pilch,
``Ground ring for the 2-D NSR string,''
Nucl.\ Phys.\ B {\bf 377}, 541 (1992)
[arXiv:hep-th/9112036].
}

\lref\KutasovQX{
D.~Kutasov, E.~J.~Martinec and N.~Seiberg,
``Ground rings and their modules in 2-D gravity with $c\le 1$ matter,''
Phys.\ Lett.\ B {\bf 276}, 437 (1992)
[arXiv:hep-th/9111048].
}

\lref\BershadskyUB{
M.~Bershadsky and D.~Kutasov,
``Scattering of open and closed strings in (1+1)-dimensions,''
Nucl.\ Phys.\ B {\bf 382}, 213 (1992)
[arXiv:hep-th/9204049].
}

\lref\PolyakovRD{
A.~M.~Polyakov,
``Quantum Geometry of Bosonic Strings,''
Phys.\ Lett.\ B {\bf 103}, 207 (1981).
}

\lref\PolyakovRE{
A.~M.~Polyakov,
``Quantum Geometry Of Fermionic Strings,''
Phys.\ Lett.\ B {\bf 103}, 211 (1981).
}

\lref\KutasovPF{
D.~Kutasov,
``Irreversibility of the renormalization group flow in two-dimensional quantum gravity,''
Mod.\ Phys.\ Lett.\ A {\bf 7}, 2943 (1992)
[arXiv:hep-th/9207064];
E.~Hsu and D.~Kutasov,
``The Gravitational Sine-Gordon model,''
Nucl.\ Phys.\ B {\bf 396}, 693 (1993)
[arXiv:hep-th/9212023].
}

\lref\SeibergEB{ N.~Seiberg, ``Notes On Quantum Liouville Theory
And Quantum Gravity,'' Prog.\ Theor.\ Phys.\ Suppl.\  {\bf 102},
319 (1990).
}

\lref\PolchinskiMH{ J.~Polchinski, ``Remarks On The Liouville
Field Theory,'' UTTG-19-90
{\it Presented at Strings '90 Conf., College Station, TX, Mar
12-17, 1990} }


\lref\MooreIR{ G.~W.~Moore, N.~Seiberg and M.~Staudacher, ``From
loops to states in 2-D quantum gravity,'' Nucl.\ Phys.\ B {\bf
362}, 665 (1991).
}

\lref\MooreAG{ G.~W.~Moore and N.~Seiberg, ``From loops to fields
in 2-D quantum gravity,'' Int.\ J.\ Mod.\ Phys.\ A {\bf 7}, 2601
(1992).
}

\lref\mooredouglas{
M.~R.~Douglas and G.~W.~Moore,
``D-branes, Quivers, and ALE Instantons,''
arXiv:hep-th/9603167.
}

\lref\ArvisTQ{
J.~F.~Arvis,
``Classical Dynamics Of The Supersymmetric Liouville Theory,''
Nucl.\ Phys.\ B {\bf 212}, 151 (1983).
}

\lref\DiVecchiaBZ{
P.~Di Vecchia, B.~Durhuus, P.~Olesen and J.~L.~Petersen,
``Fermionic Strings With Boundary Terms,''
Nucl.\ Phys.\ B {\bf 207}, 77 (1982).
}

\lref\FukudaBV{
T.~Fukuda and K.~Hosomichi,
``Super Liouville theory with boundary,''
Nucl.\ Phys.\ B {\bf 635}, 215 (2002)
[arXiv:hep-th/0202032].
}

\lref\KostovWV{
I.~K.~Kostov,
``String equation for string theory on a circle,''
Nucl.\ Phys.\ B {\bf 624}, 146 (2002)
[arXiv:hep-th/0107247].
}

\lref\MooreSF{
G.~W.~Moore,
``Double scaled field theory at c = 1,''
Nucl.\ Phys.\ B {\bf 368}, 557 (1992).
}

\lref\MooreZV{
G.~W.~Moore, M.~R.~Plesser and S.~Ramgoolam,
``Exact S matrix for 2-D string theory,''
Nucl.\ Phys.\ B {\bf 377}, 143 (1992)
[arXiv:hep-th/9111035].
}

\lref\DharGW{
A.~Dhar, G.~Mandal and S.~R.~Wadia,
``Discrete state moduli of string theory from the c=1 matrix model,''
Nucl.\ Phys.\ B {\bf 454}, 541 (1995)
[arXiv:hep-th/9507041];
A.~Dhar,
``The emergence of space-time gravitational physics
as an effective  theory from the c = 1 matrix model,''
Nucl.\ Phys.\ B {\bf 507}, 277 (1997)
[arXiv:hep-th/9705215].
}

\lref\MooreGA{
G.~W.~Moore,
``Gravitational phase transitions and the Sine-Gordon model,''
arXiv:hep-th/9203061.
}

\lref\BergmanKM{
O.~Bergman and M.~R.~Gaberdiel,
``Dualities of type 0 strings,''
JHEP {\bf 9907}, 022 (1999)
[arXiv:hep-th/9906055].
}

\lref\BilloNF{
M.~Billo, B.~Craps and F.~Roose,
``On D-branes in type 0 string theory,''
Phys.\ Lett.\ B {\bf 457}, 61 (1999)
[arXiv:hep-th/9902196].
}

\lref\BershadskyAY{
M.~Bershadsky and D.~Kutasov,
``Open string theory in (1+1)-dimensions,''
Phys.\ Lett.\ B {\bf 274}, 331 (1992)
[arXiv:hep-th/9110034].
}

\lref\BerkovitsTG{ N.~Berkovits, S.~Gukov and B.~C.~Vallilo,
``Superstrings in 2D backgrounds with R-R flux and new extremal
black  holes,'' Nucl.\ Phys.\ B {\bf 614}, 195 (2001)
[arXiv:hep-th/0107140].
}

\lref\DiFrancescoRU{ P.~Di Francesco, ``Rectangular Matrix Models
and Combinatorics of Colored Graphs,'' Nucl.\ Phys.\ B {\bf 648},
461 (2003) [arXiv:cond-mat/0208037].
}

\lref\TakayanagiSM{
T.~Takayanagi and N.~Toumbas,
``A Matrix Model Dual of Type 0B String Theory in Two Dimensions,''
arXiv:hep-th/0307083.
}

\lref\GrossAY{
D.~J.~Gross and N.~Miljkovic,
``A Nonperturbative Solution of $D = 1$ String Theory,''
Phys.\ Lett.\ B {\bf 238}, 217 (1990);
%
E.~Brezin, V.~A.~Kazakov and A.~B.~Zamolodchikov,
``Scaling Violation in a Field Theory of Closed Strings
in One Physical Dimension,''
Nucl.\ Phys.\ B {\bf 338}, 673 (1990);
%
P.~Ginsparg and J.~Zinn-Justin,
``2-D Gravity + 1-D Matter,''
Phys.\ Lett.\ B {\bf 240}, 333 (1990).
}

\lref\WittenCD{
E.~Witten,
``D-branes and K-theory,''
JHEP {\bf 9812}, 019 (1998)
[arXiv:hep-th/9810188].
}

\lref\TeschnerRV{
J.~Teschner,
``Liouville theory revisited,''
Class.\ Quant.\ Grav.\  {\bf 18}, R153 (2001)
[arXiv:hep-th/0104158].
}

\lref\FriedanGE{
D.~Friedan, E.~J.~Martinec and S.~H.~Shenker,
``Conformal Invariance, Supersymmetry And String Theory,''
Nucl.\ Phys.\ B {\bf 271}, 93 (1986).
}

\lref\PonsotNG{
B.~Ponsot and J.~Teschner,
``Boundary Liouville field theory: Boundary three point function,''
Nucl.\ Phys.\ B {\bf 622}, 309 (2002)
[arXiv:hep-th/0110244].
}

\lref\HosomichiXC{
K.~Hosomichi,
``Bulk-boundary propagator in Liouville theory on a disc,''
JHEP {\bf 0111}, 044 (2001)
[arXiv:hep-th/0108093].
}

\lref\GaberdielXM{
M.~R.~Gaberdiel, A.~Recknagel and G.~M.~Watts,
``The conformal boundary states for SU(2) at level 1,''
Nucl.\ Phys.\ B {\bf 626}, 344 (2002)
[arXiv:hep-th/0108102];
M.~R.~Gaberdiel and A.~Recknagel,
``Conformal boundary states for free bosons and fermions,''
JHEP {\bf 0111}, 016 (2001)
[arXiv:hep-th/0108238].
}

\lref\VafaQF{
C.~Vafa,
``Brane/anti-brane systems and U(N$|$M) supergroup,''
arXiv:hep-th/0101218.
}

\lref\KazakovUE{
V.~Kazakov,
``Bosonic Strings and String Field Theories
in One-Dimensional Target Space,''
LPTENS-90-30
{\it Lecture given at Cargese Workshop, Cargese, France, May 27 - Jun 1,
1990}
}

\lref\BilloTV{
M.~Billo, B.~Craps and F.~Roose,
``Ramond-Ramond couplings of non-BPS D-branes,''
JHEP {\bf 9906}, 033 (1999)
[arXiv:hep-th/9905157];
C.~Kennedy and A.~Wilkins,
``Ramond-Ramond couplings on brane-antibrane systems,''
Phys.\ Lett.\ B {\bf 464}, 206 (1999)
[arXiv:hep-th/9905195].
}

\lref\KazakovCH{ V.~A.~Kazakov and A.~A.~Migdal, ``Recent Progress
In The Theory Of Noncritical Strings,'' Nucl.\ Phys.\ B {\bf 311},
171 (1988).
}

\lref\DasKA{ S.~R.~Das and A.~Jevicki, ``String Field Theory And
Physical Interpretation Of D = 1 Strings,'' Mod.\ Phys.\ Lett.\ A
{\bf 5}, 1639 (1990).
}

\lref\DalleyBR{ S.~Dalley, C.~V.~Johnson, T.~R.~Morris and
A.~Watterstam, ``Unitary matrix models and 2-D quantum gravity,''
Mod.\ Phys.\ Lett.\ A {\bf 7}, 2753 (1992) [arXiv:hep-th/9206060].
}

\lref\DKR{
K.~Demeterfi, I.~R.~Klebanov and J.~P.~Rodrigues,
``The Exact S matrix of the deformed c = 1 matrix model,''
Phys.\ Rev.\ Lett.\  {\bf 71}, 3409 (1993)
[arXiv:hep-th/9308036].
}

\lref\PolchinskiFQ{ J.~Polchinski, ``Combinatorics Of Boundaries
In String Theory,'' Phys.\ Rev.\ D {\bf 50}, 6041 (1994)
[arXiv:hep-th/9407031].
}

\lref\Itoh{
K.~Itoh and N.~Ohta,
``BRST cohomology and physical states in 2-D supergravity coupled to c <= 1 matter,''
Nucl.\ Phys.\ B {\bf 377}, 113 (1992)
[arXiv:hep-th/9110013];
``Spectrum of two-dimensional (super)gravity,''
Prog.\ Theor.\ Phys.\ Suppl.\  {\bf 110}, 97 (1992)
[arXiv:hep-th/9201034].
}

\lref\BouwknegtVA{
P.~Bouwknegt, J.~G.~McCarthy and K.~Pilch,
``BRST analysis of physical states for 2-D (super)gravity coupled to (super)conformal matter,''
arXiv:hep-th/9110031.
}

\lref\BergmanRF{
O.~Bergman and M.~R.~Gaberdiel,
``A non-supersymmetric open-string theory and S-duality,''
Nucl.\ Phys.\ B {\bf 499}, 183 (1997)
[arXiv:hep-th/9701137].
}

\Title{\vbox{\baselineskip12pt \hbox{hep-th/0307195}
\hbox{PUPT-2090}
\hbox{EFI-03-35}
\hbox{RUNHETC-2003-23}
}}
{\vbox{\centerline{A New Hat For The $c=1$ Matrix Model}
}}
\smallskip
\centerline{M. R. Douglas,$^{1,2}$ I. R. Klebanov,$^{3,4}$
D. Kutasov,$^5$
J. Maldacena,$^3$ E. Martinec,$^5$ and N. Seiberg$^3$}
\medskip


\centerline{\it $^1$ Department of Physics, Rutgers University}
\centerline{\it Piscataway, New Jersey 08855-0849, USA}
\smallskip

\centerline{\it $^2$ I.H.E.S., Le Bois-Marie, Bures-sur-Yvette, 91440 France}
\smallskip

\centerline{\it $^3$ Institute for Advanced Study}
\centerline{\it Princeton, New Jersey 08540, USA}
\smallskip

\centerline{\it $^4$ Joseph Henry Laboratories, Princeton University}
\centerline{\it Princeton, New Jersey 08544, USA}
\smallskip

\centerline{\it $^5$ Enrico Fermi Institute, University of
Chicago} \centerline{\it Chicago, IL 60637, USA}

\bigskip
\noindent
We consider two dimensional supergravity coupled to $\hat c=1$ matter.
This system can also be interpreted as noncritical
type 0 string theory in a two dimensional target space.
After reviewing and extending the traditional descriptions of
this class of theories, we provide a matrix model description.
The 0B theory is similar to the realization
of two dimensional bosonic string theory
via matrix quantum mechanics in an inverted harmonic oscillator potential;
the difference is that we expand around a non-perturbatively stable vacuum,
where the matrix eigenvalues are equally distributed
on both sides of the potential.
The 0A theory is described by a quiver matrix model.

\Date{July 2003}

\vfil\eject


\newsec{Introduction}

Matrix quantum mechanics of an $N\times N$ Hermitian matrix
$\M(x)$  describes two-dimensional bosonic string theory (for
reviews see \refs{\igor\ginsparg-\joe}). Briefly, the large $N$
planar graph expansion of the Euclidean matrix path integral
\eqn\EQM{
  {\cal Z}=\int D^{N^2}\M(x)\exp \biggl [-\beta\int_{-\infty}^\infty
    d x~\Tr \left (\half (D_x \M )^2+ V(\M)\right)\biggr ]\ ,
} discretizes a closed string worldsheet embedded in the target
space (Euclidean) time direction parametrized by $x$ \KazakovCH; a
spatial direction grows out of the matrix eigenvalue coordinate
\DasKA. The coupling $\beta$ is the inverse Planck constant in the
matrix quantum mechanics. The continuum (``double scaling'') limit
\refs{\GrossAY}
isolates a quadratic maximum of the potential
\eqn\maxt{
  V (\M) = -{1\over 2\alpha'} \M^2
 \ . }
This formulation of 2-d string theory has long been suspected of
being an example of exact open/closed string duality. Recent work
\refs{\mgv\KlebanovKM\McGreevyEP\MartinecKA-\AlexandrovNN} has
established this idea concretely through computations of D-brane
dynamics in the two-dimensional bosonic string. This progress
builds on studies of holography in linear dilaton backgrounds
(see, e.g. \AharonyUB), and on construction of D-branes in such
backgrounds. Of particular importance for formulating a precise
open/closed string duality, as emphasized in \Pol, is the
discovery of boundary states in Liouville theory localized at
large $\phi$ \zz.

Let us summarize the basic facts about two-dimensional string
theory. The closed string background for two-dimensional strings
consists of a free scalar field $x$, with conformal invariance
maintained by coupling to a Liouville field theory \PolyakovRD,
often thought of as worldsheet gravity. The closed string
``tachyon'' is the principal effective field in spacetime; the
tachyon mass is in fact lifted to zero by a linear dilaton
background, so that the theory is perturbatively stable. There is
in addition a set of discrete physical degrees of freedom which
appear at special momenta.

The bosonic string has unstable D-branes. The open string tachyon
field on $N$ such D0 branes is a Hermitian matrix $\M (x)$. The
curvature of the potential at the quadratic maximum is given by
the open string tachyon mass-squared $m_T^2 =-1/\alpha'$, in
agreement with the matrix model value \maxt\ which was originally
established by comparison with closed string amplitudes \igor.
This serves as a consistency check on the identification of
$\M(x)$ with the tachyon field localized on the D0-branes. The
open string spectrum also includes a non-dynamical gauge field
$A$, corresponding to the vertex operator $\dot x$. It enters the
covariant derivative in \EQM:
 \eqn\cov{ D_x \M= \partial_x \M -
[A,\M] \ . } $A$ acts as a lagrange multiplier that projects onto
$SU(N)$ singlet wave functions which depend on the $N$ matrix
eigenvalues only. After a Vandermonde Jacobian is taken into
account, the eigenvalues act as free fermions \BIPZ. In the
bosonic string theory the Fermi sea is asymmetric; it fills only
one side of the upside down harmonic oscillator potential to Fermi
level $-\mu$ (as measured from the top of the potential).

The proposal is then that the matrix model description simply {\it
is} the effective dynamics of the bosonic string D0 branes, and
that the closed string state that they `decay' into is the
background described by $c=1$ matter coupled to Liouville gravity.
This proposal has been subjected to a number of nontrivial checks,
including the computation of closed string radiation from a
decaying D0 brane \refs{\mgv\KlebanovKM-\McGreevyEP}, as well as
the leading instanton corrections to the partition function
\refs{\KlebanovKM,\MartinecKA,\AlexandrovNN}.

This model is unstable nonperturbatively. Various stable
nonperturbative definitions of the theory were studied in
\refs{\MooreSF\MooreZV-\DharGW} and
were critically discussed in
\joe. One of the attractive ways to stabilize the model is to fill
the two sides of the potential to the same Fermi level. The
authors of \DharGW\
argued that the problems raised in \joe\ could
be avoided in this case.

In this paper we argue that this two-sided matrix model in fact
describes two-dimensional NSR strings. Let us describe the general
structure of this theory. The closed string background is again
described on the worldsheet by Liouville theory coupled to a free
scalar field, now with $N=1$ worldsheet supersymmetry. Here it is
natural to carry out a non-chiral GSO projection, so that we have
type 0 strings. The closed string ``tachyon'' is again lifted to
zero mass by the linear dilaton of Liouville theory, so the theory
is perturbatively stable.

The R-R spectrum depends on the GSO projection. In the type 0B
theory the \hbox{R-R} spectrum consists of a massless scalar
field. There are also discrete physical states appearing at
special momenta. Like the bosonic string, the 0B theory also has
unstable D0 branes whose worldline dynamics is described again by
Hermitian matrix quantum mechanics about an unstable quadratic
maximum. It is natural to conjecture that the same sort of matrix
path integral \EQM\ also describes the 2-d type 0B theory. The
difference is that the open string potential is a stable
double-well, with the ground state
filling up both wells with equal numbers of eigenvalues.%
\foot{A conjecture relating double-well $c=1$
matrix model to ten-dimensional superstrings was made in \mgv.}%
$^{,}$%
\foot{Equivalently, this model can be described
as quantum mechanics of a unitary matrix $U$
with potential $\Tr (U+U^\dagger)$.
After the double scaling limit is taken, both models
are described by fermions in an upside down
harmonic oscillator potential filling both
sides of the potential to Fermi level $-\mu$. }
It is natural to suspect that this theory is
non-perturbatively well-defined.
The eigenvalue density perturbations that are even and odd
under the $Z_2$ reflection symmetry of the potential
correspond to the NS-NS and R-R scalars of type 0B
theory, respectively.

In contrast, the type 0A theory
has no propagating modes in the R-R sector, but has two vector fields
(the two electric fields are examples of discrete states).
To obtain type 0A theory from the type 0B matrix model,
we need to quotient by a $Z_2$ acting in the manner familiar
from quiver gauge theories \mooredouglas.
This leads in general to a matrix
quantum mechanics of a complex $M\times N$ matrix,
with $U(M)\times U(N)$ (gauge) symmetry.

This relation between the matrix model and open strings on
unstable D0-branes of type 0 theory,%
\foot{The spectrum of D-branes
in 10-d type 0 theories was studied in
\refs{\BergmanRF\KT\BilloNF\BergmanKM-\Dudas}.}
provides a clear motivation
for this duality and allows for additional precise checks.

The plan of the paper is as follows. Section 2 introduces
two-dimensional type 0 string theory, and its worldsheet
description as $\chat=1$ superconformal field theory of a free
superfield $X$, whose lowest component is $x$, coupled to
super-Liouville theory. We discuss the possible GSO projections at
finite and infinite radius of $x$, and their spectra.

Sections 3 and 4 explore some perturbative aspects of the theory.
Section 3 investigates the ground ring structure
\refs{\WittenZD\KutasovQX-\BershadskyUB} and tree level S-matrix
of the 2-d bosonic and type 0B strings. The ground ring is a
collection of BRST invariant operators of the worldsheet theory
satisfying a closed operator algebra, which is related to the
phase space of the eigenvalues in the matrix model description.
Recent progress in (super-)Liouville theory enables one to explore
this relation in detail. Section 4 calculates the torus partition
function, which includes an important contribution from the odd
spin structure.

Sections 5, 6, 7 and 8 discuss the D-branes
of the theory from various points of view.
Section 5 contains a semiclassical description of
boundary conditions in super-Liouville theory.
In section 6 we study the minisuperspace approximation.
Section 7 discusses the boundary states of the full quantum theory,
and their relation to the minisuperspace wavefunctions.
The annulus partition function is also considered.
Section 8 discusses some aspects of the worldvolume
dynamics of D-branes and their interactions with
closed string fields.%
\foot{Sections 2-8 review known material
but also describe many new results.
The reader who is impatient to get to the matrix model
can move directly to section 9.}

The precise nature of our proposal for a type 0B matrix model is
presented in section 9,
where we check the torus worldsheet
calculation against a computation of the finite temperature matrix
model free energy. Section 10 introduces the type 0A matrix model
and performs the analogous check.  In section 11 we study the
relation of tachyon condensation and rolling matrix eigenvalues
\refs{\mgv\KlebanovKM-\McGreevyEP} through a calculation of the
disk expectation value of the ground ring generators. We also
explore the radiation of closed strings produced by the decay.
Three appendices contain useful technical results.

It is also possible to show that $\hat c< 1$ superconformal
minimal models coupled to 2-d supergravity are dual to unitary
matrix models,\foot{A suggestion for string duals of the unitary
matrix models, which involves open and closed strings, was made in
\DalleyBR.} as one might expect on the basis of gravitational RG
flow \refs{\MooreGA\KutasovPF-\KostovWV}. These matrix models were
solved in the planar limit in \GrossHE, and in the double scaling
limit in \refs{\PeriwalGF\CrnkovicMS-\Crn}. These models will be
the subject of a separate publication.

\vskip .5cm
\noindent{\it Note Added:} While completing this manuscript,
we learned of work by T. Takayanagi and N. Toumbas
\TakayanagiSM\
where the matrix model for 2d type 0 strings is
also proposed.


\newsec{The 2d Fermionic String}

Fermionic strings are described by $N=1$ supersymmetric world
sheet field theories coupled to worldsheet supergravity. The
construction of type II string theories requires the existence of
a non-anomalous chiral $(-1)^{F_L}$ symmetry of the worldsheet
theory. The generic background may only admit a nonchiral
$(-1)^F$; the use of this class of (GSO) projection gives type 0
string theory (see \PolchinskiRR\ for a review).

There are in fact two distinct choices, depending on how the
$(-1)^F$ symmetry is realized in the closed string Ramond-Ramond
(R-R) sector; the closed string spectrum admits the sectors
\eqn\zerodef{\eqalign{
    {\rm type\ 0A}\  : \quad &\qquad
    (\NS-,\NS-)\oplus (\NS+,\NS+)   \oplus(R+,R-) \oplus(R-,R+)\cr
    {\rm type\ 0B}\  : \quad &\qquad
    (\NS-,\NS-)\oplus (\NS+,\NS+)   \oplus(R+,R+) \oplus(R-,R-)
}} where $\pm$ refers to worldsheet fermion parity. As is familiar
from type II, the two theories are interchanged under orbifolding
by $(-1)^\FL$ where $\FL$ is left-moving NSR parity\foot{
Note that $(-1)^{\FL} $
reverses the sign of the left moving spin field. This
should not be confused with the worlsheet $Z_2$ operation
 $(-1)^{F_L}$
giving a
minus sign to the leftmoving worlsheet fermion $\psi_L \to - \psi_L$.
}.
Because there
are no $(\NS,R)$ or $(R,\NS)$ sectors, there are no closed strings
that are spacetime fermions. In addition to the usual NS sector
fields appearing in the $(\NS+,\NS+)$ sector, one has a closed
string tachyon $T$ appearing in the $(\NS-,\NS-)$ sector. The R-R
spectrum is doubled; in particular, at the massless level there
are two R-R gauge fields $C_{(p+1)}$, $\tilde C_{(p+1)}$
of each allowed rank $p+1$ ($p$ is even for type 0A, odd for type 0B).%
\foot{For the middle rank $p+2=d/2$, there is {\it one} gauge
field, but it has both self-dual and anti self-dual components.}
The spacetime effective action for these fields is ({\it c.f.}
\refs{\PolchinskiRR,\PolyakovJU,\KT})
 \eqn\seffact{\eqalign{
  S &= \frac{1}{2\kappa^2}\int\! d^d\!x\,\sqrt{\!-G}\Bigl[
    e^{-2\Phi} \Bigl( \coeff{10-d}{\alpha'}+
    R+4(\nabla\Phi)^2-\coeff1{2} |H_3|^2
    -\coeff12(\nabla T)^2+\frac1{\alpha'} T^2\Bigr)\cr
   &\hskip 3cm
    -\frac{1}{2}\Bigl(\sum_p |F_{p+2}|^2+|\tilde F_{p+2}|^2
    +T\,|F_{p+2}\tilde F_{p+2}|  \Bigr)
        + \dots\Bigr]\ ,
}}
where $|F_n\tilde F_n|=\frac1{n!}F^{m_1...m_n}\tilde F_{m_1...m_n}$
(there is an additional factor of $\half$ for $n=d/2$).

The specialization of this theory to $d=2$ is particularly simple.
There is no NS $B$ field, and the dilaton gravity sector of $\Phi$
and $G$ has no field theoretic degrees of freedom. A closed string
background which solves the equations of motion of \seffact\ is
(we set $\alpha'=2$ unless indicated otherwise)
 \eqn\twodeom{
  G_{\mu\nu}=\eta_{\mu\nu}\ ,\quad
  \Phi = \phi\ ,\quad
  T = \mu e^{\phi}
 }
where $\phi$ is the spatial coordinate. The (possibly Euclidean)
time coordinate will be denoted by $x$.

In two dimensions, there are no transverse string oscillations,
and longitudinal oscillations are unphysical at generic momenta.
Thus the tachyon is the only physical NS sector field.%
\foot{ When we consider operators in the spacetime theory, i.e.
non-normalizable deformations, then there are additional operators
(usually called discrete ``states'') at special values of the
momenta. For instance, when $x$ is compactified, the radius
deformation $V_{\sst G}=\partial x\bar\partial x$ corresponds to a
non-normalizable physical vertex operator on the worldsheet.} In
the R-R sector, the spectrum consists of two vector fields $C_1$,
$\tilde C_1$ for type 0A, while the type 0B theory has a scalar
$C_0$ (the self-dual part of $C_0$ comes from the $(R+,R+)$ sector
and the anti-selfdual part comes from the $(R-,R-)$ sector)
and a pair of two-forms $C_2,\tilde C_2$. Only
the scalar $C_0$ gives a field-theoretic degree of freedom. The
rest of the fields give rise to discrete states.

Let us discuss the action of the 0B theory in more detail.
Equation \seffact\ becomes
 \eqn\seffacta{\eqalign{
  S &= \int\! d\phi dt\sqrt{\!-G}\Bigl[
    {e^{-2\Phi} \over 2\kappa^2} \Bigl( \coeff{8}{\alpha'}+
    R+4(\nabla\Phi)^2
    -f_1(T)(\nabla T)^2+f_2(T) +\dots\Bigr)\cr
   &\hskip 3cm
    -{ 1\over 8\pi}f_3(T) (\nabla C_0)^2        + \dots\Bigr]\ ,
 }}
with $f_i(T)$ functions of the tachyon field $T$.  To all orders
in perturbation theory the action is invariant under shifts of the
R-R scalar $C_0$. In the matrix model this shift symmetry
corresponds to multiplying the wave functions of all the fermions
on one side of the potential by a common phase while not doing
that to the fermions on the other side of the potential.
Nonperturbatively, the fermions on the two sides affect each other
and this shift symmetry is violated by instantons.
Correspondingly, the theory is invariant only under a discrete
shift of $C_0$. This discrete shift symmetry is a gauge symmetry;
i.e.\ $C_0$ is compact.  We normalize $f_3(T=0) = 1$. We will
later argue that in this normalization the field $C_0$ is a two
dimensional field at the self dual radius. Also, we will give
evidence that $f_3(T) = e^{-2T }$ (see also \ortin ).
With this form of $f_3$ we
derive a few consequences:
 \item{1.} The effective radius of $C_0$ goes to zero as $T\to
 \infty$ -- the circle parametrized by $C_0$ is pinched to a
 point. Therefore, the theory based on
 the action \seffacta\ violates the winding symmetry in
 $C_0$ at tree level.  D-instantons which are not included in
 \seffacta\ violate also the momentum symmetry.  This can be seen
 by computing a disk amplitude with one insertion of $C_0$.
 \item{2.}  The theory has a peculiar ``S-like-duality'' under
 which $T\to -T$ and the compact boson $C_0$ is dualized.  In
 terms of the dual field $\tilde C_0$ the classical theory has
 no momentum symmetry because the field $\tilde C_0$ is pinned at
 infinity where the coefficient of its kinetic term is infinite.
 In the matrix model this S-duality corresponds to changing the
 sign of $\mu$; i.e.\ to lifting the Fermi level above the maximum
 of the potential.  In the worldsheet description of the theory
 this operation is $(-1)^{F_L}$ where $F_L$ is the worldsheet
 fermion number.
 \item{3.} The field $\chi=e^{-T}C_0$ has a canonical kinetic term
 \eqn\chiki{-{1\over 8\pi} \left(\nabla \chi + \chi \nabla T
 \right)^2}
 but the shift symmetry is more complicated
 $\chi \to \chi + e^{-T} \alpha$ for constant $\alpha$.
 We can think of $F=d\chi +\chi dT$ as a one form field strength
 of $\chi$.  Then, in the background \twodeom\ the equation of
 motion and the Bianchi identity are
 \eqn\chieom{\eqalign{
 &\left(-{\partial \over \partial \phi} + \mu
 e^{\phi}\right)F_\phi = -{\partial \over \partial t} F_t \cr
 &\left({\partial \over \partial \phi}+ \mu
 e^{\phi}\right)F_t = {\partial \over \partial t} F_\phi \ .}}
 \item{4.} Because of the coupling to $T$ for each sign of $\mu$
 there is only one  solution of the equations of motion with
 time independent $F$
 \eqn\zereC{C_0=\cases{
 \int^\phi\! d\phi' e^{2T}=\int^\phi \!d\phi' \exp\bigl(2\mu
 e^{\phi'}\bigr)& $\mu<0$\cr \cr
 t& $\mu>0$\cr}}
 or equivalently
 \eqn\zereF{F=\cases{\exp\left(\mu
 e^{\phi}\right)d\phi & $\mu<0$\cr
 \exp\left(-\mu
 e^{\phi}\right)dt & $\mu>0$\ .}}
 The other solution is unacceptable because it
 diverges at $\phi \to +\infty$. The solution for $\mu > 0$
and the solution for $\mu < 0$
 are exchanged by the S-duality we mentioned above. These
solutions
 represent one possible non-normalizable deformation of the theory,
 which is different depending on the sign of $\mu$.

 \item{5.} In the matrix model with positive $\mu$ this deformation
 corresponds to changing the two Fermi levels on the two sides of
 the potential.\foot{For negative $\mu$ these are the two distinct
 Fermi levels of the left movers and the right movers.}
 Nonperturbatively, eigenvalues can tunnel from one side of the
 potential to the other and the only stable situation is when the
 Fermi levels are the same.  This means that nonperturbatively
 this zero mode of $C_0$ is fixed (or more precisely, it is
 quantized according to the periodicity of $C_0$) and cannot be changed.

\noindent
The situation in the 0A theory is analogous.  In fact, by
compactifying the Euclidean time direction on a circle of radius
$R$, T-duality relates these two theories.  The target space action
is
 \eqn\seffactb{\eqalign{
  S &= \int\! d\phi dt\sqrt{\!-G}\Bigl[
    {e^{-2\Phi} \over 2\kappa^2} \Bigl( \coeff{8}{\alpha'}+
    R+4(\nabla\Phi)^2
    -f_1(T)(\nabla T)^2+f_2(T) +\dots\Bigr)\cr
   &\hskip 3cm
    -{ (2\pi) \alpha' \over 4} f_3(T) (F^{(+)})^2   -
    { (2\pi) \alpha' \over 4}
    f_3(-T) (F^{(-)})^2     + \dots\Bigr]\ ,
 }}
where we see two different one form gauge fields and their field
strengths $F^{(\pm)}$.  Our S-duality transformation $T\to -T$
exchanges them.  We have normalized the gauge fields so that the
coupling to a unit charge is $e^{ i \!\int \!\! A }$. The normalization
of the kinetic terms for the gauge fields in \seffactb\ follows
from T-duality and the normalization of the $C$ field, which
is argued to be at the self dual radius in appendix C.
Ordinarily the only degree of freedom in a two
dimensional gauge theory is the zero mode, and one might expect
the possibility of turning on time independent $F^{(\pm)}$.
However, in this system, because of the coupling $f_3(\pm T)$,
time independent field strength can be turned on only for that
gauge field whose coefficient $f_3(\pm T)$ is nonzero at infinity.
This means that we can only turn on $F^{(-)} = \exp(-2\mu e^\phi)$
for $\mu$ positive and $F^{(+)} = \exp(2\mu e^\phi)$ for $\mu$
negative.  This statement is T-dual to the analogous fact
of having only one zero energy solution in the 0B theory.
Nonperturbatively, this zero mode is
quantized, as in the 0B theory.
It corresponds to the background field of
D0-branes whose charge is quantized.

\subsec{Worldsheet description}

The worldsheet description of the background \twodeom\
involves two scalar superfields
\eqn\superf{\eqalign{
  \Phi &= \phi + i\theta\psi + i \bar \theta
    \bar \psi + i\theta\bar \theta F\cr
  X &= x+i\theta\chi+i\bar\theta\bar\chi+i\theta \bar\theta G\ .
}}
Our 2d supersymmetry conventions are as follows:
The covariant derivatives, supercharges and algebra
are
\eqn\supco{\eqalign{
 & D={\partial \over \partial \theta} + \theta \partial
    \quad, \qquad
 \bar D={\partial \over \partial \bar\theta} + \bar\theta \bar \partial
    \quad,\qquad
 \{D,D\}=2  \partial
    \quad, \qquad
 \{\bar D,\bar D\}=2  \bar\partial \cr
 & Q={\partial \over \partial \theta} - \theta \partial
    \quad,\qquad
 \bar Q ={\partial \over \partial
 \bar\theta} - \bar\theta \bar\partial
    \quad ,\qquad
 \{Q,Q\}=-2  \partial
    \ \ ,\qquad
 \{\bar Q,\bar Q\}=-2  \bar \partial }}
and all other (anti)commutators vanish.
We define $z=x+iy$, $\bar z=x-iy$ and therefore
$\partial=(\partial_x-i\partial_y)/2$,
$\bar\partial=(\partial_x+i\partial_y)/2$.
Finally, the integration measure is
$\int\! d^2z d^2\theta = 2\int\! dxdyd\bar\theta d\theta$.

The action for $X$ is the usual free field action
\eqn\freefield{
  S_X = \frac{1}{4\pi}\int\!d^2z d^2\theta\,D\!X\bar D \!X\ ,
} while the dynamics of $\Phi$ is governed by the super-Liouville
action \refs{\PolyakovRE\DiVecchiaBZ\DHokerZY\ArvisTQ-\ArvisKD}
\eqn\lag{
  S = \frac{1}{4\pi}\int\!d^2z d^2\theta\,
    \Bigl[ D\Phi \bar D\Phi + 2 i\mubare e^{b\Phi}\Bigr]\ .
}
There is implicitly a dilaton linear in $\Phi$, of slope
\eqn\liouparams{
  Q=  b +{1\over b}
 }
which makes its usual appearance as an improvement term in the
super stress tensor. The action \lag\ gives rise to an $N=1$
superconformal field theory with central charge
 \eqn\centchge{
  \hat c_L= 1+2 Q^2\ .
} The case of interest here will be $b=1$, $\hat c_L=9$.
Exponential operators have dimension
 \eqn\confdim{
  \Delta(e^{\alpha\Phi+ikX}) = \coeff12\,\alpha(Q-\alpha)+\coeff12 k^2\ ;
}
in particular, the Liouville interaction is scale invariant.


\subsec{Compactification}

In the Euclidean theory we may replace the
non-compact free scalar superfield theory of $X$
by any other $\chat=1$ superconformal field theory.
The set of $\chat=1$ superconformal field
theories was classified in \refs{\fran,\dgh}.
Apart from orbifolds and isolated theories which will not concern
us here, there are two lines of theories parametrized by the radius
of compactification $R$ of the lowest component of the superfield $X$.
The first line of models is called the `circle' theory;
in it, the sum over the spin structures of $\chi$, $\bar\chi$
is independent of the (left and right)
momentum $k, \bar k$ of $x$,
giving rise to a tensor product
of a compact boson and an Ising model.
The momentum takes the usual
form $(k, \bar k)=({p\over R}+{wR\over2},
{p\over R}-{wR\over2})$; hence the lattice of momenta is
$\Lambda=\{(p,w)|p,w\in\IZ\}$.
The circle model has the usual $R\rightarrow{2\over R}$ T-duality symmetry.
At the self dual point ($R=\sqrt2$) an affine
$SU(2)\times SU(2)$ symmetry appears for the boson $x$. This symmetry
does not commute with worldsheet supersymmetry and therefore is not
a symmetry of the spacetime theory.

The second line, the `super affine' theory, is obtained from the
circle by modding out by a $\IZ_2$ symmetry $(-)^\FL e^{i\pi p}$
where $(-)^\FL=1\;(-1)$ on NS-NS (R-R) states. This correlates the
sum over spin structures with the momentum and winding of $x$. To
describe the resulting spectrum it is convenient to define the
following sublattices: \eqn\lattices{
\eqalign{\Lambda^+=&\{(p,w)|p\in 2\IZ, w\in\IZ\}\cr
     \Lambda^-=&\{(p,w)|p\in 2\IZ+1, w\in\IZ\}\cr
     \Lambda^+_\delta=&\{(p,w)|p\in 2\IZ, w\in\IZ+\coeff12\}\cr
     \Lambda^-_\delta=&\{(p,w)|p\in 2\IZ+1, w\in\IZ+\coeff12\}\ .}
 }
The NS states in the super affine theory have momenta in
$\Lambda^+\cup \Lambda_\delta^-$, whereas Ramond states inhabit
$\Lambda^-\cup\Lambda^+_\delta$ \dgh. The super affine model is
self dual under $R\rightarrow{4\over R}$. At the self dual point,
$R=2$, an $SU(2)\times SU(2)$ super affine algebra appears, making
this the most symmetric point in the moduli space of $\hat c=1$
theories. These symmetries commute with worldsheet supersymmetry
so they translate into symmetries of the spacetime theory. Despite
the fact that the super affine line of theories is conveniently
described by twisting the circle theory, it is not in any sense
``less fundamental''. The super-affine theory has the following
interesting property. Recall that degenerate superconformal
representations at $\hat c=1$ occur at $k={r-s\over2}$ with
$r-s\in 2\IZ\; (2\IZ+1)$ corresponding to NS (R) degenerate
representations; we see that the self dual super affine model at
$R=2$ has the property that all the representations that occur in
it are degenerate, as is the case for the self-dual ($R=\sqrt 2$)
bosonic circle model; there is no point on the superconformal
circle line with these properties.


\subsec{Spectrum and vertex operators}

Corresponding to the two lines of SCFT's described above are four
different two-dimensional string theories
obtained by coupling $X$ \freefield\ to the super Liouville
field \lag.  Consider first the circle line.
Physical states corresponding to momentum modes of the NS ``tachyon''
field have the form (in the $(-1,-1)$ picture):
\eqn\vns{
  T_{k}^{(\pm)}=c\bar c\; \exp\left[-(\varphi+\bar\varphi)
    +ik(x_L+x_R)+(1\mp k)\phi\right]
}
where as explained above, $k=\bar k={p\over R}$, $p\in\IZ$.%
\foot{The fields $\varphi$, $\bar\varphi$ bosonize the spinor
ghosts of the fermionic string.} Note that, due to the linear
dilaton in \twodeom, the dispersion relation of the ``tachyon''
$T$ is in fact massless. The Ramond vertex has the form (in the
$(-\coeff12,-\coeff12)$ picture) \eqn\vram{
  V_{k}^{(\pm)}=c\bar c\;\exp\left[
    -{\varphi+\bar\varphi\over2}\mp{i\over2}(H+\bar H)
    +ik(x_L+x_R)+(1\mp k)\phi \right]
}
with $k={p\over R}$ as in \vns; we have bosonized
the fermions in \superf\ as usual via
$\psi+i\chi = {\sqrt2}\,e^{iH}$.
Physical winding modes $\hat T_k$,
$\hat V_k$ have the same form as \vns, \vram\ with
$x_R, \bar H\rightarrow-x_R, -\bar H$, and $k={wR\over2}$, $
w\in\IZ$.

All the NS states above have even fermion number and are physical.
In the Ramond sector there are two allowed fermion number
projections which lead to different spectra. Projecting to states
invariant under $H\rightarrow H+\pi$, $\bar H\rightarrow\bar
H-\pi$ keeps the momentum modes $V_k$ \vram\ while projecting out
the winding modes $\hat V_k$. The opposite projection $H(\bar
H)\rightarrow H(\bar H)+\pi$ has the opposite effect. These are
the type 0B and type 0A circle theories, respectively. As is
familiar from higher dimensions, they are dual to each other under
$R\rightarrow{2\over R}$. At infinite $R$ the physical spectrum of
the type 0B theory includes two (massless) field theoretic degrees
of freedom, the tachyon $T_k$ and the R-R scalar $V_k$; the type
0A theory has only the NS scalar field $T_k$.

Repeating the analysis above for the case of the super affine theory,
we are led to the following spectrum: in the NS sector we find
{\it even} momentum modes \vns\ $T_k$ with $k={2n\over R}$
and {\it integer} winding modes
$\hat T_k$ with $k={mR\over2}$.
In the Ramond sector we find {\it odd} momentum modes \vram\
$V_k$ with $k={2n+1\over R}$
and {\it half integer} windings,
$\hat V_k$ with $k=(m+\coeff12){R\over2}$.
The GSO projection in this theory
is momentum sensitive.
One gets two super affine theories,
depending on the sign of the projection:
the type 0B theory allows all the Ramond states
listed above (momentum and winding),
and the type 0A theory projects them all out
(the NS states are again physical).
Each of the two theories is separately self dual under
$R\rightarrow{4\over R}$.
At infinite radius they give rise to the same two
theories as the infinite radius circle ones.


\newsec{The ground ring and tree level scattering}

The BRST cohomology of string theory in two dimensions
includes a set of operators of dimension zero and ghost number zero,
known as the ground ring.
E. Witten \WittenZD\ proposed that the ground ring
provides insight into the relation between
the continuum formulation and the matrix model.
In this section we discuss this idea,
first in the bosonic string, and then
for the NSR case. Progress in Liouville field theory
since \WittenZD\ makes it possible
to say more about the properties of the ring
and its relation to the matrix model
(a subject to which we will return in section 11).

\subsec{The ground ring of the 2d
bosonic string and the matrix model}

We start by listing a few results on Liouville field theory which
will be useful later (see {\it e.g.} \refs{\fzz\zz-\TeschnerRV}).
In this subsection we set $\alpha'=1$.  The Liouville central
charge is \eqn\cliouv{c_L=1+6Q^2=13+6b^2+{6\over b^2}~} where we
used the parametrization \eqn\bQ{ Q=b+{1\over b}~. } The
cosmological term in the Liouville Lagrangian is:
\eqn\LIOU{
\delta \CL= \mu_0 e^{2b \phi}~.
}
A set of natural observables in the theory is
\eqn\opalph{V_\alpha(\phi)=e^{2\alpha\phi}}
whose scaling dimension is
$\Delta_\alpha=\bar\Delta_\alpha=\alpha(Q-\alpha)$.
Degenerate representations of the Virasoro algebra occur for
$\alpha=\alpha_{m,n}$, such that
\eqn\delmn{\Delta_{m,n}^{(L)}=\Delta(\alpha_{m,n})={1\over4}Q^2-
{1\over4}({m\over b}+nb)^2\ .}
The corresponding $\alpha_{m,n}$ are
\eqn\almn{\alpha_{m,n}={1\over2b}(1-m)+{b\over2}(1-n)
    \quad ;\qquad m,n=1,2,\ldots}
The null state in the representation \delmn\ occurs at level $mn$.
The first few cases are:
\eqn\nullsteq{\eqalign{
\alpha_{1,1}=&0;\;\;\partial V_{1,1}^{(L)}=0\cr
\alpha_{1,2}=&-{b\over2};\;\;(\partial^2+b^2 T^{(L)})V_{-{b\over2}}=0\cr
\alpha_{2,1}=&-{1\over2b};\;\;(\partial^2+{1\over b^2}
T^{(L)})V_{-{1\over2b}}=0\ .
}}
Here $ T^{(L)}$ is the Liouville stress tensor.

The matter theory in $c\le 1$ string theory
can be described by taking $b\to ib$
in the above formulae. Thus, one has (see \cliouv)
\eqn\cmatt{c_M=13-6b^2-{6\over b^2}\quad;\qquad c_L+c_M=26}
and the dimensions of degenerate operators $V_{m,n}^{(M)}$ are
(see \delmn)
\eqn\delmatmn{\Delta_{m,n}^{(M)}=-{1\over4}(b-{1\over b})^2+
{1\over4}({m\over b}-nb)^2\ .}
In particular, one has
\eqn\delmattliouv{\Delta_{m,n}^{(M)}+\Delta_{m,n}^{(L)}=1-mn\ .}
The ground ring operators are obtained by applying raising operators
of level $mn-1$ to $V_{m,n}^{(L)}V_{m,n}^{(M)}$. The first few
examples (corresponding to \nullsteq) are \refs{\WittenZD,\BouwknegtYG}:
\eqn\Omn{\eqalign{
\CO_{1,1}=&V_{1,1}^{(L)}V_{1,1}^{(M)}=1\cr
\CO_{1,2}=&|cb-{1\over b^2}(L_{-1}^{(L)}-L_{-1}^{(M)})|^2
V_{1,2}^{(L)}V_{1,2}^{(M)}\cr
\CO_{2,1}=&|cb-b^2(L_{-1}^{(L)}-L_{-1}^{(M)})|^2
V_{2,1}^{(L)}V_{2,1}^{(M)}\ .
}}
Note that here $b$ stands for two  unrelated quantities: the
reparametrization ghost and the parameter introduced in \bQ. Hopefully,
it is clear from context which is which.

There are also current operators obtained by tensoring standard
holomorphic vertex operators with dimension $(1,0)$ with these
antiholomorphic fields of dimension $(0,0)$ to form left moving
currents of dimension $(1,0)$ (and similarly right moving currents
of dimension $(0,1)$). These satisfy a $W_\infty$ algebra
\refs{\WittenYJ,\KlebanovUI}. These are spacetime symmetries which
reflect the fact that the dual theory can be written in terms of
free fermions.

The operators \Omn\ have $\Delta=\bar\Delta=0$ and are in the BRST cohomology
of the model. One can show that $\partial_z\CO_{m,n}$ and
$\partial_{\bar z}\CO_{m,n}$ are BRST exact. Therefore, any amplitude that
involves $\CO_{m,n}$ and other BRST invariant operators does not depend on the
position of $\CO_{m,n}$. Below we will use the freedom to move these operators
around when calculating amplitudes.

We will be mostly interested in the case $c_M=1$ corresponding to $b=1$.
In that case one has
\eqn\vcone{
\eqalign{
V_{1,2}^{(L)}V_{1,2}^{(M)}=&e^{ix-\phi}\cr
V_{2,1}^{(L)}V_{2,1}^{(M)}=&e^{-ix-\phi}\cr
}}
and the ground ring generators \Omn\ are
\eqn\Oone{\eqalign{
\CO_{1,2}=&(cb+\partial\phi+i\partial x)
(\bar c \bar b+\bar\partial\phi+i\bar\partial x)e^{ix-\phi}\cr
\CO_{2,1}=&(cb+\partial\phi-i\partial x)
(\bar c \bar b+\bar\partial\phi-i\bar\partial x)e^{-ix-\phi}\ .
}}
It is argued in \WittenZD\ that for the case
of non-compact $x$, which can be continued to Minkowski
time by replacing $x\to it$, the operators $\CO_{1,2}$,
$\CO_{2,1}$ are nothing but the phase space
variables of the inverted harmonic oscillator Hamiltonian $H=p^2-q^2$,
\eqn\Omappq{\eqalign{
\CO_{1,2}=&(q+p)e^{-t}\cr
\CO_{2,1}=&(q-p)e^{t}\ .
}}
The quantities on the r.h.s. are constants of motion in the inverted harmonic
oscillator potential. Their product  is the matrix model Hamiltonian,
\eqn\aplusminus{\CO_{1,2}\CO_{2,1}=q^2-p^2=-H\ .}
One expects \WittenZD\ that in the perturbative string sector of the theory
\eqn\expmu{\CO_{1,2}\CO_{2,1}\simeq \mu\ .}
Comparing to \aplusminus\ we see that $-\mu$ is
the level of the Fermi sea of the matrix model (measured from the top
of the potential), on  which the perturbative string excitations live.

Equation \expmu\  can be verified directly by using the fact that
the Liouville operators that enter $\CO_{1,2}$ and $\CO_{2,1}$ are
degenerate (see \nullsteq). For example, the OPE of
$V_{1,2}^{(L)}=V_{-{b\over2}}$ with any other $V_\alpha$ has the
form \eqn\vmbtwo{V_{-{b\over2}}\cdot V_\alpha=V_{\alpha-{b\over2}}
+C_-(\alpha)V_{\alpha+{b\over2}} +\ldots} where the $\cdots$
stands for Virasoro descendants of the operators on the r.h.s. of
\vmbtwo, and \refs{\DornXN,\ZamolodchikovAA}
\eqn\cccminus{C_-(\alpha)=-\mu_0{\pi\gamma(2b\alpha-1-b^2)\over
\gamma(-b^2)\gamma(2b\alpha)}} where
$\gamma(x)\!=\!\Gamma(x)/\Gamma(1-x)$. Repeated application of
$V_{-\frac b2}$ can take a vertex operator that satisfies the
bound $\alpha<\frac Q2$ \SeibergEB\ to one that violates it.  One
can still use \vmbtwo\ in this case, with the understanding that
Liouville operators satisfy the reflection property (see {\it
e.g.} \TeschnerRV)
\eqn\lreflect{\eqalign{
  V_\alpha &= S(\alpha) V_{Q-\alpha} \cr
  S(\alpha) &= \bigl(\pi\mu_0\gamma(b^2)\bigr)^{-(Q-2\alpha)/b}
    \frac{\Gamma\bigl(\frac1b(Q-2\alpha)\bigr)
    \Gamma\bigl(b(Q-2\alpha)\bigr)}%
    {\Gamma\bigl(-\frac1b(Q-2\alpha)\bigr)
    \Gamma\bigl(-b(Q-2\alpha)\bigr)}\ .
}}
The limit $b\to 1$ (or $c_M\to 1$) is singular. This can be dealt with
as in \KlebanovKM. Define
\eqn\defmn{\mu=\pi\mu_0\gamma(b^2)}
and hold $\mu$ fixed as $b\to 1$ (and $\mu_0\to\infty$). This gives
\eqn\cmnorm{C_-(\alpha)={\mu\over (2\alpha-2)^2(2\alpha-1)^2}\ .}
Using \vmbtwo, one can calculate the OPE $\CO_{1,2}\CO_{2,1}$
\Oone. The first term on the r.h.s. of \vmbtwo\ does not contribute. The
second term gives
\eqn\pertope{\CO_{1,2}(z)\CO_{2,1}(w)=\left(1-{28\over4}\right)^2
{\mu\over 9\cdot 4}=\mu\ .}
We see that the ground ring generators \Oone\ satisfy the relation \expmu, in
agreement with their identification with the matrix model objects \Omappq.

The tachyon vertex operators form a module under the
action of the ground ring.
The vertex operator of a tachyon of momentum $k$ is
 \eqn\lrtach{T_k^{(\pm)}=c\bar c e^{ikx+(2\mp k)\phi}}
where the superscript $(\pm)$ refers to the spacetime  chirality.
For positive (negative) $k$ only the $-(+)$ signs define good
spacetime operators (see the discussion following \cccminus). By
using \vmbtwo\ with $\alpha=1\mp {k\over2}$ one can check that the
action of the ground ring on the tachyon modules is
\eqn\tachrec{\eqalign{ \CO_{1,2}T_k^{(+)}=&k^2T_{k+1}^{(+)}\cr
\CO_{1,2}T_k^{(-)}=&{\mu\over (k+1)^2}T_{k+1}^{(-)}\cr
\CO_{2,1}T_k^{(-)}=&k^2T_{k-1}^{(-)}\cr
\CO_{2,1}T_k^{(+)}=&{\mu\over (k-1)^2}T_{k-1}^{(+)}\ . }} These
equations can be simplified by redefining $T_k$ as follows:
\eqn\bosredef{
  \tilde T^{(\pm)}_k = { \Gamma(\pm k) \over \Gamma(1\mp k) }T^{(\pm)}_k\ .
}
In terms of $\tilde T_k$, one finds
\eqn\tildetach{\eqalign{
\CO_{1,2}\tilde T_k^{(+)}=&-\tilde T_{k+1}^{(+)}\cr
\CO_{1,2}\tilde T_k^{(-)}=&-\mu\tilde T_{k+1}^{(-)}\cr
\CO_{2,1}\tilde T_k^{(-)}=&-\tilde T_{k-1}^{(-)}\cr
\CO_{2,1}\tilde T_k^{(+)}=&-\mu\tilde T_{k-1}^{(+)}\ .
}}
The relations \bosredef, \tildetach\
were previously derived for `bulk' correlation functions
\refs{\KutasovQX, \BershadskyUB}.
We now see that they are exact properties of the full CFT.
Note that \tildetach\
implies \pertope; this is consistent with
the fact that the tachyons $T_k^{(\pm)}$
correspond to infinitesimal deformations of the Fermi surface.

These relations lead to certain periodicity
properties of the three point
functions of tachyons \lrtach\ in two dimensional
string theory. In four and higher
point functions, \tachrec\ receive corrections
due to the presence of integrated
vertex operators, or from the spacetime point of
view due to the deformation
of the Fermi surface \pertope\ caused by
the propagation of tachyons on it.


\subsec{The ground ring of the 2d fermionic string}

In the fermionic case it is convenient to return to the
conventions $\alpha'=2$.
In this case there are two classes of observables, corresponding to
Neveu-Schwarz and Ramond vertex operators,
\eqn\nsrobs{\eqalign{
N_\alpha=&e^{\alpha\phi}\cr
R_\alpha=&\sigma e^{\alpha\phi}\cr
}}
where $\sigma$ is a spin field. Correspondingly, there
are two kinds of degenerate operators: the NS sector operators
\eqn\nsdeg{N_{m,n}=e^{\alpha_{m,n}\phi}}
with $\alpha_{m,n}$ given again by \almn, $n,m=1,2,3,
\cdots$, and $m-n\in 2Z$ (note that $n=m=1$ is the identity operator).
We also have  Ramond sector operators
\eqn\ramdeg{R_{m,n}=\sigma e^{\alpha_{m,n}\phi}}
with $m-n\in 2Z+1$. The
natural analog of the operators $V_{-{b\over2}}$,
$V_{-{1\over2b}}$ (see \nullsteq) in this case are the
Ramond operators
\eqn\ronetwo{\eqalign{
R_{1,2}=&\sigma e^{-{b\over2}\phi}\cr
R_{2,1}=&\sigma e^{-{1\over2b}\phi}\cr
}}
which satisfy the (level one) null state conditions
\eqn\nullonetwo{\eqalign{
(L_{-1}+b^2G_{-1}G_0)R_{1,2}=&0\cr
(L_{-1}+{1\over b^2}G_{-1}G_0)R_{1,2}=&0\ .
}}
There are similar degenerate operators in the matter theory
obtained by taking $b\to ib$.
As in the bosonic case, these degenerate operators
lead to nontrivial BRST cohomology
at ghost number zero. For $\hat c_M=1$
(which again corresponds to $b\to 1$) these can be
written as \BouwknegtAM:
\eqn\apm{\eqalign{
\CO_{1,2}=&\left( e^{-{\varphi\over2}+{i\over2}H}
-{c\over\sqrt2}\partial\xi e^{-{3\varphi\over2}-{i\over2}H}\right)
\left( e^{-{\bar\varphi\over2}+{i\over2}\bar H}
-{\bar c\over\sqrt2}\bar\partial\bar \xi e^{-{3\bar\varphi\over2}-{i\over2}\bar H}\right)
e^{{i\over2}x-{1\over2}\phi}\cr
\CO_{2,1}=&\left( e^{-{\varphi\over2}-{i\over2}H}
-{c\over\sqrt2}\partial\xi e^{-{3\varphi\over2}+{i\over2}H}\right)
\left( e^{-{\bar\varphi\over2}-{i\over2}\bar H}
-{\bar c\over\sqrt2}\bar\partial\bar \xi e^{-{3\bar\varphi\over2}+{i\over2}\bar H}\right)
e^{-{i\over2}x-{1\over2}\phi}\cr
}}
where $\varphi$ is the bosonized superconformal ghost and $H$ bosonizes the
worldsheet fermions $\psi$, $\chi$ as in \vram.

A few comments are in order here.
\item{(1)}
The operators $\CO_{1,2}$, $\CO_{2,1}$ are present in
the type 0B theory, where they generate the ground ring.
These operators are projected out in the type 0A theory.
There, the ring is generated by the lowest NS sector operators
$\CO_{1,3}$, $\CO_{3,1}$, and $\CO_{2,2}$.
\item{(2)}
As in the bosonic string, one can also form left moving and right
moving currents \refs{\Itoh,\BouwknegtVA,\AldazabalAE} which satisfy an interesting current
algebra.
\item{(3)}
Upon compactification, one again finds a structure similar
to the bosonic one.  As mentioned in section 2,
the most symmetric point in moduli space is the
0B super-affine theory at the self-dual radius $R=2$.
In this theory, we can define the (anti-)chiral
vertex operators
\eqn\affring{\eqalign{
  {\bf x} &= \left( e^{-{\varphi\over2}+{i\over2}H}
-{c\over\sqrt2}\partial\xi e^{-{3\varphi\over2}-{i\over2}H}\right)
    e^{{i\over2}x_L-{1\over2}\phi_L}\cr
  {\bf y} &= \left( e^{-{\varphi\over2}-{i\over2}H}
-{c\over\sqrt2}\partial\xi e^{-{3\varphi\over2}+{i\over2}H}\right)
        e^{-{i\over2}x_L-{1\over2}\phi_L}\cr
  \bar{\bf x} &= \left( e^{-{\bar\varphi\over2}+{i\over2}\bar H}
-{\bar c\over\sqrt2}\bar\partial\bar\xi e^{-{3\bar\varphi\over2}
    -{i\over2}\bar H}\right)
        e^{{i\over2}x_R-{1\over2}\phi_R}\cr
  \bar{\bf y} &= \left( e^{-{\bar\varphi\over2}-{i\over2}\bar H}
-{\bar c\over\sqrt2}\bar\partial\bar\xi
    e^{-{3\bar\varphi\over2}+{i\over2}\bar H}\right)
        e^{-{i\over2}x_R-{1\over2}\phi_R}\ ,
}}
and the ground ring is generated by the four
operators $a_1={\bf x\bar x}$, $a_2=\bf y\bar y$,
$a_3=\bf x \bar y$, and $a_4=\bf \bar x y$,
all of which are in the spectrum.  These operators
are subject to the relation $a_1a_2-a_3a_4=\sl const$.
The rich mathematical structure associated to
this ring was explored in \WittenZD.

\noindent
In order to repeat the discussion of the bosonic case, we need the
analog of \vmbtwo\ for this case \refs{\RashkovJX,\PoghosianDW}:
\eqn\roneope{\eqalign{
R_{1,2}N_\alpha=
    &R_{\alpha-{b\over2}}+C_-^{(N)}(\alpha)R_{\alpha+{b\over2}}\cr
R_{1,2} R_\alpha=
    &N_{\alpha-{b\over2}}+C_-^{(R)}(\alpha)N_{\alpha+{b\over2}}
}}
where
\eqn\cmnr{\eqalign{
  C_-^{(N)}(\alpha)=&{\mu_0
    b^2\gamma(\alpha b-\half b^2-\half)\over
    4\gamma({1-b^2\over2})\gamma(\alpha b)}\cr
  C_-^{(R)}(\alpha)=&{\mu_0 b^2\gamma(\alpha b-\half b^2)\over
    4\gamma({1-b^2\over2})\gamma(\alpha b+\half)}\ .
}}
In the limit $b\to 1$, and keeping
\eqn\musuper{\mu=\coeff14\mu_0\gamma({1+b^2\over2})} fixed, one has
\eqn\cpmbone{\eqalign{
C_-^{(N)}(\alpha)=&-{\mu\over(\alpha-1)^2}\cr
C_-^{(R)}(\alpha)=&-{\mu\over(\alpha-\half)^2}\ . }} We would next
like to compute the OPE of $\CO_{1,2}$ and $\CO_{2,1}$ using the
Liouville results \roneope, \cpmbone. As in the bosonic case, only
the term proportional to $C_-$ on the second line of \roneope\
contributes. Moreover, it is not difficult to see that in
multiplying the two lines of \apm, only the cross-terms are
non-zero: \eqn\crossterm{\CO_{1,2}\CO_{2,1}=2\cdot {c\bar c\over2}
\partial\xi\bar\partial\bar\xi e^{-{3\over2}(\varphi+\bar\varphi)
-{i\over2}(H+\bar H)+{i\over2}x -{\phi\over2}}\cdot
e^{-\half(\varphi+\bar\varphi)-{i\over2}(H+\bar H)-{i\over 2}x-
{\phi\over2}}\ .}
By using the second line of \roneope\ with \cpmbone, $C_-^{(R)}(-1/2)=-\mu$,
one finds
\eqn\opeminustwo{\CO_{1,2}\CO_{2,1}=-\mu e^{-2(\varphi+\bar\varphi)}
c\bar c\partial\xi\bar\partial\bar\xi\ .}
The r.h.s. of \opeminustwo\ is the identity operator
in disguise. More precisely,
by applying picture changing, using the term%
\foot{We use the conventions of \PolchinskiRR.}
\eqn\qbrst{Q_{\sst\rm BRST}=-\oint{dz\over 2\pi i}\gamma^2
b+\ldots} in the BRST charge, one finds that \opeminustwo\ is
equivalent to \eqn\opefinalferm{\CO_{1,2}\CO_{2,1}=\mu\ .} We next
move on to the action of the ground ring generators on the two
massless scalar fields of two dimensional (0B) string theory.
Applying the ring generators $\CO_{1,2}$, $\CO_{2,1}$ \apm\ to the
R-R vertex operator $V^{(\pm)}_k$ \vram, one finds (using \vns,
the second line of \roneope\ and \cpmbone) \eqn\pmrr{\eqalign{
\CO_{1,2}\cdot V^{(+)}_k=&T^{(+)}_{k+\half}\cr \CO_{2,1}\cdot
V^{(+)}_k=&-{\mu\over (k-\half)^2}T^{(+)}_{k-\half}\cr
\CO_{1,2}\cdot V^{(-)}_k=&-{\mu\over
(k+\half)^2}T^{(-)}_{k+\half}\cr \CO_{2,1}\cdot
V^{(-)}_k=&T^{(-)}_{k-\half}\ . }} The action of $\CO_{1,2}$ and
$\CO_{2,1}$ on $T^{(\pm)}$ can be deduced from \opefinalferm\ and
\pmrr, \eqn\pmnsns{\eqalign{ \CO_{1,2}\cdot
T^{(+)}_k=&-k^2V^{(+)}_{k+\half}\cr \CO_{2,1}\cdot T^{(+)}_k=&\mu
V^{(+)}_{k-\half}\cr \CO_{1,2}\cdot T^{(-)}_k=&\mu
V^{(-)}_{k+\half}\cr \CO_{2,1}\cdot
T^{(-)}_k=&-k^2V^{(-)}_{k-\half}\ . }}
One can diagonalize the action \pmrr, \pmnsns\ by
redefining \eqn\redef{\eqalign{ \tilde T^{(\pm)}_k &= { \Gamma(\pm
k) \over \Gamma(1\mp k) }T^{(\pm)}_k\cr \tilde V^{(\pm)}_k &= {
\Gamma(\half\pm k) \over \Gamma(\half\mp k) }
    V^{(\pm)}_k\cr
}} and changing variables to \eqn\abdef{\eqalign{
T_R^{(\pm)}(k)=&\half\Bigl(\tilde T^{(\pm)}_k+\tilde
V^{(\pm)}_k\Bigr)\cr T_L^{(\pm)}(k)=&\half\Bigl(\tilde
T^{(\pm)}_k-\tilde V^{(\pm)}_k\Bigr)\ . }} In terms of these
variables, the action \pmrr, \pmnsns\ breaks up into two copies of
the bosonic one.  That is, one has
\eqn\actab{\eqalign{
\CO_{1,2}\cdot T_R^{(+)}(k)=&T_R^{(+)}(k+\half)\cr \CO_{1,2}\cdot
T_L^{(+)}(k)=&-T_L^{(+)}(k+\half)\cr \CO_{2,1}\cdot
T_R^{(-)}(k)=&T_R^{(-)}(k-\half)\cr \CO_{2,1}\cdot
T_L^{(-)}(k)=&-T_L^{(-)}(k-\half)\ ; }} these relations, together
with \opefinalferm, specify the ring action completely.  We will
see later that $T_L$ and $T_R$ are excitations living on the two
sides of the inverted harmonic oscillator potential of the matrix
model. Equation \actab\ indicates the side of the potential
algebraically by the sign of $\CO_{1,2}$, $\CO_{2,1}$.

One can further show \refs{\DiFrancescoUD} that the scattering
amplitudes of $T_L$, $T_R$ factorize, at least at tree level. One
has \eqn\mix{ \eqalign{
&\langle\prod_{i=1}^nT_L(k_i)\prod_{j=1}^mT_R(p_j)\rangle=0\;\;(n,m\geq1)\cr
&\langle\prod_{i=1}^nT_L(k_i)\rangle=\langle\prod_{i=1}^nT_R(k_i)\rangle=
{1\over4}\langle\prod_{i=1}^n\tilde
 T^{(B)}(\sqrt{2}k_i)\rangle_B^{ }\cr} }
where $\tilde T^{(B)}$ is
the bosonic string vertex operator \bosredef, and the correlator
$\langle\cdots\rangle_B^{ }$ is computed in bosonic string theory.

The first line of \mix\ follows from the action of the ground ring
on the tachyons $T_L$, $T_R$. One can show, using similar methods
to \BershadskyUB, that bulk amplitudes satisfy the selection rule
on the first line of \mix, and then extend the result to non-bulk
amplitudes as in \DiFrancescoUD. The second line of \mix\ is the
statement that correlators of excitations living on a given side
of the potential are the same as in the bosonic string, up to the
usual rescaling of $\alpha'$ by a factor of two. The overall
factor of $1/4$ can be thought of as due to a rescaling by a
factor of two of $g_s$ between the bosonic and fermionic string.


\newsec{The torus partition function}

The one-loop string path integral provides a wealth of
information about the theory.
We will consider, following \refs{\BerKleb,\BerKlebnew},
the compactified theory with $X$ living on a circle of radius $R$.
As discussed in section 2,
the compactification of the scalar superfield $X(z, \bar z, \theta,
\bar\theta)$ can be done in two ways.
In the circle theory
one simply sums over all windings and momenta in each spin structure,
while in the super-affine theory \refs{\fran,\dgh}
one correlates the windings and momenta
with the fermion boundary conditions.

We begin however with a review of the bosonic theory.\foot{ In this
section we work with the Lagrangian ${1\over4\pi\alpha'} (dx)^2
+{1\over8\pi}(d\phi)^2+\cdots$, i.e.\ we make the scale of $x$
explicit, while fixing the scale of $\phi$. In the matrix model
the scale of $\phi$ is not visible and the curvature of the
potential sets the scale of the $x$-field. }

\subsec{Liouville on the torus}

In the torus path integral of the 2d bosonic string,
the oscillator contributions cancel
among Liouville, matter, and ghosts, leaving a zero mode sum
and an integral over the torus moduli
\eqn\Ltorus{{ {\cal Z}_1(R/\sqrt{\ap})\over V_L}=
{R \over \sqrt{\ap}}{1\over 4\pi\sqrt 2}
\int_{\cal F} {d^2\tau\over \tau_2^2}
\sum_{n, m}\exp\left(-{\pi R^2|n-m\tau|^2\over\ap\tau_2}\right)
\ ;  }
here ${\cal F}$ is the fundamental domain
for the torus modular parameter, and
$V_L=- (\ln \mu)/\sqrt 2$ is the volume of
the zero-mode of the Liouville field.

The integral in \Ltorus\ may be evaluated
by trading the summation over $(n,m)$ for $(n,m) \not = 0 $ for
extending the integration region from ${\cal F}$ to the entire strip
$-1/2 \leq \tau_1 < 1/2$ and summing only over $(n,m) = (n,0)$, $n>0$
 \polchinskitorus :
$$
\int_{\cal F} {d^2\tau\over \tau_2^2}
\sum_{n, m}\exp\left(-{\pi R^2|n-m\tau|^2\over\ap\tau_2}\right)
= \int_{\cal F} {d^2\tau\over \tau_2^2}
    + 2 \sum_{n=1}^\infty {d\tau_2\over \tau_2^2}
\exp\left(-{\pi R^2 n^2\over\ap\tau_2}\right)
\ .
$$
Evaluating these integrals we get \GrossKleb\
\eqn\torus{
  { {\cal Z}_1(R/\sqrt{\ap})} =
    - {1\over 24} \ln \mu \left ({R\over \sqrt{\ap}}
    + {\sqrt{\ap}\over R} \right) \ .
}
Another useful, though somewhat formal,
way of evaluating this integral is to first
perform a Poisson resummation:
$$
{R \over \sqrt{\ap}} \sum_{n, m}
    \exp\left(-{\pi R^2|n-m\tau|^2\over\ap\tau_2}\right)
=
\sqrt{\tau_2} \sum_{s,t\in Z}
q^{(s\sqrt{\ap}/R+tR/\sqrt{\ap})^2/4}
\bar q^{(s\sqrt{\ap}/R-tR/\sqrt{\ap})^2/4}
\ .$$
Now let us formally extend the integration region
to the entire strip \BerKlebnew.
Then we find
\eqn\formalsum{
{ {\cal Z}_1(R/\sqrt{\ap})\over V_L} =  {1\over 4\pi\sqrt 2}
\sum_s \int_0^\infty {d\tau_2\over \tau_2^{3/2} }
e^{-\pi \tau_2 s^2\ap/R^2} + (R\rightarrow {\ap\over R})
\ .
}
Since
$$ - {1\over 4\pi\sqrt 2} \int_0^\infty
{d\tau_2\over \tau_2^{3/2} } e^{-2\pi \tau_2 \omega^2 }
$$
is a proper time representation for the quantum mechanical
zero-point energy $\omega/2$, we find
$$
{ {\cal Z}_1(R/\sqrt{\ap})\over V_L} = -
\left ( {R\over \sqrt{\ap}} + {\sqrt{\ap}\over R} \right )
{1\over \sqrt 2} \sum_{s=1}^\infty s
\ .
$$
After using the standard zeta-function regularization
$\sum_{s=1}^\infty s =-1/12$, we find that the sum
over toroidal surfaces is  again \torus\ \refs{\GrossKleb,\igor}.
Although this approach is more formal,
it suggests that the term proportional to $1/R$ and $R$
arise from the momentum and winding modes, respectively.

The factor $\ln \mu$ in ${\cal Z}_1$ should be interpreted as $\ln
(\mu/\Lambda)$ with $\Lambda \gg \mu$ a cutoff at large negative
$\phi$.  ${1\over\sqrt2}\ln (\Lambda/\mu) $ is the effective
length of the Liouville direction.  $\Lambda$ is a UV cutoff on
the worldsheet and an IR cutoff in spacetime.  In the equations
below we will suppress an additive (infinite) constant
proportional to $\ln \Lambda$.


\subsec{Super-Liouville on the torus: the uncorrelated GSO projection}

First consider the standard type 0 non-chiral projection
(the `circle theory'),
with no correlation between the fermionic boundary conditions
and the soliton winding numbers.
Let $(r, s)$ label the spin structures as
$(e^{i\pi r}, e^{i\pi s})$.
The three even spin structures
have $(r, s)= (0, 1),~ (1, 0),~ (1, 1)$.
Let
${\cal D}_{r, s}$ be the corresponding
fermionic determinants divided by the
square root of the scalar determinant. Then
the partition function in the $(r, s)$ sector is
\eqn\asuper{\tilde Z_{r, s}^{(S)} (R/\sqrt{\ap}, \tau, \bar\tau)=
{R \over \sqrt{\ap}}
{1\over \sqrt{\tau_2}}|{\cal D}_{r, s}|^2
                  \sum_{m,n\in Z}  e^{-S_{m,n}}
\ .} In coupling such a matter system to supergravity we have to
multiply eq. \asuper\ by the path integrals over the super
Liouville field and over the superghost field. The contribution of
the super Liouville theory is equal to ${V_L \over {2 \pi \sqrt{2
\tau_2}}} |{\cal D}_{r, s}|^2$, where $V_L= -\ln |\mu|$;
the contribution of the
superghost sector equals ${1 \over 2\tau_2} |{\cal D}_{r,
s}|^{-4}$. One can see that the determinants due to all the
excitations again cancel out, and only the zero modes contribute
to the  full partition function. Coupling this system to
supergravity and counting each spin structure with factor $1/2$,
we find that the contribution of the even spin structures to the
genus one amplitude is \eqn\uncorr{ {\tilde {\cal Z}_{\rm even}
(R/\sqrt{\ap})\over V_L}= {1\over 2} {R \over \sqrt{\ap}} {1 \over
4\pi\sqrt 2}\int_{\cal F} {d^2 \tau \over {\tau_2 ^2}}
     \left( \sum_{(r, s)}~ \sum_{m,n\in Z} e^{-S_{m,n}} \right) =
{3\over 2} f(R/\sqrt{\ap})
\ ,
}
where
\eqn\fdef{
f (x) = {1\over 12\sqrt 2}
\left(x +{1\over x}\right)\ .
}
The peculiar factor of $3/2$ is due to
the fact that we have included 3 even spin structures.
This is inconsistent because in the type 0B string
there are twice as many fields as
in the bosonic string. The coefficient of $1/R$
is related to the $\sim T^2$ term in the thermal
free energy, which can be calculated
in 2-d free field theory and does not require a stringy
regularization. Therefore, in type 0B theory
this coefficient should be double that of the bosonic string.
Indeed, the odd spin structure
contributes in this string theory, since the zero
modes of the fermions are cancelled by those of the ghosts.
As shown in Appendix A, its contribution is
\eqn\odd{ {\tilde {\cal Z}_{\rm odd}
(R/\sqrt{\ap})\over V_L}=
\pm {1\over 24\sqrt 2}
\left ( {R\over \sqrt{\ap}} - {\sqrt{\ap}\over R} \right )
\ .
}
The sign of \odd\ is chosen such that
the coefficient of $1/R$ is that of two massless free fields
in the type 0B theory, and one such field for type 0A.
This leads to the result
\eqn\slb{\eqalign{
{\cal Z}_B &=  - {\ln|\mu|\over 12 \sqrt 2}
\left ( {R\over \sqrt{\ap}} + 2 {\sqrt{\ap}\over R} \right )\cr
{\cal Z}_A &= - {\ln|\mu|\over 12 \sqrt 2}
\left ( 2 {R\over \sqrt{\ap}} + {\sqrt{\ap}\over R} \right )\ .
}}
Note that T-duality $R\to\alpha'/R$ properly interchanges the two models.

These results can also be derived via zeta-function regularization
from the spectra of 0B and 0A circle theories discussed in section
2. For example, in the 0B case, we may use this logic to derive
$$
{ {\cal Z}_B(R/\sqrt{\ap}) \over V_L} = -
\left ( {R\over \sqrt{\ap}} + 2 {\sqrt{\ap}\over R} \right )
{1\over \sqrt 2} \sum_{s=1}^\infty s
\ ,
$$
which agrees with \slb.
The factor of 2 in the second term
is due to the doubling of momentum states.


\subsec{Correlated GSO projection}

Now consider the super-affine theory.
Here certain instanton sectors are weighted
with the opposite sign relative to others \fran,
\eqn\supaff{
\eqalign{&Z_{r, s} ^{(S)} (\tau, \bar\tau)=
{{\cal R} \over \sqrt{\ap} {\sqrt \tau_2}}
            |{\cal D}_{r, s}|^2
    \left( \sum_{m,n \in Z} e^{-S_{m,n}}
        - 2 \sum_{i,j \in Z} e^{-S_{2i+r,2j+s}} \right),\cr
&S_{m, n}={\pi {\cal R}^2\over\ap\tau_2}|n-m\tau|^2\ .\cr}
}
Counting each even spin structure with a factor of $1/2$, we get
after coupling to supergravity
\eqn\soliton{
\eqalign{{{\cal Z}_{\rm even} \over V_L} &= {1\over 2}
{{\cal R} \over \sqrt{\ap}}{1\over 4\pi\sqrt 2}
\int_{\cal F} {d^2 \tau \over {\tau_2 ^2}}
                     \left( \sum_{m,n \in Z} e^{-S_{m,n}}
                 + 2 \sum_{i, j \in Z} e^{-S_{2i ,2j}} \right)  \cr
&= {1\over 2} f ({\cal R}/\sqrt{\ap})
+  {1\over 2} f  (2{\cal R}/\sqrt{\ap})\ .\cr}
}
Therefore, for the super-affine theory we find
\eqn\saf{
  {\cal Z}_{\rm even} = -{ \ln|\mu|\over 8 \sqrt 2 }
    \left({{\cal R}\over\sqrt{\ap}}
    +{\sqrt{\ap}\over 2 {\cal R}}\right)\ .
}

This theory is a $Z_2$ orbifold of the usual type 0 theory;
and in terms of the covering space radius $ R = 2 {\cal R}$,
\eqn\safnew{ {\cal Z}_{\rm even} = - { \ln|\mu|\over 16  }
\left({ R\over\sqrt{2\ap}}+{\sqrt{2\ap}\over R}\right)\ .
 }
The duality $R \to 2 \ap/R$ is the T-duality of the super-affine
theory discussed in section 2.

We also need to include the contribution of the odd spin structure.
Consider first the super-affine theory
before coupling to supergravity. The contributions of
the NS-NS and NS-NS $(-1)^{F+\tilde F}$
spin structures to the partition function are
\fran:%
\foot{Note that the momentum number is $E$,
but the winding number is $M/2$.}
\eqn\nscont{\eqalign{& {1\over 2 |\eta(q)|^2 }
\Bigl [ |\chi_0 + \chi_{1/2} |^2 \Bigl(\sum_{\sst E,M\ {\rm even}}
+ \sum_{\sst E,M\ {\rm odd}}\Bigr)
+  |\chi_0 - \chi_{1/2} |^2 \Bigl(\sum_{\sst E,M\ {\rm even}}
- \sum_{\sst E,M\ {\rm odd}}\Bigr) \Bigr ] \cr
&
\hskip 3cm
\times q^{(E\sqrt{2 \ap}/R+MR/\sqrt{2 \ap})^2/8}
\bar q^{(E\sqrt{2\ap}/R- MR/\sqrt{2\ap})^2/8}
\ , \cr}
}
while the R-R spin structure gives
\eqn\rcont{{1\over |\eta(q)|^2 } |\chi_{1/16} |^2
\Bigl (\sum_{E\ {\rm even}\atop M\ {\rm odd}}
+ \sum_{M\ {\rm even}\atop E\ {\rm odd}}\Bigr)
\,q^{(E\sqrt{2 \ap}/R+MR/\sqrt{2 \ap})^2/8}
\,\bar q^{(E\sqrt{2\ap}/R- MR/\sqrt{2\ap})^2/8}
\ .
}
Note that each of these partition functions
is explicitly symmetric under the T-duality
$R\rightarrow 2\alpha'/R$ which interchanges $E$ and $M$.
Here $\chi_0$, $\chi_{1/2}$ and $\chi_{1/16}$
are the characters of the $c=1/2$ Majorana fermion theory.
In each spin structure they cancel
after we include the Liouville and ghost factors.
Thus, the sum over even spin structures gives
\eqn\safeven{ \eqalign{
{{\cal Z}_{\rm even}\over V_L} = & {1\over 4\pi\sqrt 2}
\int_{\cal F} {d^2 \tau \over \tau_2^{3/2} }
\left (\sum_{\sst E,M\ {\rm even}} +{1\over 2}
\sum_{E\ {\rm even}\atop M\ {\rm odd}} +
{1\over 2} \sum_{M\ {\rm even}\atop E\ {\rm odd}}\right ) \cr
&\hskip 2cm
\times q^{(E\sqrt{2 \ap}/R+MR/\sqrt{2 \ap})^2/8}
\;\bar q^{(E\sqrt{2\ap}/R- MR/\sqrt{2\ap})^2/8}
\ .\cr }
}
We see that in the NS-NS sector each state
with even $E,M$ enters with correct normalization,
while in the R-R sector there are incorrect factors of $1/2$.

Naively, to correct this
the contribution of the odd spin structure
has to be
\eqn\safodd{
\pm {1\over 4\pi\sqrt 2}
\int_{\cal F} {d^2 \tau \over \tau_2^{3/2} } \left ( {1\over 2}
\sum_{E\ {\rm even}\atop M\ {\rm odd}} -
{1\over 2} \sum_{M\ {\rm even}\atop E\ {\rm odd}}\right )
\, q^{(E\sqrt{2 \ap}/R+MR/\sqrt{2 \ap})^2/8}
\,\bar q^{(E\sqrt{2\ap}/R- MR/\sqrt{2\ap})^2/8}
\ .
}
An interesting property of this expression is that,
after Poisson resummation, it is equal to
\eqn\safoddnew{
\pm {1\over 8\pi\sqrt 2} {{\cal R}\over \sqrt{\ap} }
\int_{\cal F} {d^2 \tau \over \tau_2^2 }
\left( \sum_{m,n \in Z} e^{-S_{m,n}}
                 - 2 \sum_{i, j \in Z} e^{-S_{2i ,2j}} \right)
\ ,}
which establishes its modular invariance.
This is precisely the contribution of the $(r,s)=(0,0)$ sector in \supaff,
after we strip off the fermionic determinant. This expression is simply
the sum over solitons with phases, and evaluating
the integral in \safoddnew, we find
\eqn\oddfinone{
\pm {1\over 2} \left (f ({\cal R}/\sqrt{\ap})
-  f  (2{\cal R}/\sqrt{\ap})\right )
=\pm { 1\over 48  }
\left(-{ R\over\sqrt{2\ap}}+{\sqrt{2\ap}\over R}\right)\ .}
As shown in Appendix A,
this result is actually the one-point function of
the dilaton operator, which is $R^2 {\partial\over \partial R^2}$
acting on the contribution of the odd spin structure.
Thus,
\eqn\oddfin{
{{\cal Z}_{\rm odd}\over V_L}\sim \pm { 1\over 48  }
\left({ R\over\sqrt{2\ap}}+{\sqrt{2\ap}\over R}\right)\ .}

Using \soliton\ and \oddfin, we find the torus amplitudes in the
two theories \eqn\Aeval{ {\cal Z}_{\rm A}^{\rm super-aff.} =- {
\ln|\mu|\over 12  } \left({ R\over \sqrt{2\ap}}+{ \sqrt{2 \ap}\over
R}\right) \ } \eqn\Beval{ {\cal Z}_{\rm B}^{\rm super-aff.} =- {
\ln|\mu|\over 24  }
 \left({ R\over \sqrt{2\ap}}+{ \sqrt{2 \ap}\over R}\right) \ . }
These results can again be derived via zeta-function
regularization, using the spectra discussed in section 2. For
example, the complete physical spectrum of the super-affine A
theory contains integer windings and even momenta. This explains
\Aeval; in particular, the sum over even integers is $-1/6$, while
the sum over all integers is $-1/12$, which explains the relative
factor of 2. Similarly, \Beval\ follows from the spectrum of
super-affine B theory via zeta-function regularization.

Note that the answer in the type B theory is twice smaller than
in the type A theory,
even though it contains physical R-R modes.
The reason is that these modes are odd-integer
valued, and
in the zeta-function regularization,
the sum over odd integers is $1/12$
while the sum over all integers is $-1/12$.
Thus, in the model with physical R-R modes
the partition function is smaller.

The fact that the type B affine theory is half of the circle answer
can also be understood in the following way. As we discussed before,
the term proportional to $1/R$ in the torus partition function arises
from the thermal free energy of the fields we have in the bulk spacetime.
In the affine B theory we have two massless scalar fields, but one of them
has odd boundary conditions as we go around the thermal circle. This
thermal partition function can be easily computed and indeed
it matches with the $1/R$ term in \Beval . The term
proportional to $R$ can be obtained by $T$ duality.


\newsec{Semiclassical Picture of D-branes in Super-Liouville}

Having discussed the perturbative structure of 2d fermionic string
theory, we now turn to the structure of its D-branes.
We begin with a general discussion of boundary conditions
in Lagrangian field theory with $N=1$ worldsheet supersymmetry.

The Lagrangian density is a superfield $\CL$ which has a component
expansion as in \superf.  The action is
$\int\! d^2\!z d^2\theta \CL$.
The supersymmetry variation of a superfield $\Phi$ is
given by its commutator with $\zeta Q + \bar \zeta \bar Q$
(equation \supco),
yielding the transformation laws
 \eqn\sust{\eqalign{
 &\delta\phi=i\zeta \psi + i \bar\zeta\bar \psi\cr
 &\delta\psi=-i\zeta \partial \phi +\bar\zeta F\cr
  &\delta\bar \psi=-\zeta F - i\bar \zeta \bar \partial \phi \cr
 &\delta F=-\zeta \partial \bar \psi +\bar \zeta \bar \partial
 \psi\ .}}
The supersymmetry variation of $\int\! d^2\!\theta\, \CL$ is
then a total derivative
 \eqn\suvar{\delta \int\! d^2\theta\CL= -\zeta \partial
 \CL\big|_{\bar \theta} + \bar \zeta \bar\partial \CL\big|_\theta.}
If the worldsheet has a boundary at $z=\bar z$, the surface term
from \suvar\ is
 \eqn\susur{-i(\zeta \CL\big|_{\bar \theta} + \bar \zeta
 \CL\big|_\theta)\big|_{y=0}.}
If we add to the action an integral along the boundary
 \eqn\boua{i\eta \oint_{y=0} dx \CL(\theta=\bar\theta=0),
 \qquad  \eta=\pm 1\ ,}
then its variation $i \eta (\zeta \CL\big|_\theta +
\bar\zeta\CL\big|_{\bar\theta})$ cancels \susur\ for
$\zeta=\eta\bar\zeta$; i.e.\ only $Q+\eta \bar Q$ is preserved. It
is important that without the boundary interaction \boua\ no
supersymmetry is preserved.

In the presence of the boundary, the preserved super-translation
operator is
\eqn\supertrans{
  D_t=D+\eta\bar D\quad,\qquad D_t^2=\partial_x\ ;
}
the conjugate coordinate is $\theta_t=\frac12(\theta+\eta\bar\theta)$.
Similarly, $D_n=D-\eta\bar D$ is the superderivative
in the normal direction.

We should make some comments about the parameter $\eta$.  In the
type II theory we gauge the worldsheet $Z_2$ R-symmetry
under which $Q\to -Q$ and $\bar Q$ is invariant. This is a symmetry under
which the Lagrangian $\CL$ and $\theta$ are odd, but $\bar \theta$
is even and therefore $\int d^2\theta \CL$ is invariant. Since
this symmetry is gauged, we sum over worldsheets obtained by the
action of this group.  This sum is included in the sum over spin
structures.  The parameter $\eta$ is odd under the symmetry, and
therefore, whenever there is a boundary, we must sum over
worldsheets with different values of $\eta$.  In the type 0 theory
this symmetry is not gauged, and it might not even be a global
symmetry on the worldsheet. Therefore, we do not sum over $\eta$
and different values of $\eta$ correspond to distinct D-branes. If
it is a worldsheet global symmetry, as in the flat ten dimensional
background, the D-branes with $\eta=\pm 1$ are related by a local
$Z_2$ spacetime symmetry.  In other cases, where there is no such
symmetry, the two values of $\eta$ correspond to two different branes.
It might happen that one of them is infinitely heavy and
then should be thought of as absent.

Now let us turn to the specific case of super-Liouville theory.
The Lagrangian density in superspace, $\CL$, is given by equation
\lag.  The Liouville interaction in \lag\ breaks the $Z_2$ chiral
R-symmetry mentioned above. Therefore, the sign of the real
constant $\mubare$ can be changed by a field redefinition, but the
sign of $\eta\mubare$ cannot be changed. In order not to clutter
the equations, we will let $\mubare$ be positive.

The action is
 \eqn\action{\eqalign{
 S=&\int d^2z d^2\theta \CL + i\eta\oint_{y=0} dx
 \CL(\theta=\bar\theta=0)\cr
 =& {1\over 4\pi}\int d^2z \left(\partial \phi \bar
 \partial\phi +\psi\bar\partial \psi +\bar\psi\partial
 \bar\psi - F^2 -2\mubare b e^{b\phi} F + 2i\mubare b^2e^{b\phi}
 \psi\bar\psi \right) \cr
 &\qquad - {\eta \over 2\pi} \oint_{y=0} dx
 ({i\over 2}\psi\bar\psi + \mubare e^{b\phi}).}}
It is interesting that the theory must have a boundary
cosmological constant $\rho=-{\mubare\eta\over 2\pi}$
\refs{\ArvisTQ,\DiVecchiaBZ}.
The variation of the fields
leads to the bulk equations of motion
 \eqn\buleom{\eqalign{
 &\partial\bar\partial \phi=\mubare^2 b^3 e^{2 b\phi}
 +i\mubare b^3e^{b\phi}\psi\bar\psi \cr
 &\bar \partial\psi =-i\mubare b^2 e^{b\phi} \bar\psi \cr
 & \partial\bar\psi = i\mubare b^2 e^{b\phi} \psi \cr
 &F=-\mubare b  e^{b\phi}}}
(we used the last equation to simplify the first one).  The
boundary variation is proportional to
 \eqn\bouvare{i (\psi -\eta\bar\psi)(\delta \psi +\eta
 \delta \bar \psi) +(\partial_y \phi + 2 \eta\mubare b
 e^{b\phi}) \delta \phi.}

There can be two kinds of boundary conditions:
 \item{1.} Fixed boundary conditions correspond to imposing
\eqn\dbc{
 \phi(y=0)=\phi_0\quad,\qquad
    \psi(y=0)+\eta \bar \psi(y=0)=0\ .
}
 These boundary conditions are not conformally invariant except for
 $\phi_0=\pm\infty$ since the Liouville field shifts under scale
transformations.
 {}From a target space point of view this can be understood
 by remembering that the tension of such a brane is proportional
 to ${1 \over g_s(\phi_0)}= e^{-{Q\over 2} \phi_0}$, and therefore
 the stable finite tension brane is at $\phi_0 \to \infty$.
 Therefore we limit the discussion to this case.
 Such a brane which is localized at infinity
 is the supersymmetric version of the
 ZZ brane \refs{\zz,\FukudaBV,\korean}.
 Semiclassically, the worldsheet looks
 asymptotically near $y=0$ as $AdS_2$
 \eqn\zzc{ e^{2b\phi}=-{1\over \mubare^2 b^4 (z-\bar z)^2}; \qquad
 \psi=\bar\psi=0; \qquad F=-\mubare b e^{b\phi}=-{1\over 2 b y}}
 (recall that we took $\mubare$ to be positive), which is a solution
 of the bulk equations of motion \buleom.
The supersymmetry variation \sust\ of \zzc\ is
\eqn\suszzc{\eqalign{
 &\delta\phi=0\cr
 &\delta\psi= {\zeta- \bar \zeta \over 2by}\cr
 &\delta\bar\psi= {\zeta-\bar \zeta\over 2by}\cr
 &\delta F=0.
}}
The variation vanishes for $\zeta=\bar\zeta$ and therefore $Q+\bar Q$
is unbroken.  This corresponds to $\eta=1$ above.  More
generally, it is $\mu_0\eta >0$.  We see that that
semiclassical ZZ branes prefer one sign of $\eta$.

 \item{2.} Free boundary conditions lead to the supersymmetric
 version of the FZZT branes \refs{\fzz,\teschner}.  The boundary equations
 of motion are
 \eqn\nbc{\eqalign{&\psi -\eta\bar\psi\big|_{y=0}=0\cr
 &\partial_y \phi + 2\eta\mubare b e^{b\phi}\big|_{y=0}=0,}}
 where we recognize the contribution of the boundary cosmological
 constant.  The classical Euclidean equations of motion
 have no regular solutions with these boundary conditions.%
\foot{The solution is essentially \zzc, so both $\partial_y\phi$
and $e^{b\phi}$ diverge at the boundary.  In particular, the
length of the boundary is infinite.  The choice $\eta=1$ barely
misses being a solution; the boundary cosmological constant is at
the critical value -- if it were infinitesimally larger, regular
solutions would exist.  Bulk vertex operators modify the classical
solution, but it still has infinite boundary length.  For
$\eta=-1$ there is no classical solution.}

\noindent
The two types of boundary condition \dbc\ and \nbc\
are simply expressed in terms of the tangential
and normal super-derivatives at the boundary:
\eqn\superbc{\eqalign{
   D_t\Phi&=(D+\eta\bar D)\Phi=0 \qquad {\rm Dirichlet} \cr
   D_n\Phi&=(D-\eta\bar D)\Phi=0 \qquad {\rm Neumann}
}}
which are equivalent to \dbc, \nbc\ modulo the equations of motion.
Here, one is instructed to take the derivatives,
and then set $z=\bar z$, $\theta=\eta\bar\theta$.

An open string ``tachyon'' interaction can be added to the action
of the FZZT branes to make regular the classical solution with
Neumann boundary conditions. It is convenient to write it with the
aid of a fermionic boundary superfield $\Gamma=\gamma+i\theta_t
f$, which realizes a Chan-Paton Hilbert space:
\eqn\openta{\eqalign{
    S_{\rm bry} &=
    \frac{1}{2\pi}\oint\! dxd\theta_t\, \Bigl(\Gamma D_t\Gamma
        + 2i\mub \Gamma e^{\frac b2 \Phi}\Bigr) \cr
    &= \frac{1}{2\pi}\oint\!dx\Bigl[
        \gamma\partial_x\gamma- f^2
        -\mub\Bigl(b \gamma(\psi+\eta\bar\psi) e^{\frac b2\phi}
            + 2fe^{\frac b2\phi}\Bigr)\Bigr] \cr
    &= \frac{1}{2\pi}\oint\!dx\Bigl[\gamma\partial_x\gamma
        -\mub\, b\, \gamma(\psi+\eta\bar\psi) e^{\frac b2\phi}
        +\mubsq\, e^{b\phi}\Bigr]
 }}
where in the last line we have eliminated the auxiliary field $f$.
Note that $\mub$ is odd under $(-1)^\FL$. The quantization of the
fermion $\gamma$ realizes it as a Pauli matrix acting on a
two-dimensional Hilbert space living on the boundary. Besides
introducing this Chan-Paton space, another effect of the boundary
interaction is to shift the boundary cosmological constant
$-\eta\mubare$ introduced by the bulk cosmological term by
$-\eta\mubare\to-\eta\mubare+\mubsq$. The introduction of $\mub$
changes the boundary conditions in such a way that (when it is
sufficiently large) classical solutions exist; they have finite
$\phi$ on the boundary, and hence finite boundary length.

In the matrix model realization of bosonic Liouville theory, the
boundary cosmological constant can be identified with the
continuum limit of the coupling to the (redundant) operator which
shifts the matrix by a constant \refs{\MartinecHT,\MooreAG}. The
analogous KPZ scaling of $\mub$ in the fermionic string motivates
us to identify it with the analogous redundant coupling in the
two-sided matrix model sketched in the introduction.

The open string ``tachyon'' interaction \openta\ is not present
on ``stable'' Dp-branes ($p$ even in type 0A, odd in type 0B),
due to the GSO projection.  It is present on brane-antibrane pairs,
and on ``unstable'' branes ($p$ odd in type 0A, even in type 0B).%
\foot{The reason for the quotes is that the linear dilaton can in some
cases stabilize a brane by lifting the open string tachyon to zero
mass.}

Let us now list the D-branes of the 2d fermionic string, described
by Liouville theory coupled to a free scalar superfield,
each obeying either Dirichlet or Neumann boundary conditions \superbc.
In the type 0B theory, D(-1) branes and D1 branes are stable.
The D(-1) branes source the R-R scalar $C_0$.  The D1-branes
produce tadpoles for the R-R two-forms which cannot be cancelled.%
\foot{Except perhaps if we include orientifold planes; we leave
this interesting possibility for future work.} Thus we
consider in Lorentzian spacetime only D1-$\overline{\rm D1}$
pairs, or D0 branes. The latter come in two varieties: Dynamical
branes, which are Neumann in $X$ and Dirichlet in $\Phi$; and
spacelike branes, D in $X$ and N in $\Phi$. It was argued in
\KlebanovKM\ that the spacelike D0 in the bosonic string is an
observable $W(\mub,x) = {\Tr}\log [M(x)-\mub ]$ in the matrix path
integral of the open string tachyon on dynamical D0's.
One expects that the spacelike  D0 brane of the type 0 theory
corresponds to similar observables in the two-sided matrix model.

For type 0A the situation is as follows. The stable D0 branes
source a R-R gauge field.  However, in this case the
resulting tadpole merely leads to
a constant R-R electric field; after cancelling this energy,
a sensible theory remains behind. There is a discrete
family of theories labelled by the net D0 charge. A spacelike
D0-$\overline{\rm D0}$ pair gives a macroscopic loop observable of
the matrix model (the boundary interaction \openta\ is naturally
expressed in terms of a complex fermionic superfield $\Gamma$).
There are also sphaleron-like D($-1$)'s in the theory.

The type 0A D1 brane, as well as the type 0B D1-$\overline{\rm D1}$
pair, do not carry R-R charge, but the open string ``tachyons''
on them are in fact massless due to the contribution of
the linear dilaton. Consideration of this open-closed string
theory lies beyond the scope of our discussion here.


\newsec{Minisuperspace wavefunctions}

The minisuperspace approximation truncates the dynamics
to the zero modes on the strip or the cylinder.
In bosonic Liouville theory,
this truncation serves as an important source of intuition
about the dynamics of the full theory.
We will now perform the analogous truncation of the fermionic string.

In the NS sector, there are no fermion zero modes,
and the dynamics is much the same as in the bosonic theory.
In the R sector, one has supersymmetric Liouville quantum mechanics.
More specifically,
we have ${\cal N}=2$ supersymmetric quantum
mechanics with Euclidean time $\tau$ (which differs by a factor of
2 from its value above).
The supersymmetry operators and supercovariant derivatives are
 \eqn\sderiv{\eqalign{
 D=&{\partial \over \partial \theta} + \theta {\partial
 \over \partial \tau}; \qquad
 \bar D= {\partial \over \partial \bar \theta} +  \bar
 \theta {\partial \over \partial \tau} \cr
 Q=&{\partial \over \partial \theta} -\theta {\partial
 \over \partial \tau}; \qquad
 \bar Q= {\partial \over \partial \bar \theta} - \bar
 \theta {\partial \over \partial \tau}
}}
Taking the superpotential%
\foot{In this section only, we ignore the distinction between the
bare cosmological constant $\mubare$ and the physical quantity
$\mu$ \musuper.} $ W(\Phi) = \mu e^{b\Phi}$, the action is
\eqn\actionaq{\eqalign{
  S=&\int d\tau d^2\theta \left({1\over 2} D\Phi \bar D \Phi +i
  W(\Phi)\right)  \cr
 =& \int d\tau \left({1\over 2}({d\over d\tau} \phi)^2 +{1\over 2}
  \psi{d\over d\tau} \psi +{1\over 2} \bar\psi{d\over d\tau}
  { \bar\psi} - {1\over 2} F^2 -  \mu b e^{b\phi}F + i\mu
  b^2e^{b\phi}  \psi\bar\psi \right)\cr
}}
and supersymmetry variation
 \eqn\sustq{\eqalign{
 &\delta\phi=i\zeta \psi + i \bar\zeta\bar \psi\cr
 &\delta\psi=-i\zeta {d\over d \tau} \phi
 +\bar\zeta F\cr
  &\delta\bar \psi=-\zeta F - i\bar \zeta
  {d\over d \tau}  \phi \cr
 &\delta F=- \zeta {d\over d \tau}  \bar \psi
 +\bar \zeta {d\over d \tau}  \psi.}}
After rotating to Lorentzian time and performing canonical
quantization the supercharges are $Q=-p\psi +W'(\phi)\bar\psi$,
$\bar Q=-p\bar \psi -W'(\phi) \psi$. The two fermions $\psi$ and
$\bar\psi$ are represented by two dimensional matrices, say
$\psi={1\over \sqrt 2} \sigma_1$,  $\bar \psi={1\over \sqrt 2}
\sigma_2$, and $\psi \bar \psi={i \over 2} \sigma_3= {i \over 2}
(-1)^F$.  They can be represented as
 \eqn\qmat{\eqalign{
 &{1\over  \sqrt 2} ( Q - i \bar Q)=i\pmatrix{
 0& {\partial \over \partial \phi} + W'(\phi)\cr
 0&0}=i\pmatrix{
 0& {\partial \over \partial \phi} +  b \mu e^{b\phi}\cr
 0&0}\cr
 & {1\over  \sqrt 2} ( Q + i\bar Q) =-i \pmatrix{
 0&0\cr
 -{\partial \over \partial \phi} + W'(\phi)&0}=-i \pmatrix{
 0&0\cr
 -{\partial \over \partial \phi} + b \mu e^{b\phi}&0}}}
For every positive energy $E={1\over 2} b^2 p^2>0$, there are two
states $|p+ \rangle=\pmatrix {\Psi_{p+}(\phi)\cr 0}$ and
$|p-\rangle=\pmatrix {0\cr \Psi_{p-}(\phi)}$ which satisfy
 \eqn\twoeq{\eqalign{
 &\left({\partial \over \partial \phi} + b \mu
 e^{b\phi}\right)\Psi_{p-}(\phi) = b p\Psi_{p+}(\phi) \cr
 &\left(-{\partial \over \partial \phi} +  b \mu
 e^{b\phi}\right)\Psi_{p+}(\phi) = b p\Psi_{p-}(\phi)\ ;}}
thus
\eqn\oneeq{
\left(-{\partial^2 \over \partial \phi^2} \pm  b^2
 \mu e^{b\phi} +  b^2 \mu^2 e^{2b\phi}\right)\Psi_{p\pm}(\phi)
 = b^2 p^2 \Psi_{p\pm}(\phi).
 }
Recall our convention that $\mu$ is positive.  Then, in terms of
$z= \mu e^{b\phi}$ we can write equations \twoeq, \oneeq\ as
\eqn\twoeqtwo{\eqalign{
 &\left(z{\partial \over \partial z} + z\right)\Psi_{p-} =
  p\Psi_{p+} \cr
 &\left(-z{\partial \over \partial z} +  z\right)\Psi_{p+} =
 p\Psi_{p-}\cr
 &\left(-\left(z{\partial \over \partial z}\right)^2 \pm  z +
 z^2 -p^2 \right)\Psi_{p\pm}=0\ .
}}
for the R minisuperspace dynamics;
for the NS sector we have simply
\eqn\nsmss{
 \left(-\left(z{\partial \over \partial z}\right)^2  +
 z^2 -p^2 \right)\Psi_{p0}=0\ .
}
The latter is identical to the corresponding equation in
bosonic Liouville theory (the theory with $\psi=\bar\psi=0$)
\refs{\SeibergEB,\PolchinskiMH,\MooreIR}. It lacks the term
proportional to $z$ which arises from the fermions.

We impose that the wave functions $\Psi_{p\pm}$, $\Psi_{p0}$ decay
deep under the Liouville potential ($\phi \to +\infty$).  These
equations with these boundary conditions are solved by Whittaker
functions or Bessel functions
\eqn\witso{\eqalign{
 &\Psi_{p+}={p \over \sqrt{2z}} W_{\lambda=-{1\over 2},
 \nu=ip}(2z)=-i \sqrt{z \over 2\pi} (K_{ip + {1\over 2}}(z)
  - K_{ip - {1\over 2}}(z)  )\cr
 &\qquad ={p e^{-z}\over 2 z}  \left(1 - {1 + p^2 \over 2 z }
 + { (p^2+1)(p^2+4)\over 8 z^2} -{ (p^2+1)(p^2+4)(p^2+9)\over 48
 z^3}+ \CO({1\over z^4}) \right)\cr
 &\Psi_{p-}={1 \over \sqrt{2z}} W_{\lambda={1\over 2},
 \nu=ip}(2z) = \sqrt{z \over 2\pi}
 (K_{ip + {1\over 2}}(z)+ K_{ip - {1\over 2}}(z)) \cr
 &\qquad = e^{-z} \left(1 -{ p^2 \over 2 z } + {p^2(p^2+1)
 \over 8  z^2}  -{p^2 (p^2+1)(p^2+4)\over 48 z^3}+ \CO({1\over
 z^4})  \right)\cr
 &\Psi_{p0}={1\over \sqrt{2z}} W_{\lambda=0,
 \nu=ip}(2z)={1\over \sqrt \pi} K_{\nu=ip}(z).
}}
We would like to make a few comments about these functions:
 \item{1.} All the wave functions decay at
 large $\ell=e^{b\phi}$ as $e^{-z}= e^{-{\mu\ell }}$.
 For small $z=\mu\ell$ and generic $p$ all these functions
 are linear combinations of terms of the form $z^{ip}(1 + \CO(z))$
 and $z^{-ip}(1 +\CO(z))$.  These represent incoming and outgoing
 waves in the Liouville coordinate $\phi$.
 \item{2.} For $p=iN$ with integer $N$ the Ramond functions
  $\Psi_{iN,\pm}(z)$ are elementary
  functions.  This is analogous to the fact that for
  $p=i(N+{1\over 2})$ the NS functions $\Psi_{i(N-{1\over 2}),
  0}(z)$ are elementary.
 \item{4.} For $p=i(N+{1\over 2})$ the Ramond functions exhibit
 a resonance phenomenon between the growing solution
 $z^{|N+{1\over 2}|}(1 +\CO(z))$ and the
 decaying solution $z^{-|N+{1\over 2}|}(1 +\CO(z))$.  These appear
 as corrections proportional to $\log z$ to the decaying solution.
 This is analogous to a similar phenomenon in the NS sector for
 $p=iN$.  These logarithms are related to the ``leg poles''
 \redef\ and to the discrete states of $\hat c=1$ \DiFrancescoUD.
 \item{5.}  In the NS sector, the delta function normalizable
 spectrum includes all positive energy states but there is no zero
 energy state.  The zero energy eigenfunction $\Psi_{p0}={1
 \over \sqrt {2z} } W_{\lambda=0, \nu=0}(2z)={1\over \sqrt \pi}
 K_{\nu=0}(z)$ satisfies the
 correct boundary conditions at $\phi\to +\infty$ but it grows
 linearly in $\phi \sim \log z$ near $\phi \to -\infty$, and
 therefore it is not normalizable.  (This is an example of the
 resonance phenomenon we mentioned above.)  Surprisingly, the
 situation is different in the Ramond sector, which does not
 exhibit resonances at integer $ip$.  Restoring the possibility of
 $\mu$ of different signs, we find a single Ramond state for $p=0$
\eqn\bouwa{
  |p=0\rangle=\cases{
    |+ \rangle=\pmatrix {\Psi_{0+}(\phi)=e^{W(\phi)}
        =e^{\mu e^{b\phi}}\cr 0} & $\mu <0$ \cr
    {\ }^{\ }_{\ } &\  \cr
        |-\rangle=\pmatrix {0\cr
        \Psi_{0-}(\phi)=e^{-W(\phi)}
        =e^{-\mu e^{b\phi}}} & $ \mu > 0$.
}}
 This wave function is (delta-function) normalizable and represents
 a supersymmetric zero energy ground state.

\noindent
For the application to $\hat c=1$ strings, we should append
to the minisuperspace quantum mechanics the dynamics
of the free coordinate $X$ \superf, whose supercharges
analogous to \qmat\ act in a two-dimensional Hilbert
space representation of the fermions $\chi$ and $\bar\chi$.
The full fermion Hilbert space is then the four-dimensional
tensor product space, with
$Q=Q_\phi\otimes \One + \One\otimes Q_x$, and so on.
The choice of fermion number projection
$(-1)^F=\sigma_3\otimes(\pm\sigma_3)$ determines
whether we are in type 0B or type 0A.

Equations \twoeq\ are similar to \chieom\ which were derived from
the spacetime Lagrangian.  This provides a motivation for our
assertion in section 2 that $f_3(T)$ of \seffacta\ is proportional
to $e^{-2T}$.

Since the wave functions $\Psi_{p\pm}$ satisfy the same equations
as the components of the field strength $F=d\chi + \chi dT =
e^{-T} d(e^T \chi)$ (see discussion after \chiki), they are
identified with them rather than with the fundamental fields $C_0$
or $\chi$. This one-form $F$ is used in vertex operators in the
$(-{1\over 2},-{1\over 2})$ picture, while $\chi$ is used in
$(-{1\over 2},-{3\over 2})$ or $(-{3\over 2},-{1\over 2})$
picture.  Therefore, even though the zero energy solution
$e^{-|\mu|\ell}$ is a delta function normalizable state in
Liouville, the corresponding wave function in string theory, which
is computed using the field $\chi$ rather than $F$ is not
normalizable.  Therefore, it corresponds to a microscopic operator
in the terminology of~\SeibergEB.

More generally, if the theory we consider includes in addition to
super-Liouville also another superconformal matter field theory
with central charge $\hat c$, we can try to form supersymmetric
states out of different states in the superconformal field theory.
If the state in the conformal field theory is a Ramond primary
with dimension $\Delta>{\hat c/ 16}$, the condition for unbroken
supersymmetry states that the total energy of the state vanishes.
This means that in the Liouville part of the theory we need a
state with negative energy (imaginary $p$ in \twoeq, \oneeq) whose
wave function is not normalizable \SeibergEB. Also, the
supercharge $Q+\bar Q$ does not annihilate the Liouville part of
the wave function nor the matter part of the wave function.
Instead it leads to a relation between them. This condition is the
spacetime Dirac equation. In our case the differential operator in
the Liouville direction in this ``Dirac equation'' is
${\partial\over \partial \phi} \pm \mu b e^{b\phi}$ of \twoeq.
Therefore, in terms of ${1\over 2} \nu^2 b^2= \Delta -{\hat c\over
16}$ the wave function is $\Psi_{i\nu \pm}(\phi)$ where the sign
is determined according to the fermion number of the matter wave
function and whether we study the 0A or the 0B theory. For $\phi
\to -\infty$ it behaves like $\Psi_{i\nu\pm}\sim e^{-\nu b\phi}$
with $\nu>0$. The $\phi$ dependence of the corresponding vertex
operator is determined from the asymptotic behavior as $\phi \to
-\infty$ of $V\sim g_s(\phi)\Psi(\phi) \sim e^{({Q\over 2}-\nu
b)\phi}$; $\nu b$ is the momentum in the $\phi$ direction.

\subsec{The transform to free fields}

One of the intriguing features of bosonic gravity is the observation
of \refs{\MooreIR,\MooreAG}
that the minisuperspace wavefunctions are related
to the mode functions of eigenvalue collective field theory
via an integral transform
\eqn\jevtrans{
\eqalign{K_{iE}(\mu\ell)=&\int_0^\infty d\tau\;
    \bigl(e^{-\mu\ell\ch\tau}\bigr)\ \cos\,E\tau\cr
    \frac{\pi\cos\,E\tau}{E\;\sh\,\pi E}=&\int_0^\infty \frac{d\ell}\ell
    \;\bigl(e^{-\mu\ell\ch\tau}\bigr)\ K_{iE}(\mu\ell)\ .}
}
It was noted that this transform bears a striking similarity to the Backlund
transform that converts Liouville theory into free field theory.
In retrospect this is not so surprising, since the kernel
of the full-fledged 1$+$1 field theory functional integral transform
\eqn\nkernel{W[\phi,\psi]=\exp\left[\smallint d\sigma
    (\phi\psi'-\mu e^{\gamma\varphi/2}\sh\psi)\right] }
has the property that it converts the Liouville Hamiltonian in $\phi$
into the free field Hamiltonian in $\psi$, hence the full 2d field theory
Wheeler-deWitt equation into a free field one.%
\foot{The reason
$\sh\psi$ appears here and $\ch\tau$ appears in \jevtrans\
is due to the relation between Minkowski and Euclidean
2d Liouville theory.}
Thus one may regard \jevtrans\ as the minisuperspace
truncation of the Backlund transform.

In the fermionic string the super-Backlund transformation
\DHokerZY\
has the same property, that it converts the interacting Liouville
super-WdW equation into a free field one.  Hence we should search for
an integral transform
that turns the Ramond wavefunctions into sines and cosines.
The answer is
\eqn\rjevtrans{
\eqalign{\sqrt{\mu\ell}\bigl(K_{\half+iE}+K_{\half-iE}\bigr)
    (\mu\ell)=&\int_0^\infty d\tau\;
    \bigl((\mu\ell)^\half
        e^{-\mu\ell\ch\tau}\ch\coeff\tau2\bigr)
    \ \cos E\tau\cr
    \frac{\pi\cos E\tau}{\ch\,\pi E}=&\int_0^\infty\frac{d\ell}\ell\;
    \bigl((\mu\ell)^\half
        e^{-\mu\ell\ch\tau}\ch\coeff\tau2\bigr)
    \ \sqrt{\mu\ell}\bigl(K_{\half+iE}+K_{\half-iE}\bigr)(\mu\ell)\cr
  -i\sqrt{\mu\ell}\bigl(K_{\half+iE}-K_{\half-iE}\bigr)
    (\mu\ell)=&\int_0^\infty d\tau\;
    \bigl((\mu\ell)^\half
        e^{-\mu\ell\ch\tau}\sh\coeff\tau2\bigr)
    \ \sin E\tau\cr
    \frac{\pi\sin E\tau}{\ch\,\pi E}=&\int_0^\infty\frac{d\ell}\ell\;
    \bigl((\mu\ell)^\half
        e^{-\mu\ell\ch\tau}\sh\coeff\tau2\bigr)
    (-i)\sqrt{\mu\ell}\bigl(K_{\half+iE}-K_{\half-iE}\bigr)(\mu\ell)\ .}
}
Let us now derive this result from the minisuperspace dynamics
for super-Liouville.

One way to view the transform \jevtrans\ is that
in the minisuperspace approximation,
the `boundary state wavefunction' is
$\Psi_{\!B}=exp[-\rho e^{b\phi}]$;
it implements the Neumann boundary condition \nbc,
since $\dot\phi=\partial_\phi=-b\rho e^{b\phi}$
when acting on the boundary state wavefunction.
The `disk one-point function' is the inner product
of this wavefunction with the minisuperspace
wavefunction $\Psi_{p0}$ of a `closed string primary'
of the NS sector of the fermionic string, or of the bosonic string.
This gives the transform in the second line of \jevtrans,
if we set $\rho=\mu\cosh\tau$.

In the R sector, the boundary state wavefunction
includes the boundary (zero-mode) part
of the bulk Liouville action \action,
as well as the boundary interaction \openta:
\eqn\psibdy{
  \Psi_{\!B}(\phi,\psi,\bar\psi) =
    (\psi-\eta\bar\psi)\,
    \exp\Bigl[\mub\, b\,\gamma(\psi+\eta\bar\psi)e^{\frac b2\phi}
    -(-\eta\mu+\mubsq)e^{b\phi}\Bigr]
}
The fermion factor in front of the exponential
implements the boundary condition \nbc\ on the fermions.
We can also organize the Liouville wavefunction \witso\ as
$\Psi_p=\Psi_{p,-\eta}(\phi)+(\psi+\eta\bar\psi)\Psi_{p,\eta}(\phi)$
and write the overlap as
\eqn\innerprod{
  \langle \Psi_p | \Psi_{\!B} \rangle =
    \int d\phi d\psi d\bar\psi d\gamma\;
        \Psi_p^\dagger(\phi,\psi,\bar\psi)
        \Psi_{\!B}(\phi,\psi,\bar\psi)\ ;
}
after integrating over Grassmann variables the result is
\eqn\rjevtrue{
   \int \frac{d\ell}{\ell} 2\eta\mub \sqrt\ell\,
    \exp[-(-\eta\mu+\mubsq)\ell]\,
    \Psi_{p,-\eta}^\dagger(\phi,\psi,\bar\psi)\ .
}
Plugging in \witso, and defining $\tau$ via
\eqn\taudef{\eqalign{
  \rho \equiv |\mu|\cosh\tau &= -\eta\mu+\mubsq \cr
  \mub &= \pm\sqrt{|\mu|\cosh\tau+\mu\eta}\ ,
}}
we recognize the second and fourth lines of \rjevtrans.
Note that for $\eta\mu<0$,
the effective boundary cosmological constant $\rho$
is real and positive all the way down to $\mub=0$,
leading to a sensible transform and
wavefunctions concentrated at finite semiclassical boundary loop length.
In contrast, for $\eta\mu>0$ one must turn on a finite $\mubsq>2|\mu|$
for the integral transform to converge.


\newsec{Quantum super-Liouville theory}

The conformal bootstrap for supersymmetric Liouville theory
has been analyzed in
\refs{\RashkovJX,\PoghosianDW}, and
the boundary bootstrap is treated in \refs{\FukudaBV,\korean}.
The latter works find the exact disk one-point functions
and annulus partition functions; these are most conveniently
expressed in terms of the boundary state wavefunctions
introduced in \refs{\fzz,\teschner,\zz}
in the context of bosonic Liouville theory.


\subsec{Neumann boundaries}

The quantum version of the super-FZZT (Neumann) boundary state
has the boundary state wavefunction \refs{\FukudaBV,\korean}
\eqn\bsn{\eqalign{
  \Psi^\NS_{\eta}(\nu,s) &=
    \cos(2\pi\nu s)
    \Bigl[\frac{\Gamma(1+i\nu b)\Gamma(1+i\nu/b)}%
        {(2\pi)^{1/2}\; (-i\nu)} (2\mukpz)^{-i\nu/b}\Bigr] \cr
  \Psi^\RR_{\eta=+{\rm sgn}(\mu)}(\nu,s) &=
    \cos(2\pi\nu s)
    \Bigl[\frac{\Gamma(\coeff12+i\nu b)\Gamma(\coeff12+i\nu/b)}%
        {(2\pi)^{1/2} } (2\mukpz)^{-i\nu/b}\Bigr] \cr
  \Psi^\RR_{\eta=-{\rm sgn}(\mu)}(\nu,s) &=
    \sin(2\pi\nu s)
    \Bigl[\frac{\Gamma(\coeff12+i\nu b)\Gamma(\coeff12+i\nu/b)}%
        {(2\pi)^{1/2} } (2\mukpz)^{-i\nu/b}\Bigr] \ ,
 }}
where $\mu$ is given by \musuper; $\eta$ is the sign \boua\ in the
fermion boundary condition. In the limit $b\to 0$, these
wavefunctions precisely reproduce (up to phases) the $\tau$-space
wavefunctions \jevtrans, \rjevtrans\ of the minisuperspace
approximation, provided we identify $E=\nu/b$, \hbox{$\tau=2\pi sb$}. Away
from this limit, one cannot ignore the contribution of the gamma
function $\Gamma(1+i\nu b)$, whose presence is required by the
$b\to 1/b$ symmetry of quantum Liouville field theory
\ZamolodchikovAA. This is the main difference between the
minisuperspace approximation to the wavefunctions and those of the
full 2d field theory.  This factor can be absorbed in an $s$
independent redefinition of the wave functions.

The boundary state wavefunction \bsn\ determines
the one-point function of closed string vertex operators
on the disk with Neumann boundary conditions
in the Liouville direction,
when combined with the
wavefunction of the matter system.
With Dirichlet boundary conditions for matter,
the analogous quantity in the bosonic
matrix model has the interpretation of a wavefunction
of a string state \refs{\MooreIR,\MooreAG},
given by the correlation function of a local operator
with a macroscopic loop;
it is natural to propose a similar interpretation here.

R-R closed string vertex operators couple to `unstable' D-branes
such as the type 0B D0-brane only in the presence of an open
string tachyon \BilloTV\ (as for example in section 11 below); the
couplings vanish as the tachyon background is turned off.

The wavefunctions \bsn\ also enter into the annulus partition function.
In the bosonic theory this amplitude is related to
the correlation function of two macroscopic loops
in the matrix model \MooreAG.
The super-Liouville annulus partition function has the form
\eqn\Zann{
  {\cal Z}_{L}^a(\eta,s;\eta',s') = \int d\nu\,
    \Psi^a_{\eta}(\nu,s)\Psi^a_{\eta'}(-\nu,s')
    \chi_\nu^{a,\eta\eta'}(q)\ ,
} where $\chi_\nu^{a,\pm}(q)$ is the character of a nondegenerate
superconformal representation of Liouville momentum $\frac
Q2+i\nu$; $a=\NS, \RR$ is the fermion boundary condition in the
closed string channel and $\eta\eta'=\pm$ is the fermion boundary
condition in the open string channel, $\eta\eta'=+$ for NS and
$\eta\eta'=-$ for R open strings.

Consider the $\hat c=1$ model with $x$ non-compact.
Combining the Liouville partition function with
the corresponding partition functions for the ghosts
and the non-compact boson $x$
obeying Dirichlet boundary conditions, one finds that
the oscillator contributions cancel, leaving the
integral over the zero modes.  We omit the details,
since they are essentially identical to the calculation in
\MartinecKA.  The annulus amplitude for spacelike D0's is
\eqn\twoloop{ \eqalign{
  {\cal Z}_{\NS}(s,s';p) &=
     \int\!d\nu\;
         \frac{\cos(2\pi \nu s)\cdot\cos(2\pi\nu s')}%
         {\sinh^2(\pi \nu)\;(\nu^2+\frac{\alpha'}{2}p^2)} \cr
  {\cal Z}_{\RR}(s,s';p) &=
    \int\!d\nu\;
         \frac{\sin(2\pi\nu s)\cdot \sin(2\pi\nu s')}
             {\cosh^2(\pi \nu)\;(\nu^2+\frac{\alpha'}{2}p^2)} \ .
 }}
(implicitly there is a regulator to cut off the small $\nu$
divergence in the NS integral). We have assumed here $\eta=-1$;
for $\eta=+1$, the sines are replaced by cosines in the Ramond
partition function.

These amplitudes yield off-shell information
about the spacetime theory.
For example, they have
poles at the locations corresponding to the discrete states
in the appropriate sector, see the discussion after
equation \witso.


\subsec{Dirichlet boundaries}

There are also super-ZZ (Dirichlet) boundary states
associated to degenerate representations of the superconformal
algebra.  These are labelled by a pair of integers $(m,n)$,
with $m-n$ even(odd) for NS(R) representations.
Their wavefunctions are \refs{\FukudaBV,\korean}
\eqn\bsd{\eqalign{
  \Psi_{\!\NS}(\nu;m,n) &=
    2\sinh(\pi m\nu/b)\sinh(\pi n\nu b)
        \Bigl[\frac{\Gamma(1\!+\!i\nu b)\Gamma(1\!+\!i\nu/b)}%
                {(2\pi)^{1/2}\; (-i\nu)} (2\mukpz)^{-i\nu/b}\Bigr] \cr
  \Psi_{\!\RR}(\nu;m,n) &=
    -i^{m+n}
    \,2\sin[\pi m(\coeff12\!+\!i\nu/b]\sin[\pi n(\coeff12\!+\!i\nu b)]
        \Bigl[\frac{\Gamma(\coeff12\!+\!i\nu b)\Gamma(\coeff12\!+\!i\nu/b)}%
                {(2\pi)^{1/2} } (2\mukpz)^{-i\nu/b}\Bigr]
 }}
Normalized disk one-point functions are \eqn\normonept{
  U_{\NS,\RR}(\nu;m,n)=\frac{\Psi_{\NS,\RR}(\nu;m,n)}%
    {\Psi_\NS(iQ/2;1,1)}\ .
 }
In the semiclassical limit $b\to 0$, the normalized one-point
function of $e^{b\Phi}$ ({\it i.e.} $i\nu=\half (b-b^{-1})$)
reproduces the semiclassical geometry \zzc, provided $m=1$. As in
\zz, we expect the diagrammatic expansion in Liouville theory will
match only the properties of the $m\!=\!n\!=\!1$ boundary state in
this limit.

A key feature of the ZZ branes is that only a finite list
of (degenerate) boundary operators couple to them --
the $(m,n)$ degenerate boundary operator can create
open strings stretched between the $(m',n')$ and $(m'',n'')$
boundary states only if the $(m,n)$ degenerate representation
appears in the fusion of $(m',n')$ and $(m'',n'')$,
{\it i.e.} only if
\hbox{$m\in\{|m'\!-\!m''|\!+\!1,\dots,m'\!+m''\!-\!1\}$},
and similarly for $n$.
In particular, only the identity operator couples to the $(1,1)$ brane.
This means that the effective open string dynamics on a collection
of such branes involves only the open string tachyon
and gauge field, which depend only on the time coordinate $x$.

It is argued in \FukudaBV\ that degenerate
boundary states exist only for $\eta = (-1)^{m-n}$.
For $m\!=\!n\!=\!1$, this agrees with the
value determined from the semiclassical analysis of
Dirichlet boundaries in section 5. In general rational conformal
field theories the sign of $\eta$ for the boundary state is determined
by whether the label of the Cardy state is an NS or Ramond operator.
Since the $(1,1)$ ZZ brane is labeled by an NS operator (the identity)
then we see that it has $\eta =1$.
A consequence of this property is that
the matrix model of tachyons on $N$ $(1,1)$ ZZ branes
involves only one type of brane.

The ZZ branes are the appropriate boundary states which describe,
in quantum Liouville theory, dynamical type 0 D0-branes.
Below, we will restrict our consideration to the simplest
possibility, namely that all such D0 branes are built
out of the $(m,n)=(1,1)$ boundary state.


\newsec{Spacetime Effective Dynamics of Type 0 D-Branes}

D0 branes in type 0A couple to only one of the two R-R vector
fields, which are governed by the action \seffactb.  The relevant
part of that action is\foot{Below we do not have to use the
asserted form of $f_3 =e^{-2T}$, but only its expansion in powers
of $T$.  For large and negative $\phi$ where $T$ is small $f_3 =1
-2T+\cdots$.}
 \eqn\Aeffnew{
 S=- \int d^2 x \sqrt{G} \left (
    {2 \pi \alpha' \over 4} f_3(T)(F^{(+)}_{\mu\nu})^2 +
    {2 \pi \alpha'\over 4}  f_3(-T) (F^{(-)}_{\mu\nu})^2
\right ) \ . } Thus, as $T$ grows, so does the asymmetry between
the two gauge fields: one of them becomes more weakly coupled
while the other becomes more strongly coupled. We are interested
in $T=\mu e^{\phi}$ which corresponds to the Liouville theory.  In
the absence of a background value of $T$, there are two types of
stable D0-branes: those charged under $F^{(+)}_{\mu\nu}$ and those
charged under $F^{(-)}_{\mu\nu}$. The branes charged under each
gauge field are branes with the two possible signs for $\eta$.
As we saw above the ZZ brane, with $(n,m) =(1,1)$, has $\eta=1$ and
therefore is charged only under one gauge field. In fact
it is charged under $F^{(-)}$.\foot{
It is amusing to note that the first term of a coupling of the form $
\int dt [e^{-\Phi} (1 + T +\cdots) + C^{(\pm)} +\cdots] $ on the
worldvolume of the brane would give the right mass for the ZZ brane,
and also for the bosonic string ZZ brane. }

We can make an analogous argument in the world sheet language.
The origin of the doubling of D-brane spectrum
in conventional type 0 theories is that one can impose
fermion boundary conditions $\psi_L = \psi_R$ or $\psi_L=- \psi_R$.
The two types of D-branes are
related by the transformation
$\psi_L \rightarrow - \psi_L$. However, after the Liouville potential is
turned on, this transformation is no longer a symmetry.
So, the existence of one type of D-brane no longer
guarantees the existence of the other.
This is a way of seeing from the spacetime point of
view the fact that each ZZ boundary state comes in only one variety.

Analogous arguments applied to type 0B theory show that there is
only one type of D-instanton. It corresponds in the matrix model
to an eigenvalue tunnelling from the left well to the right well,
while an anti-instanton corresponds to tunnelling in the opposite
direction. An unstable D0-brane in 0B  is obtained by starting with D0
brane-antibrane pair of the 0A theory and applying a $Z_2$
projection. In this way we obtain an uncharged unstable D0-brane.
It should correspond to an eigenvalue at the unstable symmetric
point of the double-well potential. If we have $N$ such unstable
D0 branes we have a $U(N)$ theory with the tachyon in the adjoint
of $U(N)$.

In the type 0A theory we can have $N$ D0 branes and $M$ anti-D0
branes, and the theory on the branes is $U(N) \times U(M)$
Yang-Mills quantum mechanics with a complex tachyon in the
bifundamental. Let $N-M=q>0$.  There will be $q$ D0 branes that
cannot decay. In the eigenvalue space of the matrix, the $q$
D0-branes are localized near the origin. In the Liouville
coordinate $\phi$, they are near $\phi=\infty$. Since the branes
are charged, they lead to $q$ units of a constant background flux
of one of the R-R vector field strengths. For positive $\mu$ it is
$F^{(-)}$ in \Aeffnew. Ignoring the NS tachyon field, the
effective Lagrangian of the system is
 \eqn\effoA{e^{-2\Phi}\left(R+4(\partial \Phi)^2 +4\right) -
 {2 \pi \alpha' \over 4} (F^{(-)})^2 +  q F^{(-)}.}

The coupling $qF^{(-)}$ is like a $\theta$ angle for $F^{(-)}$.
Nevertheless, its value cannot be changed by the standard process
of pair creation.  The reason for that is that the mass of the
D0-branes depends on $\phi$.  Therefore, it takes infinitely more
energy to create a pair and to separate it to infinity than is
being gained by screening the background charge. $q$ is an integer
and it represents the net number (branes minus antibranes)
of ZZ D0 branes at $\phi = \infty$.

Integrating out the gauge field, we find the Lagrangian
 \eqn\effoAe{e^{-2\Phi}\left(R+4(\partial \Phi)^2 +4\right)
 - { q^2 \over 4 \pi \alpha'} .}
This Lagrangian and its classical solutions were studied in
\BerkovitsTG.

In our problem there is also an NS tachyon field $T$.  It is
sourced both by the linear dilaton and the background R-R field
strength $F^{(-)}$. For example, as in \Aeffnew, we have a
coupling of the form ${1\over 4} e^{2T}  (F^{(-)})^2 $. Near
$\phi=-\infty$ the Lagrangian \effoAe\ can be used because the
string coupling is small and the tachyon and the backreaction are
exponentially suppressed. However, for finite $\phi$ the
corrections to the leading order Lagrangian \effoAe\ are
important.

Since in most of spacetime the Lagrangian \effoAe\ can be used, we
can conclude that the $q$ dependence of the torus amplitude is given
by
\eqn\zoneq{
{ \cal Z}_q =  { q^2 \over 4 \pi \sqrt{ 2 \alpha'} }
 \ln (|\mu|/\Lambda)
 ~,}
  where $\Lambda$ is a
cutoff on the Liouville direction $\phi$, and the factor of
$\ln (\Lambda/|\mu|)$ is the effective length of this direction.  It is
cut off at $\ln |\mu|$ by the coupling to the tachyon which we
neglected in \effoAe. The coefficient of this term can be exactly
calculable using this Lagrangian by properly normalizing the
various terms.  Note that this contribution is infinite and
therefore the one loop contribution to the D-brane mass is
infinite.  Therefore, unlike ordinary D-branes in the critical
string, such D0-branes are not finite energy excitations of the
theory.  They constitute different sectors which are separated by
infinite energy.

The net charge $q$ is a parameter in the theory, which corresponds
to a background R-R field in the worldsheet action. However, since
it is quantized, it cannot be changed in a continuous fashion.


\newsec{The Matrix Model for Type 0B Strings}

The type 0B version of the ZZ boundary state
\refs{\zz,\FukudaBV,\korean} contains an open string tachyon of
mass $m_T^2 =-1/(2 \alpha')$. The matrix model dual to type 0B
string theory should correspond to the dynamics of $N\to\infty$
such D0 branes. It is described by the double-scaled matrix
quantum mechanics \EQM, with two important modifications. First,
$\alpha'$ has to be replaced by $2\alpha'$ to obtain the curvature
at the top corresponding to the NSR open string tachyon. Second,
the NSR tachyon effective action must be symmetric under the $Z_2$
symmetry $\M \rightarrow - \M$ which, in the world sheet language,
is the spacetime $(-1)^{\FL}$. This is the operation that reverses
the sign of R-R states. Hence, the Fermi sea fills the potential
symmetrically on both sides. This two-cut hermitian matrix model
is equivalent to the double scaling limit of the quantum mechanics
of a unitary matrix whose potential $\sim \cos (\lambda)$; the
eigenvalue distribution is automatically symmetric.

In analogy with the bosonic string case,
we conjecture an exact duality between the
double-well hermitian (or, equivalently, the unitary)
matrix model describing open strings on unstable D0-branes of type 0B theory,
and the closed type 0B strings.
In the double-scaling limit, this implies that closed type 0B strings
are described by $N\to \infty$ eigenvalues moving in the potential
\eqn\epotl{
  V(\lambda)= -{1\over 4\alpha'} \lambda^2\ ,
} and Fermi sea filling both sides symmetrically to Fermi level
$-\mu$ as measured from the top of the potential. Thus, for
$\mu>0$ the Fermi level lies below the top, while for $\mu<0$ it
is above the top. The latter possibility is absent in the matrix
model for the 2-d bosonic string, but is present for the 0B
theory.

This conjecture has immediate consequences
for perturbative closed string dynamics of type 0B strings.
The Fermi sea fluctuations, $T_L (k)$ and $T_R(k)$,
of the left and right sides
of the barrier are perturbatively independent.
Therefore, connected correlation functions involving
both $T_L (k)$ and $T_R(k)$ vanish.
Correlation functions involving either only the left or
only the right modes are the same as
those given by the bosonic matrix model
where only one side of the barrier is filled,
up to the rescaling $\alpha' \rightarrow 2\alpha'$
(or equivalently, the rescaling of momenta $k\rightarrow \sqrt 2  k$).

In section 3,
we saw that the type 0B excitations exhibit
precisely the same structure.
In particular, the first line of equation \mix\
implies that the modes $T_L$ and $T_R$ defined
in \abdef\ decouple on the sphere.  The second line
of \mix\ shows that the tree level scattering amplitudes
of $T_L$ and $T_R$ agree with those of the bosonic string.
This provides a strong argument in favor of the conjecture.

In the matrix model, one can form symmetric and antisymmetric
combinations, \hbox{$T_L\pm T_R$}. These combinations are related
to the natural observables of the type 0B string as follows: The
symmetric combination
\eqn\symtach{ T(k) = e^{i \delta_{NS}(k) }
\left  [T_R (k) + T_L (k)\right ] } is related to the NS-NS
tachyon \vns\ while the antisymmetric combination
\eqn\asymscal{
V(k) =e^{i \delta_{\RR}(k)}
\left [T_R (k) - T_L (k)\right ]\
. }
is related to the R-R scalar \vram. The phase factors in these
equations are given by \eqn\phasesde{
 e^{i \delta_{NS}(k) } =
{ \Gamma(i k \sqrt{\alpha'/2} ) \over \Gamma(-i k\sqrt{\alpha'/2}
) }
    \quad,\qquad
 e^{i \delta_{\RR}(k) } =
{ \Gamma(\half + i k\sqrt{\alpha'/2} )
    \over \Gamma(\half - i k\sqrt{\alpha'/2} ) }\ ,
}
see equation \redef\
($T$ is normalized slightly differently there).

Another argument in favor of our proposal
is related to the fact that, in super-Liouville theory,
both signs of $\mu$ are admissible.
As we showed in section 5,
the transformation $\mu\rightarrow -\mu$ is equivalent to
$\psi_L\rightarrow - \psi_L$.
Similarly, in the quantum mechanics
of a unitary matrix, both signs of $\mu$ are admissible and
the perturbative expansion of the theory is singular as $|\mu|\rightarrow 0$
in both cases.\foot{ The nonperturbative answers are non-singular
as $\mu \to 0 $.}
In the double scaling limit, when
the potential is an inverted parabola, there is a simple transformation
relating theories with opposite signs of $\mu$ \refs{\igor,\joe}:
Interchange of the coordinate $\lambda$ with conjugate
momentum $p=-{i\over \beta}{\partial\over\partial\lambda}$,
accompanied by particle-hole conjugation.
This is evident from the fact that the Fermi surface
$p^2-\lambda^2 = - 2\mu$ becomes transformed into
$\lambda^2- p^2 =- 2\mu$.
Furthermore, the second-quantized Hamiltonian
\eqn\secone{ {\hat H} = \int d\lambda \left\{ {1 \over 2\beta^2 } {\partial
\Psi^{\dagger}(\lambda)  \over
\partial \lambda }
{\partial \Psi (\lambda) \over \partial \lambda} -{\lambda^2\over 2}
\Psi^\dagger (\lambda) \Psi(\lambda)
    + \mu \Psi^\dagger(\lambda) \Psi(\lambda) \right\}
\,\,,
}
becomes in momentum space
\eqn\sectwo{ {\hat H} = - \int dp \left\{ {1 \over 2\beta^2 } {\partial
\tilde \Psi^{\dagger}(p)  \over
\partial p }
{\partial \tilde \Psi (p) \over \partial p} -{p^2\over 2} \tilde
\Psi^\dagger (p) \tilde \Psi(p) - \mu \tilde \Psi^\dagger (p)
\tilde \Psi (p) \right\} \,\,. }

In the Liouville theory the corresponding transformation $\psi_L
\to - \psi_L$ changes the relative sign of the two R-R states, the
selfdual scalar from the $(R+,R+)$ sector and the anti-selfdual
scalar from the $(R-,R-)$ sector. This implies that the left
moving part of the R-R scalar changes a sign relative to the right
moving part. Thus, this is the same as a spacetime T-duality
(electric-magnetic duality) of the massless R-R scalar $C_0$. This
is the S-duality transformation we discussed above.

As we discussed above, filling the two sides asymmetrically
corresponds to adding a  constant gradient for the RR scalar at
$\phi = - \infty$ (see also \mgv). This gradient is in the time or
$\phi$ direction depending on the sign of $\mu$. This corresponds
to the fact that we are either changing the Fermi levels on the
two sides of the potential for $\mu > 0$, or we are changing the
Fermi levels of the left and right movers for $\mu < 0$.\foot{This
statement is true perturbatively.  Nonperturbatively, due to
eigenvalue tunnelling ({\it i.e.} D-instantons), for $\mu>0$ there
is both an asymmetric filling of the two sides and an
exponentially small flux from one side to the other; for $\mu<0$
there is both a net flux of eigenvalues in one direction, and an
exponentially small asymmetry in the filling of the two sides.}

In the remainder of this section
we carry out another sensitive check comparing
the matrix model at finite temperature
to the compact $\hat c=1$ theory coupled to supergravity.


\subsec{Matrix Quantum Mechanics at Finite Temperature}

Consider the matrix quantum mechanics at finite temperature
$T= 1/(2\pi R)$ and its connection
with Liouville and super-Liouville theories.
Again we first review the bosonic case.

Compactified bosonic two-dimensional  string is described by a
Euclidean path integral for a Hermitian $N\times N$ matrix
\eqn\EQM{ {\cal Z}_R =\int D^{N^2}\!\M(x)\exp \biggl
[-\beta\int_0^{2\pi R}dx~\Tr
  \left (\half (D_x \M )^2-{1\over 2\alpha'} \M^2\right
)\biggr ]\ . } The gauge field $A$ acts as a lagrange multiplier
that projects onto $SU(N)$ singlet wave functions. In the
Hamiltonian language, \EQM\ is a path integral representation for
the thermal partition function \eqn\therm {{\cal Z}_R = \Tr
e^{-2\pi R\beta H}\ , } where the trace runs over singlet states
only. Since the singlet wave functions depend only on the matrix
eigenvalues, which act as free fermions \BIPZ, evaluation of the
path integral reduces to studying free fermions in the upside down
harmonic oscillator potential \maxt\ at finite temperature
\GrossKleb. This problem is exactly solvable, and the free energy
as a function of the original ``cosmological constant'' $\Delta$
is specified by the equations \eqn\oldrev{ {\partial {\cal F}\over
\partial\Delta } = \mu\ ,\qquad {\partial\Delta\over \partial \mu}
= \tilde\rho (\mu)\ , } where \eqn\Rduality{ {\partial\tilde
\rho\over \partial \mu} = {\sqrt{\alpha'}\over \pi \mu}  {\rm Im}
\int_0^\infty dt e^{-it}~ {t/(2\beta\mu\sqrt{\alpha'})\over\sinh
[t/(2\beta\mu\sqrt{\alpha'})]}~ {t/ (2 \beta\mu R)\over\sinh [t/(2
\beta\mu R)]}\ . } The equations \oldrev\ are suggestive of a
Legendre transform where $\mu$ is a variable conjugate to
$\Delta$. Let us define the Legendre transform of ${\cal
F}(\Delta)$,
 \eqn\leg{ \Gamma (\mu) = \Delta \mu - {\cal
 F}(\Delta)\ ,}
which satisfies \GrossKleb
 \eqn\oldGK{ {\partial^2
\Gamma (\mu)\over \partial \mu^2 } = \tilde\rho (\mu)\ . }
 This
Legendre transformation arises because the sum over surfaces
corresponds to fixed $N$ while $\Gamma$ is the thermodynamic
potential for fixed chemical potential $\mu$. Another interpretation
of this Legendre transform is that it arises in the $c=1$ matrix
model with trace-squared terms \GubKleb, which was argued to
describe Liouville theory perturbed by the operator $\mu
e^{2\phi}$ (as opposed to the operator $\phi e^{2\phi}$). In such
a model the sum over surfaces is given by $I=2 \pi R \beta^2
\Gamma (\mu)$, and $\mu$ is therefore naturally interpreted as the
cosmological constant. The double scaling limit is taken as
$\mu\rightarrow 0$, $\beta\rightarrow\infty$, with  $\beta\mu$
kept fixed. This double-scaled parameter, proportional to $1/g_s$,
corresponds to the parameter denoted by $\mu$ in Liouville theory.
Since the double scaling limit is taken at fixed $\beta\mu$ and at
infinite $N$, we do not need to be concerned about the consistency
of adding a finite number of fermions to the system. Creation of a
D-brane strictly speaking corresponds to exciting a fermion from
the Fermi level to the top of the potential. Alternatively, we may
simply add a fermion near the top of the potential with the
understanding that it is borrowed from the infinite pool at large
$\lambda$.

Both \Rduality\ and
the sum over surfaces are symmetric under the T-duality
\eqn\Tdual{
R\rightarrow \alpha'/R\ , \qquad \beta \mu
    \rightarrow {R\over \sqrt{\alpha'}} \beta \mu
\ .}
In particular, if we divide the answer by 2,
which corresponds to filling the
potential on one side, then we find the answer
matches the bosonic Liouville theory path integral calculation
of section 4.
Thus the fact that in the bosonic theory the Fermi sea fills
only one side of the potential is
crucial for obtaining exactly the same factor
as in the Liouville calculation.


\subsec{The type 0B model at finite temperature}

According to our conjecture, in order
to adapt the matrix model to describe the type 0B
theory, all we need to do is send
$\alpha'\rightarrow 2\alpha'$ and also fill both sides of the potential
with fermions up to the same Fermi level $-\mu$.
Therefore, the non-perturbative
sum over surfaces is given by
$I_B=2 \pi R \beta^2 \Gamma_B (\mu)$, where
\eqn\oldGKB{
{\partial^2 \Gamma_B (\mu)\over \partial \mu^2 } = \tilde\rho_B (\mu)\ ,
}
and
\eqn\RdualityB{
{\partial\tilde \rho_B\over \partial \mu}
= {\sqrt{2\alpha'}\over \pi \mu}  {\rm Im}
\int_0^\infty dt e^{-it}~
{t/(2\beta\mu\sqrt{2\alpha'})\over\sinh [t/(2\beta\mu\sqrt{2\alpha'})]}~
{t/ (2 \beta\mu R)\over\sinh [t/(2 \beta\mu R)]}\ .
}
This is the exact non-perturbative expression. See appendix B for
a derivation.

After these modifications, the term in the matrix model free energy that
should correspond to the sum over tori in the NSR string becomes
\eqn\torusNSR{
  {\cal Z}_{MB}= - {\ln |\mu| \over 12}
    \left ( {R\over \sqrt{2\ap}} + {\sqrt{2\ap}\over R} \right ) \ .
}
which agrees with \slb .
The theory has two
types of excitations: those symmetric under the $Z_2$
will be identified with states in the NS-NS sector,
and those antisymmetric under the $Z_2$ with states in the R-R sector.

The winding operators are related to inserting Wilson lines in the
matrix quantum mechanics description. In the type 0B model we have a
single $U(N)$ gauge group and correspondingly we have only one
type of winding modes in the CFT (the ones coming from the NS
sector). On the other hand, in the type 0A matrix model we have a
$U(N) \times U(N)$ gauge group and two possible Wilson lines. The
symmetric and antisymmetric combination of Wilson lines
corresponds to the NS and R-R winding modes that we have in the 0A
string theory.


\newsec{The Matrix Model for Type 0A}

The type 0B matrix model contains a gauge field and a scalar:
$(A_0,\M)$, both of which are hermitian matrices. This is the
field theory on a set of unstable zero branes of the 0B theory.
The $Z_2$ action $(-1)^{\FL}$ on the 0B theory maps $(A_0, \M) \to
(A_0, - \M)$.  Quotienting by this symmetry,
we get the stable branes of type 0A theory,
which contain no tachyon. In addition we can embed the
$Z_2$ into the gauge group (see \mooredouglas). Suppose that we
start with $2N +q$ D0-branes in the 0B theory. A particular
embedding would lead to the identification \eqn\part{
 ( A_0 , \M) \sim \tilde \sigma_3 ( A_0, - \M) \tilde \sigma_3
}
where $\tilde \sigma_3$ is ${\sl diag}(1,...,1;-1,...,-1)$,
with $N$ 1's and $(N+q)$ $-1$'s.
The states that are invariant have the form
\eqn\invstaes{
 A_0 = \pmatrix{ A_0 & 0 \cr 0 & \tilde A_0 } ~,~~~~~~~
\phi = \pmatrix{ 0 & t \cr  t^\dagger & 0}
}
where $A_0$, $\tilde A_0$ are hermitian and $t$ is a complex matrix.
In other words, the gauge group is $U(N) \times U(N+q)$ with
$t$ in the bifundamental.

This is the matrix model that gives  the 0A theory. It is a
$ U(N) \times U(N+q) $ theory with a tachyon
field $t$ in the bifundamental.  The Lagrangian is
\eqn\lagr{
  L =  Tr[ (D_0 t)^\dagger D_0 t + { 1\over 2 \alpha'} t^\dagger t ]\ ;
}
the mass of the tachyon is still $m^2 = -{ 1 \over 2 \alpha'}$ as
in all fermionic string theories.
Complex square and rectangular matrix models have been extensively
studied.  For a recent paper and a list of earlier references see
\DiFrancescoRU.

This model can be thought of as having $N+q$ D-branes and $N$
anti-D-branes.  Therefore, it describes the type 0A background
with $q$ D0-branes.

The fact that the ground ring generators $(q\pm p)e^{\mp t}$
\Omappq, \apm\ are projected out in the worldsheet description of
type 0A is consistent with the fact that the eigenvalues
themselves are not gauge invariant. Instead, we must form products
$[q\bar q+p\bar p\pm(q\bar p+p\bar q)]e^{\mp 2t}$ and $q\bar
q-p\bar p$ which are independent of the phase of the eigenvalues;
these correspond to the NS sector generators of the type 0A ground
ring $\CO_{1,3}$, $\CO_{3,1}$, and $\CO_{2,2}$ discussed in
section 3.

In order to analyze this model it is instructive to look first at
the $N=1$, $q=0$,  case.  Here only the difference of gauge fields
couples to $t$ and this just removes the phase of the field $t$.
It is convenient however to quantize the complex field $t$ on the
plane and then impose the condition that its wavefunctions are
$U(1)$ invariant. This $U(1)$ is the $U(1)$ of angular momentum on
the plane. We have two harmonic oscillators of frequency $w^2 =
m^2$. Suppose for a moment that $w^2$ were positive. In this case
we can quantize the system in terms of creation and annihilation
operators $a_{\pm}$ with definite angular momentum. States with
zero angular momentum have the same number of $a^\dagger_+$ and
$a^\dagger_-$. So the spectrum of $U(1)$ invariant states is given
by $\epsilon_n = w(1 + 2 n)$. The important point to notice is
that this gives a result that is a factor of two bigger than the
result we would have obtained in the corresponding computation for
the $N=1$ hermitian matrix model.

Now let us consider the case $N>1$. For simplicity we first
continue to take $q=0$. As discussed in
\refs{\DiFrancescoRU\Morris-\Dalley} we can first gauge fix the
matrix $t(\tau)$ to a real diagonal matrix with positive entries
at every time $\tau$.  The ghosts of this gauge fixing lead to a
measure factor
 \eqn\meass{
 \prod_i d \lambda_i \lambda_i \prod_{i<j} ( \lambda_i^2 -
 \lambda_j^2)^2
 }
at every time $\tau$.  This measure factor implies that we have
$N$ fermions each of them moving on a plane with zero angular
momentum. The $\lambda_i$ is just the radius in each plane. The
integral over the vector fields $A_0(\tau)$ almost completely
cancels this measure factor.  More precisely, if we discretize the
time direction, the measure factor \meass\ from the ghosts appears
on the sites; the gauge fields are link variables, and
integrating over them
cancels the measure factor \meass\ everywhere except at
the end points of the time evolution.  We end up with a factor of
 \eqn\factow{\prod_i\sqrt{\lambda_i} \prod_{i<j} ( \lambda_i^2 -
 \lambda_j^2)}
in the wave function in the initial and final state.  The factor
of $\sqrt{\lambda_i}$ is characteristic of motion in two
dimensions and the factor of $\prod_{i<j} ( \lambda_i^2 -
\lambda_j^2)$ makes the eigenvalues fermions. In other words, the
wavefunction $\chi = \prod_{i<j} ( \lambda_i^2 - \lambda_j^2)
\Psi$ obeys the single particle equation
 \eqn\singlepart{ (  -  { 1 \over 2\lambda_i } {\partial \over
\partial \lambda_i } \lambda_i { \partial \over
\partial \lambda_i }  - { 1 \over 4 \alpha'} \lambda_i^2 ) \chi = E \chi\ .
}

Let us now consider $q>0$.  As in
\refs{\DiFrancescoRU\Morris-\Dalley}, \meass\ becomes
 \eqn\meassa{
 \prod_i d \lambda_i \lambda_i^{1+2q} \prod_{i<j} ( \lambda_i^2 -
 \lambda_j^2)^2.
 }
Again, integrating out the gauge fields cancels this measure
factor except at the end points of the time evolution.  The
initial and final wave functions have a factor of $\lambda_i^{{1
\over 2} +q } \prod_{i<j} ( \lambda_i^2 - \lambda_j^2)$.  The
second factor $\prod_{i<j} ( \lambda_i^2 - \lambda_j^2)$ again
turns the eigenvalues into fermions.  The first factor
$\lambda_i^{{1 \over 2} +q }$ can be given two different physical
interpretations.  First, we can think of $\lambda_i$ as being the
radius of a motion in $2 + 2q$ dimensions.  Alternatively, as for
$q=0$, we can keep the motion in two dimensions, but state that
the angular momentum is not zero but it is instead $q$.
Mathematically, the extra factor of $\lambda_i^q$ has the effect
of pushing the eigenvalues away from the origin.  The two
different physical interpretations (higher dimensional motion or
two dimensional motion with nonzero angular momentum) have the
same effect.

Now the wavefunction $\chi = \prod_{i<j} ( \lambda_i^2 -
\lambda_j^2) \Psi$ obeys the single particle equation
 \eqn\singlepartq{ \eqalign{
    &(  -  { 1 \over 2\lambda_i^{1+2q} } {\partial \over
 \partial \lambda_i } \lambda_i^{1+2q} { \partial \over
 \partial \lambda_i }
  - { 1 \over 4 \alpha'} \lambda_i^2 ) \lambda^{-q} \chi
  = E \lambda^{-q}\chi \cr
 &(  -  { 1 \over 2\lambda_i } {\partial \over
 \partial \lambda_i } \lambda_i { \partial \over
 \partial \lambda_i } + {q^2 \over 2 \lambda_i^2}
  - { 1 \over 4 \alpha'} \lambda_i^2 ) \chi = E \chi
 }}
where the first (second) equations are more natural in the first
(second) physical interpretation above.

The system of $N$ eigenvalues moving in two dimensions with
angular momentum $q$ also arises from another matrix model.  We
can start with a $U(N)\times U(N)$ matrix model with
bi-fundamentals $t$ as in \lagr\ and add a Chern-Simons term
 \eqn\CSter{q\int d\tau \Tr (A-\tilde A)}
where $A$ and $\tilde A$ are the gauge fields of the two $U(N)$
factors.  We can gauge fix $t$ to a diagonal matrix with
eigenvalues $\lambda_i$, but we do not yet gauge fix the phases of
$\lambda_i$. As above, the measure factor is $\Delta^2 =
\prod_{i<j \leq N} (|\lambda_i|^2 - |\lambda_j|^2 )^2$ and it
makes the eigenvalues into fermions. We integrate out all the
gauge fields but keep the diagonal elements of $A-\tilde A$ and
denote them by $a_i$. This leads to the Lagrangian
 \eqn\evlag{ \sum_i \left( |(\partial_\tau + i a_i) \lambda_i |^2
 + q a_i - V(|\lambda_i|^2)\right).}
We now fix axial gauge $a_i=0$.  Gauss' law which is the equation
of motion of $a_i$ states that the angular momentum of $\lambda_i$
is $q$.  Therefore, the resulting quantum mechanics of the
eigenvalues is exactly as we found above when we started with the
$U(N)\times U(N+q) $ gauge theory.

We conclude that a system of $q$ D0-branes in the 0A theory can be
described either by a  $U(N)\times U(N+q) $ quiver theory or by a
$U(N)\times U(N) $ quiver theory with a Chern-Simons term \CSter\
with coefficient $q$.  The latter construction makes it clear that
the flux due to the $q$ D0-branes is represented in the system by
the operator $q\int d\tau \Tr (A- \tilde A)$.  This coupling is
similar to the way the flux is introduced in the spacetime
Lagrangian \effoA\ $q\int\! d\tau d\phi F^{(-)}_{\tau \phi} = q\int\!
d\tau \!\left(A^{(-)}_\tau (\phi\!=\!\infty) - A^{(-)}_\tau
(\phi\!=\!-\infty)\right)$.


\subsec{The type 0A theory at finite temperature}

The free energy for this case can be computed as reviewed around
\Rduality . We  use the trick reviewed in \igor , which consists
in doing the computation for a right-side up harmonic oscillator
and then reversing $w^2$. The only difference relative to the type
0B matrix model is the doubling of energies. This amounts to
replacing $\beta \mu \rightarrow \beta\mu/2$  in the first
fraction in the integrand of \RdualityB, and introducing an
overall factor of $1/2$.  Using this trick it is possible to
compute the thermal partition function for the $q=0$ case. The
general $q$ case can be treated by computing the exact density of
states as is spelled out in appendix B. This gives \eqn\oldGKA{
{\partial^2 \Gamma_A (\mu)\over \partial \mu^2 } = \tilde\rho_A
(\mu)\ , } where \eqn\RdualityA{ {\partial\tilde \rho_A\over
\partial \mu} = {\sqrt{\alpha'/2}\over \pi \mu}  {\rm Im}
\int_0^\infty dt e^{-it}~ {t/(2\beta\mu\sqrt{\alpha'/2})\over\sinh
[t/(2\beta\mu\sqrt{\alpha'/2})]}~ {t/ (2 \beta\mu R)\over\sinh
[t/(2 \beta\mu R)]}\, e^{ - q t/( 2 \beta |\mu| \sqrt{\alpha'/2
})} \ . } 
Again, this is the exact non-perturbative expression. The
one-loop term is
\eqn\oneloopa{ {\cal Z}_{M A} =  {
\ln |\mu|} \left[
 - { 1 \over 24} \left(4  { R \over \sqrt{2 \alpha'} }
+ { \sqrt{2 \alpha'} \over R} \right)
 + { 1\over 2} q^2 { R \over \sqrt{ 2 \alpha' }} \right]
} The term proportional to $q^2$ agrees precisely with
 the contribution of the
R-R field strength to the ground state energy of the system \zoneq
, after we multiply by $(2 \pi R)$. We also see that the $q=0$
torus contribution agrees with the result of the continuum
calculation, \slb. It is self-dual under $ R \to {\alpha'\over
2R}$. This is actually a duality of the full $q=0$ answer,
provided that we also change the coupling $ \beta \mu \to { 2 R
\over \sqrt{2 \alpha'} } \beta \mu $.

An important check on the matrix models is that T-duality along
the compact direction relates 0A and 0B theories. Indeed, for
$q=0$, the full $\Gamma_A$ is obtained from $\Gamma_B$ given in
\oldGKB, \RdualityB\ through the T-duality transformation $R\to
\alpha'/R$, $ \beta \mu \to { R \over \sqrt{2 \alpha'} } \beta \mu
$.

An amusing limit that one could take is the 't Hooft limit where
\eqn\fixedlim{
q \to \infty, ~~ |\mu| \to  \infty ~,~~~~~ \lambda = {q \over
2 \beta |\mu| \sqrt{ \alpha'/2} }  =
{\rm fixed}
}
In this case the free energy \RdualityA, becomes
\eqn\freeen{
 \partial_w^2 ( 2 \pi R \Gamma )  = - { R \over \sqrt{ \alpha'/2}} q^2
 { 1 \over 4} \ln(1+ w^2)
} where $w = 1/\lambda $. It scales as $q^2$ as we expect in the
't Hooft limit. It would be nice to understand what the dual
string background is. It would also be nice to see if one can
write a decoupled matrix model in the limit \fixedlim. We could
also rescale $R$, as the same time that we take \fixedlim\ and
then we get a more interesting function of $R$ in the limit.
Setting $ \hat R = 2 \beta |\mu| R $ and keeping $\hat R$ fixed in
the limit we obtain
 \eqn\obtain{ \lambda
\partial_\lambda \tilde \rho = - { \sqrt{ \alpha'/2} \over \pi} {
\rm Im} \int_0^\infty dt e^{ - i t } { t/\hat R \over \sinh(
t/\hat R ) } e^{ - \lambda t}
 }
There are various other interesting limits that will probably yield
interesting relationships.

\subsec{Affine A theory}

In section 4.3 we discussed the theories that arise by acting with
 $(-1)^{\FL}$ when we go around the circle, where $\FL$ is the
spacetime left-fermion number. In the 0A matrix model, this corresponds
to the operation exchanging the two gauge groups. So let us set
$q=0$. When we go around the circle we impose the boundary condition
\eqn\bcimp{
 A_0(\tau + \pi R) = \tilde A_0(\tau) ~,~~~~\tilde A_{0}(\tau + \pi R)
= A_0(\tau)~,~~~~~~t(\tau + \pi R) = t^\dagger(\tau)
}

We can compute the partition function by imposing the $\dot A_0 =
\dot { \tilde A}_0 =0$ gauge. Then we are left only with the zero
mode of $A_0=\tilde A_0$ by \bcimp . We then expand the $t$ field
in fourier modes $t \sim \sum_n t_n e^{ i n t/R}$. Then \bcimp\
implies that the even modes are hermitian and the odd modes are
anti-hermitian, $ t_n^\dagger = (-1)^n t_{-n}$. So the computation
has the same structure as the computation we would do if we were
doing the 0B model. So we get the same as the 0B answer. In
particular the term corresponding to the one loop string amplitude
agrees with \Aeval .


\newsec{ D-brane decay }

In \KlebanovKM\ it was argued that the D-brane corresponding to $(1,1)$ ZZ
boundary conditions and Neumann boundary conditions in the $c=1$ direction
$x$, with a boundary perturbation
\eqn\boundpert{
  \delta S=\lambda\oint d\sigma \cos x\ ,
}
corresponds to an eigenvalue in the matrix model
executing the motion%
\foot{The relative factor of two in the coefficient of $\mu$
with respect to the corresponding formula in \KlebanovKM\ is due
to the fact that we normalize the matrix model Hamiltonian differently; see
\aplusminus.}
\eqn\qxeucl{q=\sqrt{\mu}\sin\pi\lambda\cos x\ .}
The Minkowski analog of this solution (the rolling tachyon with $x=it$)
corresponds to an eigenvalue coming in from $q=\infty$,
reaching a turning point at $q=\sqrt\mu\sin\pi\lambda$,
and then going back out to infinity.

Following the discussion (and conventions) of section 3,
this can now be verified in Liouville
theory by computing the (normalized) expectation values of
the ground ring generators
$\CO_{1,2}$ and $\CO_{1,2}$ in this state,%
\foot{An alternative way
of thinking about this calculation, in analogy to \pertope,
is that one is computing the limit $z\to\bar z$ of the operator $\CO_{1,2}$.
In this limit, the operator collides with its image and we are computing the
coefficient of the identity in this OPE.}
and comparing to \Omappq.
Consider for example $\langle\CO_{1,2}\rangle$.
The matter part of the expectation value is \Callan
\eqn\expmatt{\langle V_{1,2}^{(M)}(z) \rangle=\langle e^{ix(z)}\rangle =
{\sin\pi\lambda\over |z-\bar z|^{1/2}}\ ;}
the Liouville expectation value is \zz
\eqn\vbminustwo{\langle V_{-{b\over2}}(z)\rangle={U(-b/2)\over
|z-\bar z|^{2\Delta(-b/2)}}\ ,}
where
\eqn\ubminus{U(-b/2)=\mu^{1/2}{\Gamma(2+b^2)\Gamma(1+{1\over b^2})\over
\Gamma(2+2b^2)\Gamma(2+{1\over b^2})}={1\over6}\sqrt{\mu}\ ;}
the last equality holds in the limit $b\to 1$.
The one point function of $\CO_{1,2}$ has two contributions:
\eqn\eexxpp{\langle\CO_{1,2}\rangle=
\langle cb\bar c\bar b V_{-{b\over2}}e^{ix}\rangle+
{1\over b^4}\langle |L_{-1}^{(L)}-L_{-1}^{(M)}|^2V_{-{b\over2}}e^{ix}\rangle
\ .}
Evaluating the two terms using eqs. \expmatt\ -- \ubminus,
we find
\eqn\finoonetwo{\langle\CO_{1,2}\rangle=
(1-{28\over4}) U(-{b\over2})\sin\pi\lambda=-\sqrt{\mu}\sin\pi\lambda\ .}
A similar calculation leads to the same result for $\langle\CO_{2,1}\rangle$.
Using the map \Omappq\ to relate these quantities to matrix model variables,
we find
\eqn\qmatmod{q=-\sqrt{\mu}\sin\pi\lambda \cosh t}
or after analytic continuation $t\to ix$, \qxeucl\ (up to a
transformation $q\to -q$).
Comparing to \aplusminus, we see that the energy
of the solution (as measured from the top of the potential)
is $E=-\mu\sin^2\pi\lambda$.  The filled Fermi sea
corresponds to $\lambda=\frac12$.

Thus in the bosonic string the ZZ $(1,1)$ D-brane with a marginal
perturbation \boundpert\ corresponds to an eigenvalue rolling according to
\qxeucl, or its Minkowski continuation. We next discuss the analogous
question in the type 0B case.


\subsec{D-brane decay in type 0B}

We would like to compute $\langle\CO_{1,2}\rangle$ on the disk
(where $\CO_{1,2}$ is given in equation \apm),
with ZZ-like boundary conditions for Liouville,
and the supersymmetric analog \eqn\susyroll{
  \lambda \oint \!dx d\theta_t \, \cos X
}
of the ``rolling tachyon'' \boundpert\ for $x$.
We adopt the conventions of section 3.
It is useful to think
of the calculation as the limit of the operator $\CO_{1,2}(z,\bar z)$ as
$z\to \bar z$. In this limit, $\CO_{1,2}$ collides with its image
and makes a boundary perturbation. We are interested in the coefficient
of the identity in this OPE.

The limit $z\to \bar z$ of $\CO_{1,2}$ receives contributions only from
the cross-terms on the first line of \apm. One has
\eqn\llkkmm{
  \lim_{z\to\bar z}\CO_{1,2}(z)\simeq -2{c\over\sqrt2}\partial
    \xi e^{-{3\varphi\over2}-{i\over2}H}
    e^{-{\bar\varphi\over2}+{i\over2}\bar H}e^{{i\over2}
    x-{1\over2}\phi}\ .
} The OPE of the matter part of the CFT gives a factor of ${\sl
sin}\,\pi\lambda$ \SenAN,
while the Liouville
contribution is $U^{(R)}(\alpha=-\half)$, where \korean
\eqn\uralpha{U^{(R)}(\alpha)=
  {(2\mu)^{-\alpha}\over \bigl(\Gamma({3\over2}-\alpha)\bigr)^2}\ .
}
Thus, we conclude that
\eqn\limtheta{\lim_{z\to\bar z}\CO_{1,2}(z,\bar z)
    =\sqrt{2}c\partial\xi e^{-2\varphi}
    \sin\pi\lambda\sqrt{2\mu}\ .}
Just like in the discussion following \opeminustwo, the boundary operator
$c\partial\xi e^{-2\varphi}$ is a picture changed version of the identity
operator.%
\foot{The picture changed version of \limtheta\
is half of what one naively gets by looking at the coefficient
of $1/(z-w)$ in the OPE of $J_{\sst\rm BRST}$ with $\xi$ times \limtheta.}
Finally, we have
\eqn\expoonetwo{
  \langle\CO_{1,2}\rangle = \sqrt{\mu}\sin\pi\lambda\ .
}
Again, one finds the same answer for $\langle\CO_{2,1}\rangle$, and by
using the identification of $\CO_{1,2}$, $\CO_{2,1}$ with matrix model
variables, one sees that the D-branes in question correspond to rolling
of eigenvalues either to the left or to the right of the top of the potential,
\eqn\superroll{
  q=\sqrt\mu\sin\pi\lambda\,\cosh t\ ,
}
with the direction of rolling correlated with the sign of $\lambda$.


\subsec{Closed string radiation from the decay}

We can also compute the radiation of closed strings
from the decaying D-brane, analogous to
the computation in \KlebanovKM.%
\foot{Recall again our conventions
\liouparams, \centchge, \confdim\ for super-Liouville theory.
In particular, recall that $\hat c=1$ matter
corresponds to $Q=2$, $b=1$, $\hat c_L = 9 $.
The cosmological constant goes as $e^{b \phi}$ and the
normalizable operators have $\alpha = { Q\over 2} + i P$.
These conventions correspond to the standard CFT conventions
of $\alpha' =2$. }
The unnormalized super-Liouville disk one point function
of the observables \nsrobs\
on $(1,1)$ branes is given by \bsd\ (up to a numerical factor
independent of $\nu$ and $b$).
Consider first the NS sector; this one point function is
\eqn\superl{
\langle N_{\alpha} \rangle = { {\rm const} \over
P \,\Gamma( - i P/b) \,\Gamma( - i P/b) } \mu^{ - iP/b }\ .
}
For $b=1$ this becomes
\eqn\superlus{
\langle N_{\alpha} \rangle \sim { \sinh \pi P   } { \Gamma( iP)
\over \Gamma( - iP) } \mu^{- iP}\ .
}
The one point function in the time direction is
given by the Fourier transform of the NS decay profile
\eqn\rhons{
\rho_{\NS}^{\ }(t) = \left[
    {1 \over 1 + \sin^2 \lambda\, e^{t} } +
    {1 \over  1 +  \sin^2 \lambda \,e^{-t} } - 1 \right]
 = { 1 - \sin^4 \lambda \over 1 + \sin^4 \lambda + 2 \sin^2 \lambda \cosh t }
}
the result is
\eqn\timedir{
  \langle e^{-i Et } \rangle =  e^{ - i E \log \sin^2\! \pi \lambda }
    { \pi \over \sinh \pi E }\ ,
 }
where we used the so called ``Hartle-Hawking'' contour \llm\ (see
also \gaiotto). When we compute the 1-point function we can do it
in the $(-1,-1)$ picture and we get the total amplitude \eqn\ampl{
{\cal A}_{\NS} \sim   e^{ - i E \log \sin^2\! \pi \lambda } {
\Gamma( iP) \over \Gamma( - iP) } \mu^{- iP}\ . } The result is
similar to the bosonic case.

One can also compute the Ramond one point functions.
For the Liouville sector we find
\eqn\ramone{
\langle R_\alpha \rangle =
 {{\rm const} \over \Gamma( -i P + 1/2)\, \Gamma( - i P + 1/2) } \mu^{- iP}
\sim \cosh \pi P { \Gamma(iP + 1/2) \over \Gamma( - iP +1/2) }
\mu^{-iP}\ , }
where $R_\alpha$ is defined in \nsrobs.
For the time direction, the Ramond sector amplitude
is given by the Fourier transform of the Ramond decay profile \SenAN
\eqn\rhoram{ \rho_R^{\ }(t) = \sin \pi \lambda
  \left[ {e^{ t / 2} \over 1 + \sin^2\! \pi \lambda \,e^t} +
  {e^{- t /2} \over 1 - \sin^2\! \pi \lambda \,e^{ - t} }
    \right]\ .
}
This Fourier transform is given by
\eqn\amplit{
 { \sin \pi \lambda
\over | \sin \pi \lambda | }
 e^{-i E \log \sin^2\!\pi \lambda } { 1 \over
\cosh \pi E }\ .
}
Thus the final total amplitude is
\eqn\rampl{
{\cal A}_{\RR} \sim {  \sin \pi \lambda
\over | \sin \pi \lambda | }  e^{ - i E \log \sin^2\! \pi \lambda }
{ \Gamma(\half+ iP)
\over \Gamma(\half - iP) } \mu^{- iP}\ .
 }
We have left undetermined the overall normalization of the
amplitudes \ampl, \rampl; these are computed in appendix C. We see
that when we express the amplitudes in terms of the fields
$T_{R,L}(p)$ defined in \symtach--\phasesde, we find that if the
tachyon rolls to the right ($\lambda
>0$) then only $T_R$ radiation is produced, while if it rolls to
the left ($\lambda <0$) then only $T_L$ radiation is produced.


\subsec{States below the Fermi level\foot{We thank A. Sen and
L. Susskind for raising this question.}}

We have argued that a single fermion eigenvalue whose energy
lies between the Fermi surface and the top of
the potential barrier can be described, in the bosonic string,
in terms of a ZZ brane tensored with the matter boundary state
of \refs{\Callan,\SenNU,\SenIN,\SenAN} with
$ 0 \leq   \lambda \leq 1/2 $
(in the fermionic string there is an analogous construction, and
$\lambda\in [-\half,\half]$).
These are fermions whose energies are\foot{For the discussion of this
subsection, it will be convenient to measure energies relative to the Fermi
surface.}
\eqn\energies{
  E  = \mu \cos^2 \pi \lambda
}
(again, we assume $\mu>0$).
One can similarly describe fermions
with energies above the oscillator barrier, $E>\mu$,
by setting $ \lambda = i s$, and continuing $t\to t+{i\pi\over2}$; see
\refs{\SenNU,\SenIN}.
Under this analytic continuation,
the decay profiles \rhons, \rhoram\ remain real,
and the trajectory \superroll\ given by the ground ring computation
is the correct one.

Continuation of $\lambda$ to complex values amounts to performing
a rotation of the left and right movers in the time direction
by an element of $SL(2,C)$.
This operation was shown in
\GaberdielXM\
to lead to boundary states obeying the Cardy condition.
For generic $SL(2,C)$ rotations, the annulus partition function
of the Euclidean theory is complex; however, for
$\lambda\in [0,\frac12]$, and $\lambda\in \frac12+i\IR$, the annulus
amplitude is real.%
\foot{Again, for the fermionic string one should allow
$\lambda\in \pm\half+i\IR$ to get trajectories
on either side of the potential.}
This suggests that these latter
boundary states can be given a physical interpretation.
Indeed, the energy \energies\ and classical motion \qmatmod, \superroll\
for these values of $\lambda$ correspond to
eigenvalue trajectories below the Fermi surface.

The D-brane states discussed above describe eigenvalues
following various classical trajectories.
These trajectories are sharply defined in the $g_s\to 0$ limit.
At finite $g_s$, the phase space coordinates $p$ and $q$
cease to commute, and one should quantize the eigenvalue dynamics,
and construct wavefunctions $\psi(q,t)$
for each energy.
The wavefunctions associated to the
trajectories correspond to the coefficients
of creation operators of particles for $E>0$, and
annihilation operators of holes for $E<0$,
in the second quantized theory.

The conjugate wavefunctions are associated to the
conjugate operators.  The creation of a hole
or the destruction of a particle carries the
opposite sign of the energy.  It is natural
to mimic this effect in the classical limit
by reversing the sign of the boundary state,
which produces an extra
minus sign in the expression for the energy \energies.
Thus the creation operators of holes are associated
with the D-brane boundary states with
$\lambda\in \pm\half+i\IR$;
similarly, the annihilation operators of particles
are associated with $\lambda\in [-\half,\half]$.
In both cases one puts a minus sign
in the boundary state wavefunction.

The extra minus sign in the Cardy state implies that the one point
functions for the tachyon and the RR scalar will have a minus
sign compared to \ampl, \rampl . This is precisely what we need
since for example the branes for real $\lambda$ correspond to fermions
$e^{ i 2 \sqrt{\pi} T_{L,R} } $ while the holes
for $\lambda\in \pm\half+i\IR$ will give us
$e^{ - i 2 \sqrt{\pi} T_{L,R} }$.

It is amusing to note that if we consider a standard brane and the
brane that results from reversing the sign of its boundary state,
then the action for the open strings on the wrong sign brane has
an overall minus sign and the open strings stretched between the
branes become fermions, since the one loop diagram changes sign.
Perhaps it means that the gauge theory on $N$ branes with the
correct sign and $M$ branes with the wrong sign is based on
the supergroup $U(N|M)$.%
\foot{Such theories were studied in \VafaQF\ in the context of
topological strings, where they were interpreted as branes and
anti-branes. Here we interpret them as fermions and holes. This
should not be confused with the matrix model for type 0A theory
introduced above, where all the kinetic terms are positive.} The
holes indeed have negative energies and they ``fill up''  the
region above the Fermi sea.  Only holes below the Fermi sea are
allowed physical excitations.

\bigskip\bigskip\vskip 2cm

\noindent
{\bf Acknowledgments:}

We would like to thank B. Craps, J. McGreevy, K. Okuyama, A. Sen,
L. Susskind, H. Verlinde and E. Witten for useful discussions. The
research of MRD is supported in part by DOE grant
DE-FG02-96ER40959. The research of DK and EM is supported in part
by DOE grant DE-FG02-90ER40560. The research of JM and NS is
supported in part by DOE grant DE-FG02-90ER40542. The research of
IRK is supported in part by NSF grants PHY-9802484 and
PHY-0243680. Any opinions, findings, and conclusions or
recommendations expressed in this material are those of the
authors and do not necessarily reflect the views of the National
Science Foundation.


\appendix{A}{The Odd Spin Structure}

In this Appendix we show how the answer \odd\ arises from an
explicit path integral calculation. The odd spin structure
$(r,s)=(0,0)$ is subtle due to the presence of supermoduli and
fermionic and superghost zero modes. There are different ways to
organize the calculation.

\subsec{Method 1}

We fix the superconformal gauge on the torus in the odd spin
structure. There is a $\gamma$ zero mode (and a $\bar \gamma$ zero
mode) which is associated with the existence of a covariantly
constant spinor on the torus and a $\beta$ zero mode (and a $\bar
\beta$ zero mode) which is associated to the fact that there is
still a gravitino component that cannot be gauged away in the
superconformal gauge.

In addition  there are four fermion zero modes, two from each of
the two Majorana fermions we have in the theory, $\chi $, $\bar
\chi $, $\psi$ and $\bar\psi$.

The zero modes of the Liouville fermions $\psi$ and $\bar\psi$
have a geometric interpretation.  They are associated with the
conformal Killing spinor on the torus, exactly like the $\gamma$
and $\bar \gamma$ zero modes.  Therefore, the ``zero'' due to the
Liouville fermion zero modes exactly cancels the ``infinity'' due
to the $\gamma$ and $\bar \gamma$ zero modes. This is analogous to
absorbing $c$ \refs{\ZamolodchikovVX,\SeibergEB} and $\gamma$
\KutasovUA\ zero modes associated with conformal Killing vectors
and conformal Killing spinors in Liouville theory on the sphere.

The $\beta$ and $\bar \beta$ zero modes lead to an insertion of
the left and right moving supercharges
 \eqn\inserts{G(z) \bar G(\bar w) = (\chi  \partial x (z)+ \psi
 \partial \phi(z))(\bar\chi  \bar\partial x (\bar w)+ \bar\psi
 \bar \partial \phi(\bar w))}
where $\phi$ is the Liouville field.  The expectation value of the
operator \inserts\ is independent of $z$ and $\bar w$ because a
derivative with respect to them turns into a derivative with
respect to the modular parameter $\tau$ which integrates to zero.
The insertions \inserts\ absorb the remaining $\chi $ and $\bar
\chi $ fermion zero modes.  Now that all the fermion zero modes
have been absorbed, the fermion determinants can be easily
computed.  The remaining nontrivial part of the computation is
proportional to
 \eqn\nontr{\langle\partial x (z)\bar \partial x (\bar w)\rangle_x}
where the expectation value is in the functional integral over $x$
only. This expectation value has two contributions. At separated
points it is $-{\pi \over \tau_2}$, and there is a contact term
proportional to $\delta^{(2)}(z-w)$. We set $z=w$ in \nontr\ and
integrate over $z$.  Since the bosonic action is proportional to
$\partial x\bar \partial x$,
 \eqn\choco{{\cal Z}_{\rm odd}\sim \langle\partial x (z) \bar
 \partial  x (\bar w)\rangle _x= R{\partial \over \partial R }
 \langle 1 \rangle_x \sim R{\partial \over \partial R } \ln |\mu|
 \left ( {R\over \sqrt{\ap}} + {\sqrt{\ap}\over R} \right )=\ln
 |\mu| \left ( {R\over \sqrt{\ap}} - {\sqrt{\ap}\over R} \right )}
The overall normalization of ${\cal Z}_{\rm odd}$ can be
determined by matching its large $R$ behavior with the
value determined by the spectrum at infinite $R$. It follows that
${\cal Z}_{\rm odd}$ has the value \odd, as we conjectured.

\subsec{Method 2}

An alternative computation of ${\cal Z}_{\rm odd}$,
which relies heavily on the treatment of
super-Liouville path integral in \BerKlebnew,
is based on inserting a discrete state vertex operator
in the $(-1,-1)$ picture,
 \eqn\dilaton{ e^{\varphi + \bar \varphi} \chi
 \bar \chi  \  }
where $e^\varphi$ can be thought of as $\delta (\gamma)$.  On the
one hand, this insertion will soak up the $\gamma,\bar\gamma,\chi
,\bar\chi $ zero modes in a natural way. On the other hand, since
in the $(0,0)$ picture this operator is $ R^2 \partial x\bar
\partial x $, the 1-point function is simply $R^2
(\partial/\partial R^2)$ acting on the torus path integral. This
will allow us to extract the torus path integral in a simple way.

For the odd spin structure one cannot completely gauge
away the gravitino field.
The partition function may be written as an integral over the supermoduli
space.
Let us represent the supertorus as the quotient of the superspace $(z, \theta)$
under the supertranslations
\eqn\storus{
(z, \theta) \sim (z+1, \theta) \sim (z+\tau+\lambda\theta, \theta+\lambda)\ ,
}
where $\lambda$ is the odd coordinate of the supermoduli space.
A variation with respect to $\lambda$ gives
$$
{{\partial} \over {\partial \lambda}}\langle Z\rangle =
\vev{\bigl ( \sqrt{2\pi i}G_0 +
       2\pi i\lambda (L_0 - {\hat c \over 16})\bigr )Z}\ .
$$
This equation
can be easily integrated, and one finds that the shift
into an odd direction of supermoduli space is generated by
$e^{\sqrt{2\pi i}\lambda G_0}$. The matter partition function for the $(+,+)$
sector can be represented as the trace
$$\Tr q^{L_0 -{\hat c}_m/16}
               \xi^{\lambda G_0}
 \bar q^{\bar L_0 -{\hat c}_m/16}
              \bar \xi^{\bar \lambda \bar G_0}
                                   (-1)^F~,
$$
where $\xi=e^{\sqrt{2\pi i}}$. Similar representations can be obtained
for the super \L\ and superghost partition functions.
The bottom component of the trace
$(\lambda=\bar \lambda=0)$ is given by the Witten index. The top component
determines the dependence on the supermoduli.

The dependence of the
super-Liouville action  on the supermodulus $\lambda$ is
\eqn\laction{
S_L=S(\Phi)-
      {i \over 2\pi \tau_2} \int d^2 \sigma \lbrack   \lambda \psi\partial\phi
               - \bar\lambda \bar \psi\bar\partial\phi\rbrack
             - {{\lambda \bar\lambda} \over {4\pi \tau_2 ^2}}
                                          \int d^2 \sigma \psi \bar\psi\ ,
}
where $\psi(w)$ is the superpartner of $\phi(w)$.
This is the standard super Liouville action in the presence
of constant gravitino background $\lambda / \tau_2$.

Now consider a superconformal matter system coupled to the
supergravity theory. We will calculate the fixed ``area'' genus
one path integral, and subsequently Laplace transform it to obtain
a function of $\Delta$ (this is another way to see the appearance
of the volume factor $V_L$ due to the Liouville zero mode). Let us
first perform the path integral over the super-Liouville field
$\Phi$,
$$Z_L(A, \tau, \lambda)=
 \int d{\phi}_0 d{\psi}_0 d\bar\psi_0 (d \tilde \Phi) e^{-S}
        \delta({1\over 4\pi}
\int d^2 z d^2\theta O_{\rm min}e^{\beta_{\rm min} \Phi}-A)\ , $$
where we have decomposed $\Phi$ into the zero modes and the
remaining modes as
$$\Phi=\phi_0+\theta\psi_0+\bar\theta\bar\psi_0+\tilde\Phi\ .
$$
As in the bosonic Liouville model, the integration over the
bosonic zero mode $\phi_0$
removes the delta function that fixes the
area and reduces the theory to free field path integrals,
$$ Z_L(A, \tau, \lambda)=
    A^{-1} {\beta}_{\rm min}^{-1} (2\pi\sqrt{2\tau_2})^{-1}
       \int d {\psi}_0 d\bar\psi_0 (d \tilde \Phi) e^{-S_L}\ .
 $$
Thus, $Z_L(A,\tau, \lambda)\sim W_L+\lambda\bar\lambda B(A, \tau)$,
where $W_L$ is the Witten index.
For the super-Liouville model, $W_L$ vanishes. In the path
integral formalism, this is due to the integration
over the fermion zero modes $\psi_0$
and $\bar\psi_0$ because $S(\Phi)$ does not
depend on them. Therefore,
\eqn\cont{
 \eqalign{&Z_L(A, \tau, \lambda)=  A^{-1} {\beta}_{\rm min}^{-1}
                     (2\pi\sqrt{2\tau_2})^{-1}
   \lambda \bar\lambda\int d
{\psi}_0 d\bar\psi_0 (d \tilde \Phi) e^{-S(\tilde\Phi)}\cr
&{1 \over {4 \pi^2 \tau_2 ^2}} \bigl(- \int d^2 \sigma_1 \psi \partial \phi
                \int d^2 \sigma_2 \bar\psi\bar\partial \phi +
                      \pi \int d^2 \sigma \psi \bar \psi \bigr)\ . \cr }
} Now recall that
\eqn\pro{\vev{\partial\phi(w_1)\bar\partial\phi(\bar w_2)}=\pi
\delta^2(\sigma_1 - \sigma_2)- {\pi\over\tau_2}\ . } Therefore,
the contact part of the first term in \cont\ completely cancels
the second term in \cont. The remainder reduces to
$$ Z_L(A, \tau, \lambda)= {\lambda \bar\lambda \over
                {4\pi^2 A |{\beta}_{\rm min}| (2\tau_2)^{3/2} } }
                     \int d {\psi}_0 d\bar\psi_0  \psi_0\bar\psi_0
        \int [d \tilde \Phi]
                    e^{-S(\tilde\Phi)}
\ .$$
Comparing with the operator formalism, one finds that the integral over
the fermionic zero mode should be normalized
as $\int d\psi_0~\psi_0=\sqrt{2\pi}$.
Therefore,  we find that
$$Z_L(A, \tau, \lambda) =-  {\lambda \bar{\lambda}
             \over |\beta_{\rm min}| A 2\pi (2\tau_2)^{3/2}}
\ .$$
We have soaked up the Liouville fermion zero modes. Also,
since the Witten index of the super Liouville
theory is equal to zero, the Liouville path integral is
proportional to $\lambda \bar \lambda$.

Now we couple the super-Liouville path integral
to the $\hat c=1$ circle theory.
To soak up the remaining zero modes, we insert the operator \dilaton.
In the resulting path integral,
all the non-zero modes cancel, and the result is
\eqn\onepoint{
\langle D\rangle \sim \int d \lambda d\bar\lambda \lambda \bar\lambda
{R \over \sqrt{\ap}}{1\over 4\pi\sqrt 2 A}
\int_{\cal F} {d^2\tau\over \tau_2^2}
\sum_{n, m}\exp\left(-{\pi R^2|n-m\tau|^2\over\ap\tau_2}\right)
}
Performing the integral over supermoduli space,
and integrating over the area, we
find
\eqn\newpoint{
\langle D\rangle =R^2 {\partial {\cal Z}_{\rm odd}\over \partial R^2} \sim
- {1\over 12\sqrt 2} \ln |\mu|
\left ( {R\over \sqrt{\ap}} + {\sqrt{\ap}\over R} \right )
}
It follows that ${\cal Z}_{\rm odd}$ is proportional to
\odd, as we conjectured.%
\foot{There could be an additive constant in ${\cal Z}_{\rm odd}$
since we are evaluating its derivative.   However, it can be
argued away because, from the perspective of the effective target
space dynamics, the free energy should have the form ${\cal F}(T)
= a + b T^2$; the additive constant would correspond to a term
linear in T.}


\appendix{B}{Density of states}

In this appendix we compute directly the density of states. The
approach is similar to methods developed in 
\refs{\GrossAY,\KazakovUE,\DKR}. The inverted
harmonic oscillator potentials that we consider throughout this
paper have continuous spectra. The density of states has,
therefore, an infinite volume factor and a finite piece.  We are
interested in the finite piece, given by \eqn\densi{ \rho(e) = { 1
\over 2 \pi} { \partial \delta(e) \over \partial e} } where
$\delta$ is the phase shift for the scattering of a wave from the
potential.

We now proceed to compute  $\delta$ for the cases of interest.
The wave equation we need to solve is
\eqn\waveeq{
 - { 1 \over \lambda^{d-1} } \partial_\lambda ( \lambda^{d-1} \partial_\lambda
\psi )  - { \lambda^2 \over 2 \alpha' }  =2 e \psi
}
where $d=1$ for 0B and $d = 2 + 2 q$ for 0A.

Redefine  variables
$ \lambda = ( {\alpha' \over 2} )^{1/4} x $
and $ a = -  \sqrt{ 2\alpha'}\, e$. We then get the
equation
\eqn\equint{
 { 1 \over x^{d-1}} \partial_x ( x^{d-1} \partial_x \psi) +
{ x^2 \over 4} \psi = a \psi
}
Writing $z = i x^2/2$ and $\psi =z^{-d/4} M $,
one finds that $M$ obeys the equation
\eqn\mequ{
\partial_z^2 M + \left( - {1 \over 4} + { i{a \over 2} \over z} +
{ {1\over 4} - \kappa^2  \over z^2 } \right) M =0
}
where
\eqn\valuemu{
\kappa = {1 \over 2} ({ d\over 2 } - 1 )
} ($\kappa=-\frac14$ for type 0B, $\kappa = q/2$ for type 0A). Its
solutions are the Whittaker functions $M_{i {a \over
2},\kappa}(z)$, $M_{i {a \over 2},-\kappa}(z)$. The phase shift
for the Whittaker function is given by \eqn\phaseshi{ e^{i
\delta(a , \kappa) } = { \Gamma( \kappa + { 1 \over 2}- i { a
\over 2} ) \over  \Gamma( \kappa + { 1 \over 2} + i { a \over 2}
)} } where we have neglected possible $a$ independent terms in the
phase shift.

In the 0B case, the two solutions with $ \kappa = \pm 1/4$
correspond to even and odd wave functions. To get the total
density of states we should sum the contributions from both
solutions. Using gamma functions identities this gives \eqn\zerob{
e^{i \delta_+ + i \delta_-} = { \Gamma( {1\over 4} + { 1 \over 2}-
i { a \over 2} ) \over  \Gamma( {1\over 4}+ { 1 \over 2} + i { a
\over 2} )} { \Gamma( -{1\over 4} + { 1 \over 2}- i { a \over 2} )
\over \Gamma( -{1\over 4}+ { 1 \over 2} + i { a \over 2} )} \sim {
\Gamma( { 1 \over 2}- i { a } ) \over  \Gamma(  { 1 \over 2} + i {
a } )} } where in the last equality we neglected a term linear in
$a$ in the phase, which contributes just a constant to the density
of states. The individual phase shifts $\delta_\pm$ for even and
odd wave functions were computed in \KazakovUE.

In the 0A case we find \eqn\zeroa{ e^{i \delta} ={ \Gamma( {q\over
2} + { 1 \over 2}- i { a \over 2} ) \over  \Gamma( {q\over 2}+ { 1
\over 2} + i { a \over 2} )} }
Note that the 0A result for $q=0$
is the same as the 0B result \zerob\ up to $a \to a/2$, which is a
result we derived in an alternative way in the main text.

It is convenient to write the density of states as
\eqn\densityof{
{ 1\over 2 \pi} \partial_a \delta
= { 1 \over 4 \pi} \int_0^\infty dt \,  \sin { a \over 2} t \, { t/2
\over \sinh t/2} \,  e^{ - q t/2}
}
for the 0A case.

After remembering the definitions of $a$ and putting in the
thermal density factors we can write the expressions for the
density of states given in the main text. Remembering that \densityof\
is an exact formula, we conclude that the formulae in
the text are exact.


\appendix{C}{Normalization of disk one-point functions}

Let us review the bosonic string case first.  In \KlebanovKM\ it
has been found that the emitted state has the form $ |\Psi \rangle
= e^{ i \alpha \varphi_L(0)} | 0 \rangle $, where $\varphi_L$ is
the left moving part of a scalar field with the usual CFT
normalization ($ S = { 1 \over 8 \pi } \int ( \partial \varphi )^2
$). Our goal is to determine the value of the coefficient
$\alpha$.

The norm of the emitted state is
\eqn\norm{
\langle \psi | \psi \rangle
 = e^{  \alpha^2   \langle \varphi_L(0^+) \varphi_L(0^-) \rangle }
 = e^{\alpha^2 \log(1/\epsilon),}
 }
where $\epsilon$ is a short distance cutoff arising from
$\varphi_L(x) \varphi_L(0) \sim - \log x $.

It was shown in \llm\ that the log of the norm of the emitted
state is equal to the one loop partition function for an array of
D-instantons, where strings connect only instantons at positive
Euclidean time with instantons at negative Euclidean time. The
divergence comes from the D instantons closest to $t=0$. These are
separated by a critical distance so that the stretched open
strings are massless. This partition function is IR divergent in
the open string channel:
 \eqn\partf{ log( \langle \psi | \psi \rangle ) = Z_1 = 2 \int
 { d t \over 2 t } Tr e^{ - 2 \pi L_0} = \int { d t \over t}
 ( e^{ 2 \pi t} -1 ) e^{- 2 \pi t}   \sim   \log({1\over
 \epsilon}) \ , }
where the first factor of 2 comes from the two orientations of the
stretched open strings. Note that in the open string string
channel the cutoff is $t \roughly< 1/\epsilon$ since it arises
from separating the D-instantons by an extra amount $\epsilon$ in
Euclidean time.\foot{ So the total separation is $d = d_c +
\epsilon$, where $d_c$ is the critical distance. Then the exponent
$ - 2 \pi t$ in
\partf\ becomes $ 2 \pi t d^2/d_c^2 \sim 2 \pi (1 + 2 \epsilon/d_c )$.
Then the integral over $t$ produces $\log(1/\epsilon)$. Note that
even in 26 dimensions  the one loop partition function for two
D-instantons has a short distance singularity like
$\log(1/\epsilon)$. This suggests that maybe even in 26 dimensions
we could think of them as fermions in some sense.} Equating the
divergence from the two points of view we get
 \eqn\divcoef{\alpha^2 = 1. }

Now let us consider the type 0 string. We have shown that the
final state has the form \eqn\answerform{
 |\Psi_\pm \rangle =
e^{ i ( \alpha_t \varphi_t  \pm \alpha_v \varphi_v ) } | 0 \rangle
\ , } where $\pm$ refers to the sign of $\lambda$, and
$\varphi_{t,v}$ are the canonically normalized tachyon and RR
scalar fields. We are going to determine $\alpha_{t,v}$.

Let us start with an unstable D0 brane of 0B theory. An $SU(2)$
rotation on the boundary state yields a D($-1$) brane and and anti
D($-1$) brane separated by a critical distance. This configuration
is again related to the computation of the norm of the state. More
precisely, let us consider the inner product of the states
$\langle \Psi_+ | \Psi_- \rangle $. This is  computed by
considering two D($-1$) branes of the same charge. This does not
give a divergence from the string theory point of view. (Note that
this is a result valid for tachyon decay in 10 dimensions also).
On the other hand, from \answerform\ the coefficient of the
divergence is  of the form $ \alpha_t^2 - \alpha_v^2 $. This
implies that $\alpha_t = \alpha_v$ (the sign is already taken into
account in \answerform).

Now let us compute the norm of  $ \langle \Psi_+ | \Psi_+ \rangle
$. Using the expression in terms of free fields in \answerform\ we
find that it diverges with a coefficient equal  to ${ \alpha_t^2 +
\alpha_v^2 } $. In string theory this divergence appears when a
D($-1$) and an anti D($-1$) brane are separated by a critical
distance. This divergence has the same form as in
\partf , since it comes from the ground state of open strings stretching
between the branes (which is massless at this distance). We
conclude that \eqn\final{ \alpha_t = \alpha_v = { 1 \over \sqrt{2}
} \ . }

Note that this computation also gives the coupling of the
D-instanton to the Ramond-Ramond scalar. If we define the
compactification radius to be the minimum allowed by the coupling
to the D-instantons then we see that the scalar is at the
self-dual radius. In other words, we are finding that the
D-instantons contribute with a phase $e^{ \pm i C/\sqrt{2} } $.
This radius agrees with the claim that the zero mode of the RR
scalar corresponds to the relative phase of the left and right
wavefunctions.

The above computation applies to type 0B theory. If we consider
the 0A, then we find that in order to compute the norm we need to
consider two unstable D($-1$) branes that are separated in the
Euclidean time direction by a critical distance. Again we obtain
the same divergence
\partf . On the other hand, here only one field is present,
so the matching is the same as in the bosonic theory.

\listrefs
\bye